\documentclass[a4paper,fleqn,usenatbib]{mnras}

\usepackage[T1]{fontenc}
\usepackage{ae,aecompl}

\usepackage{graphicx}	
\usepackage{amsmath}	
\usepackage{amssymb}	
\usepackage{caption}
\usepackage[caption=false]{subfig}
\usepackage[flushleft]{threeparttable}

\newcommand{\kepler}{{\it Kepler\ }}

\title[Long period rotators]{Long rotation period main-sequence stars from \kepler SAP light curves}

\author[Kaiming Cui et al.]{
Kaiming Cui,$^{1,2}$\thanks{E-mail: ckm@nao.cas.cn}
Jifeng Liu,$^{1,2,3}$
Shuhong Yang,$^{4,2}$
Qing Gao,$ ^{1} $
Huiqin Yang,$ ^{1} $
\newauthor
Roberto Soria,$ ^{2,1} $
Lin He,$ ^{1,5} $
Song Wang,$ ^{1} $
Yu Bai,$ ^{1} $
and Fan Yang$ ^{1,2} $
\\
$^{1}$Key Laboratory of Optical Astronomy, National Astronomical Observatories, Chinese Academy of Sciences, Beijing 100101, China\\
$^{2}$School of Astronomy and Space Sciences, University of Chinese Academy of Sciences, Beijing 100049, China\\
$^{3}$WHU-NAOC Joint Center for Astronomy, Wuhan University, Wuhan, Hubei 430072, China\\
$^{4}$Key Laboratory of Solar Activity, National Astronomical Observatories, Chinese Academy of Sciences, Beijing 100101, China\\
$^{5}$School of Astronomy and Space Science, Nanjing University, Nanjing 210093, China
}

\date{Accepted XXX. Received YYY; in original form ZZZ}

\pubyear{2019}

\begin{document}
\label{firstpage}
\pagerange{\pageref{firstpage}--\pageref{lastpage}}
  \maketitle

\begin{abstract}
Stellar rotation plays a key role in stellar activity. 
The rotation period could be detected through light curve variations caused by starspots. 
\kepler provides two types of light curves, one is the Pre-search Data Conditioning (PDC) light curves, the other is the Simple Aperture Photometer (SAP) light curves. 
Compared with the PDC light curves, the SAP light curves keep the long-term trend, relatively suitable for searches of long period signals. 
However, SAP data are inflicted by some artefacts such as quarterly rolls and instrumental errors, making it difficult to find the physical periods in the SAP light curves. 
We explore a systematic approach based on the light curve pre-processing, period detection and candidate selection. 
We also develop a simulated light curve test to estimate our detection limits for the SAP-like LCs. 
After applying our method to the raw SAP light curves, we found more than 1000 main-sequence stars with the period longer than 30 days, 165 are newly discovered. 
Considering the potential flaw of the SAP, we also inspect the newly found objects with photometry methods, and most of our periodical signals are confirmed.

\end{abstract}

\begin{keywords}
stars: rotation  --- starspots --- stars: statistics
\end{keywords}



\section{Introduction} \label{sec:intro}
Stellar rotation is a fundamental factor that affects stellar evolution and stellar activities. \citep[e.g.,][]{Kraft1967, Zahn1992, 1997ARA&A..35..557P} The stellar age and the magnetic activity caused by dynamo processes also depend on rotation. \citet{Wilson1963}, \citet{Wilson1964} and \citet{Skumanich1972} found the relationship among rotation, chromospheric activity and age. 
\citet{Barnes2003} and \citet{Barnes2007} put forward
the gyrochronology, an empirical formulation among the rotation period, colour and age, is often used as a method to estimate the age of a star, because the rotation rate slows down with age, like what is believed to happen on the Sun.
Subsequent studies extended the relationship to the evolved stars \citep{Barnes2010} and older cluster stars \citep{Meibom2015}, but \citet{Angus2015} and \citet{Saders2016} also find some abnormal \kepler field stars are inconsistent with the formulation.

The relationship between the rotation period and many activity indicators have been extensively studied. 
\citet{Pallavicini1981} first found that the X-ray luminosity is highly correlated to rotational velocity for cool stars. 
\citet{Noyes1984} studied the relationship between the rotation period and averaged Ca II H\&K emission ratio. Similar work, which was applied on X-ray brightness of different late-type stars, got similar results --- stars with lower Rossby number (i.e., the ratio of the rotation period and the convective turnover time) have more extensive corona and stronger X-ray emission \citep{Wright2011}. \citet{Reiners2014} extend the Rossby number to the combination of the rotation period and stellar radius. With the development of photometry,  \citet{Strassmeier2009}, \citet{Affer2009}, \citet{Basri2010} and \citet{Basri2011} compared stellar activity with the solar activity using photometric variability. M dwarfs are usually more active than earlier type stars, thus their relationship between activity proxy and rotation rate has been well-studied \citep[e.g.,][]{Wright2016, Yang2017, Wright2018}. 


Detecting the rotational broadening of spectra lines is a classical method to measure the rotation rate \citep[e.g.,][]{Fekel1997, Kaler}. The time series of photometry could also reveal the rotation period \citep[e.g.,][]{Alekseev1998, Norton2007, Hartman2010, Irwin2011, Newton2016}.
However, old stars have low-amplitude and slow surface movement, and it is difficult to detect their rotation periods through spectroscopy or ground-based photometry.

High-quality photometric data obtained by \kepler mission \citep{Borucki2010, Koch2010} can be used to study the surface rotation period and magnetic activity of dwarfs directly, especially for slow rotators. In particular, the solar-type stars have a similar physical condition with the Sun, which is studied by helioseismology \citep{Christensen-Dalsgaard2010}.
The power spectrum of a light curve (hereafter LC) contains information about the stellar interior and the surface rotation for a specific star, therefore the asteroseismology could also constrain the stellar rotation period \citep{Gizon2002}.

Many rotators have been found from \kepler data. For example, \citet{Nielsen2013} measured nearly 12,000 periods from Q2--Q9 data, and \citet{Reinhold2013} found ~24,000 periods using Q3. They both used the Lomb-Scargle methods to derive the periods of stars. Moreover, \cite{McQuillan2013, McQuillan2014} built an automated version of autocorrelation function (ACF) method and the detected rotation periods that range from 4 hours to 69 days in 34,000 main sequence stars in the \kepler field. \citet{Reinhold2015} re-analysed their previously studied results based on different quarters and different approaches. \citet{Garcia2014} used both of Pre-search Data Conditioning (PDC) and the Simple Aperture Photometer (SAP) data to look for rotators from pulsating solar-like stars and analysed their activities. They found 321 stars with surface modulation periods.

\begin{figure*}
	\includegraphics[width=0.9\textwidth]{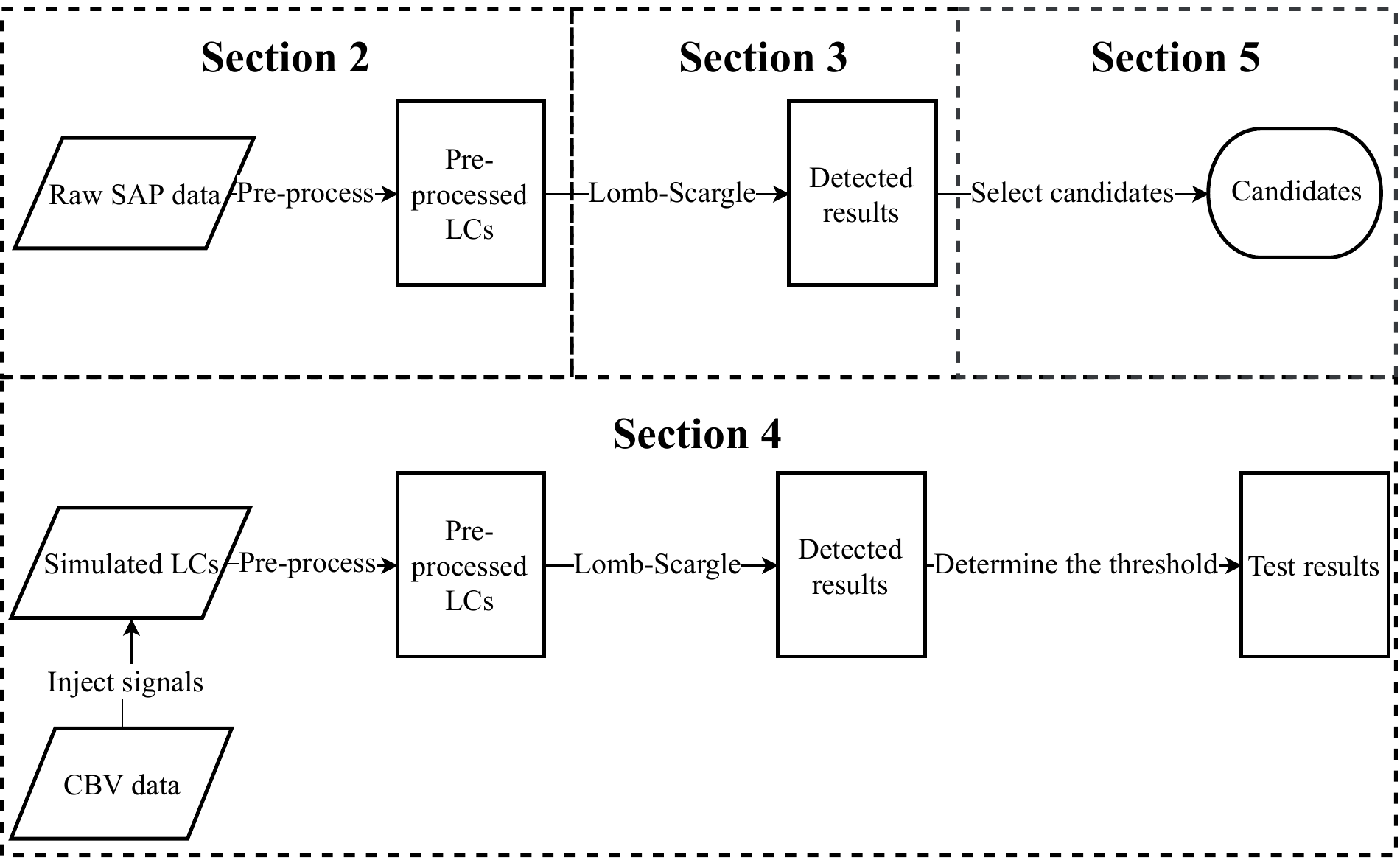}
	\caption{General flowchart of the Section \ref{sec:lc_pre}--\ref{sec:can_sele}. The dashed square areas indicate different sections. \label{fig: flowchart_whole}}
\end{figure*}
In this paper, we describe our period detection and selection work with the SAP data, and inspect the candidates to obtain the reliable results. In Section \ref{sec:lc_pre}, we give an overview of the characteristics of the SAP data and pre-process the SAP LCs to avoid some instrumental trends.
Then we build our systematic period detection in Section \ref{sec:detection}. To estimate the completeness and reliability of our method, we also develop a simulated data test in Section \ref{sec:mock_test}. Then we applied our method to the real data and select candidates in Section \ref{sec:can_sele}. 
Since looking for long rotation period in the SAP LCs is complicated, we plot a general flowchart (see Figure \ref{fig: flowchart_whole}) to illustrate our work flow for the above sections.
Then we summarise our work in Section \ref{sec:summary}. 

\begin{figure*}
	\includegraphics[width=\columnwidth]{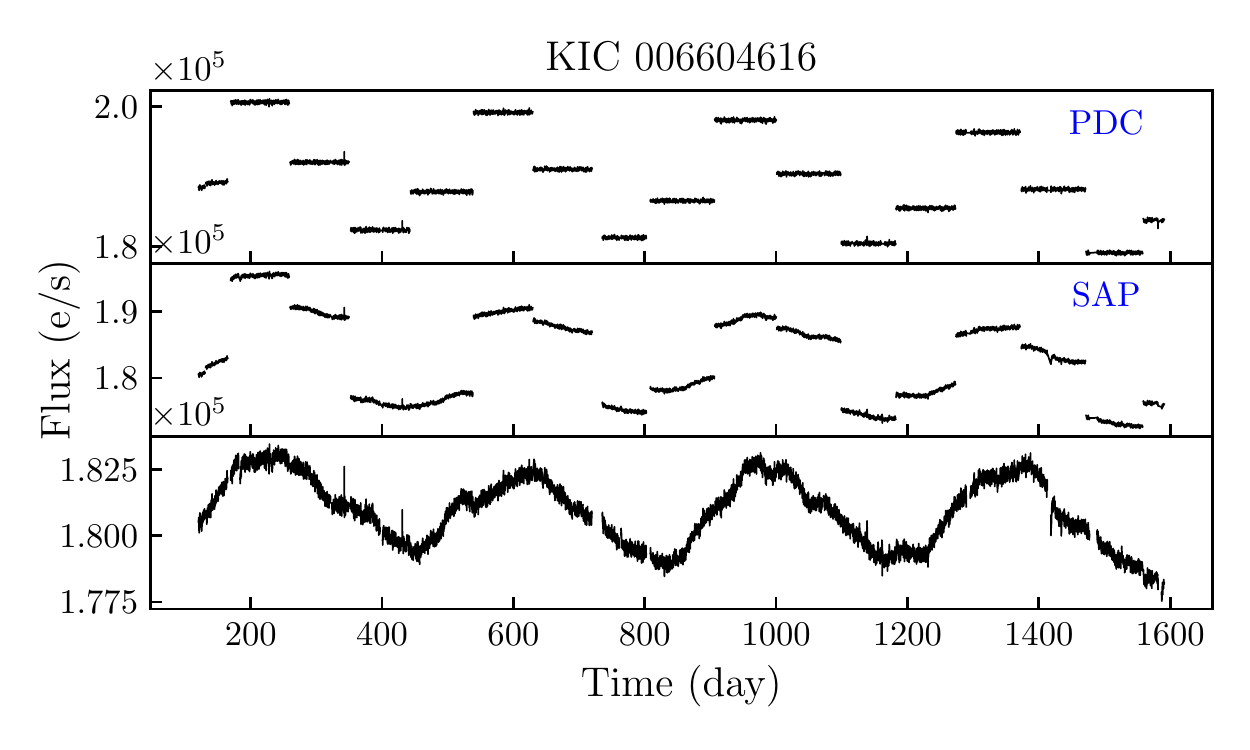}
	\includegraphics[width=\columnwidth]{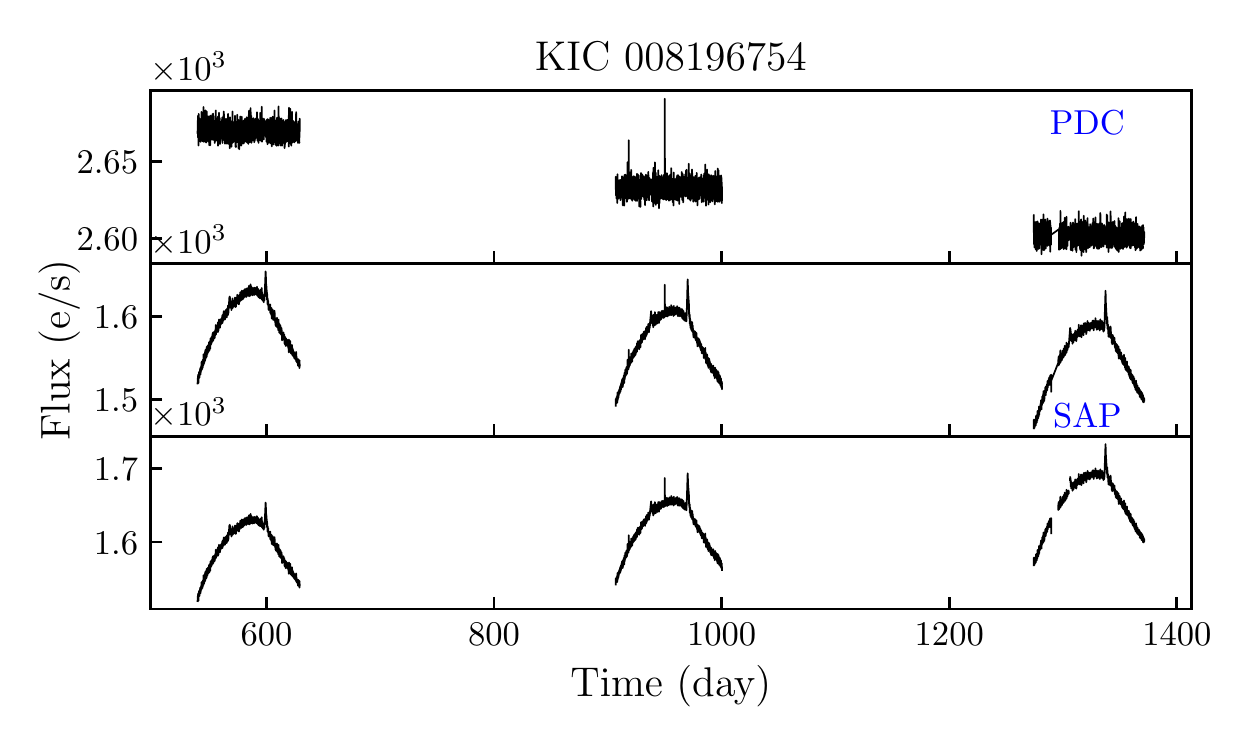}
	\includegraphics[width=\columnwidth]{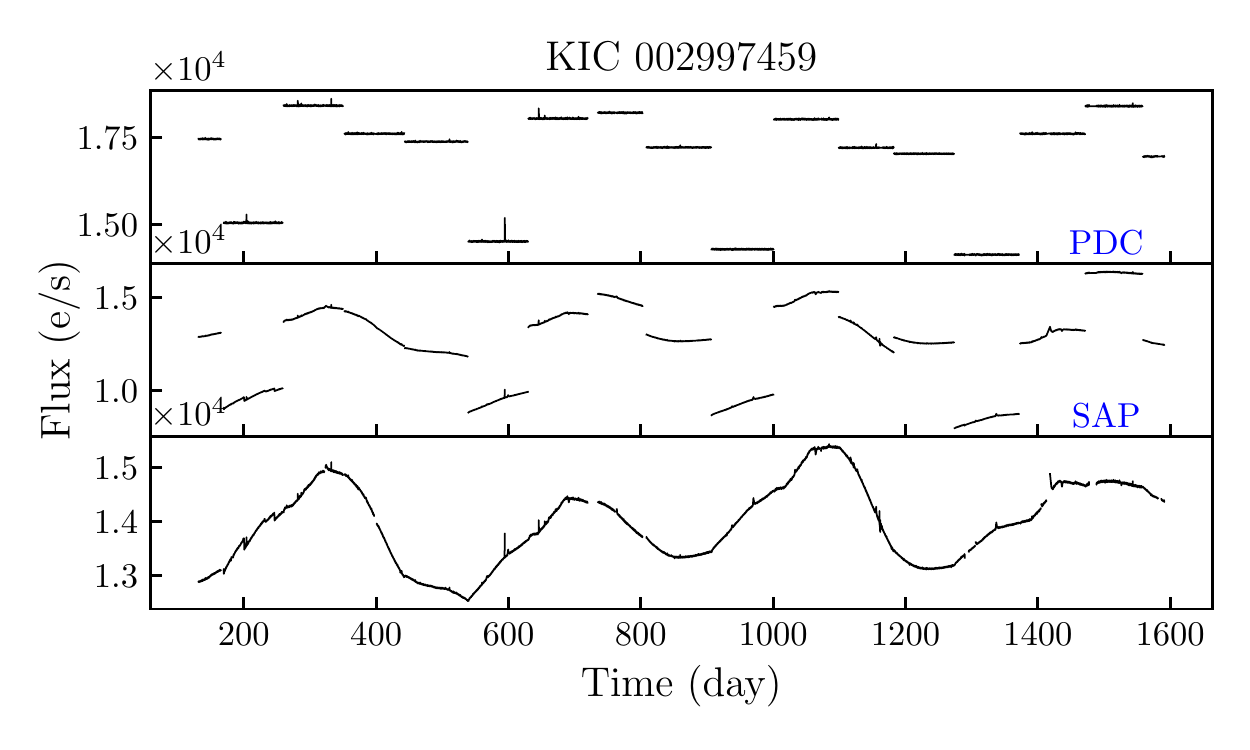}		\includegraphics[width=\columnwidth]{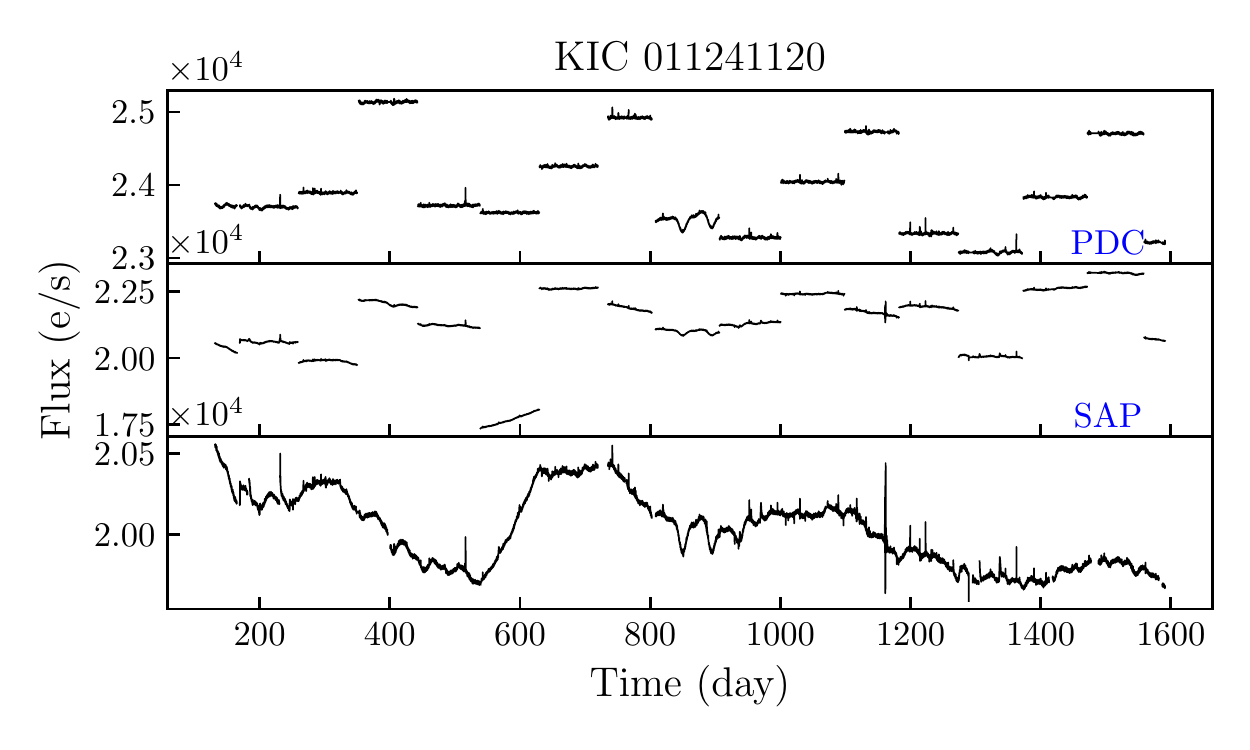}
	\caption{Four typical SAP LCs with long-term trends. In every sub-figure, from top to bottom: PDC LC, SAP LC and concatenated SAP LC. \label{fig: sap-examples}}
\end{figure*}

\section{Light Curve Preparation and Pre-Processing}\label{sec:lc_pre}
\begin{figure*}
	\includegraphics[width=\textwidth]{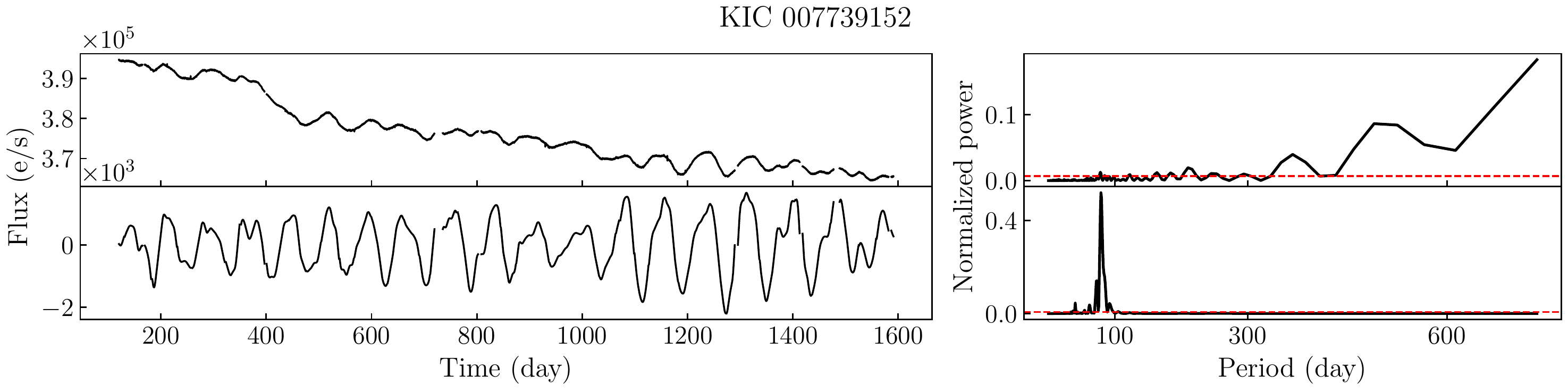}\\
	\includegraphics[width=\textwidth]{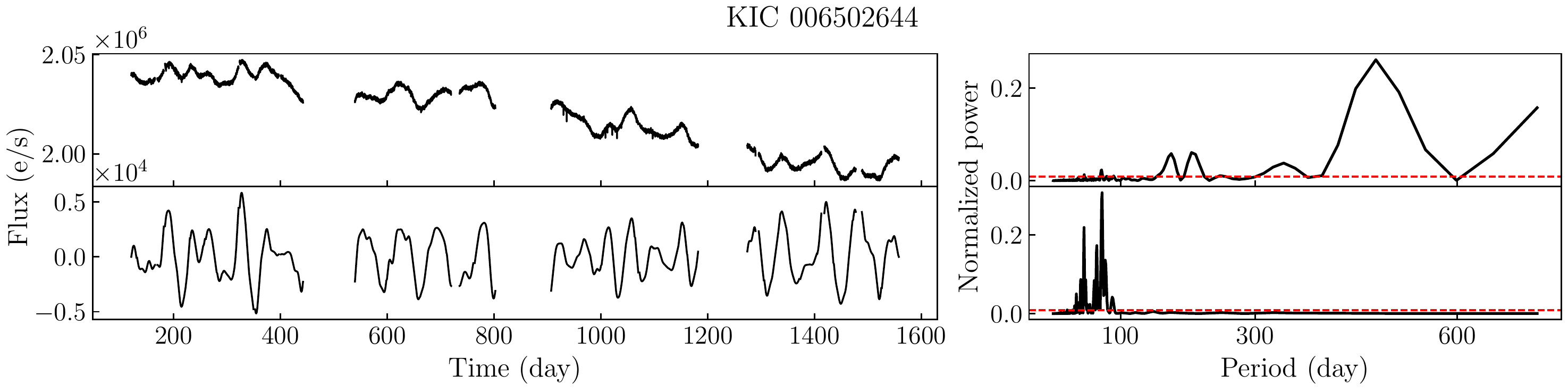}\\
	\includegraphics[width=\textwidth]{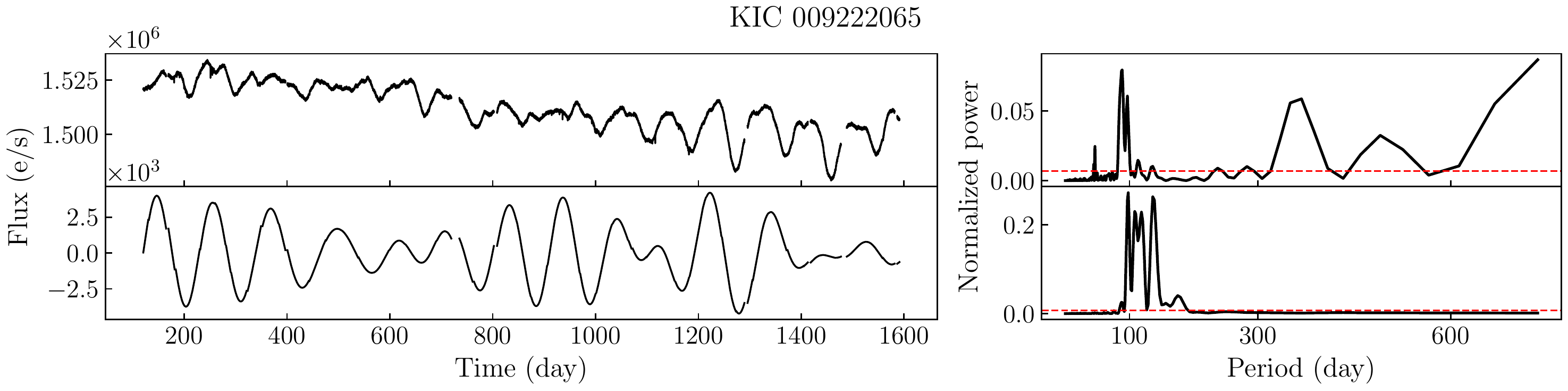}\\
	\includegraphics[width=\textwidth]{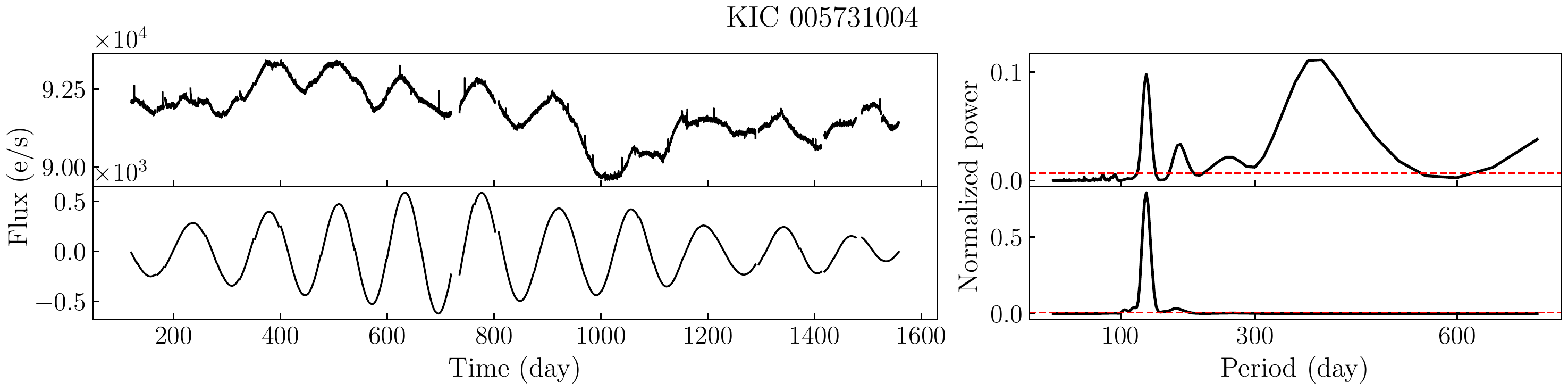}

	\caption{Four example figures of long period objects. The top two sub-figures use the first filter, the bottom tow sub-figures use the second filter. In every sub-figure, the upper panel shows the concatenated SAP LC, and the bottom panel shows the band-pass filtered LC. The Lomb-Scargle periodograms are on the right side correspondingly. The dashed red line is the significance threshold given by letting FAP (False Alarm Probability) equals to $ 10^{-4} $. \label{fig: example-lc}}
\end{figure*}

\kepler released two types of LCs. One is the SAP 
LCs, which are computed by summing the readout value of cosmic ray-cleaned, background-subtracted pixels within the optimal photometric aperture \citep{jenkins2017kepler}. The SAP LCs retain the long-term trends, which need to be adjusted by removing systematic errors (e.g., quarterly rolls and unstable amplifier electronics) manually \citep{Revalski2014}. 
The other type of LC production is PDC, which are generated from the SAP data by the \kepler Science Operations Centre with complicated methods to correct the effect of systematic errors. However, previous studies have reported that some astrophysical signals were removed, but some instrumental trends are still visible. \citep[e.g.,][]{Balona2013, InstrumentVanCleve2016}. Moreover, on some occasions, PDC LCs keep the long-term varieties partially, but the signals may be incomplete (Figure \ref{fig: example-lc} shows some examples). Therefore, it is impossible to study the evolution of starspots for long period rotators without SAP LCs. Some examples of the SAP and PDC LCs are plotted in Figure \ref{fig: sap-examples}. Given the above comparisons, we choose the long cadence SAP LCs in \kepler Mission Data Release 25 (DR25) to search for more examples of long period signals. These data are downloaded from the \kepler public archive \footnote{\url{http://archive.stsci.edu/kepler}}. 

In the early work on PDC data processing, the Cotrending Basis Vector (CBV) data were generated by Singular Value Decomposition of 50 per cent most correlated stars for each channel \citep{VanCleve2016}. Some previous work \citep[e.g.,][]{Roettenbacher2013, Aigrain2017} used it to inspect the systematic errors.
Although the systematic trends extracted by CBV data are incomplete, we use it as a supplementary measurement to generate some simulated data and check the artefacts in our work (discussed in Section \ref{sec:mock_test} and \ref{sec:can_sele}).

\begin{figure}
    \centering
	\includegraphics[width=\columnwidth]{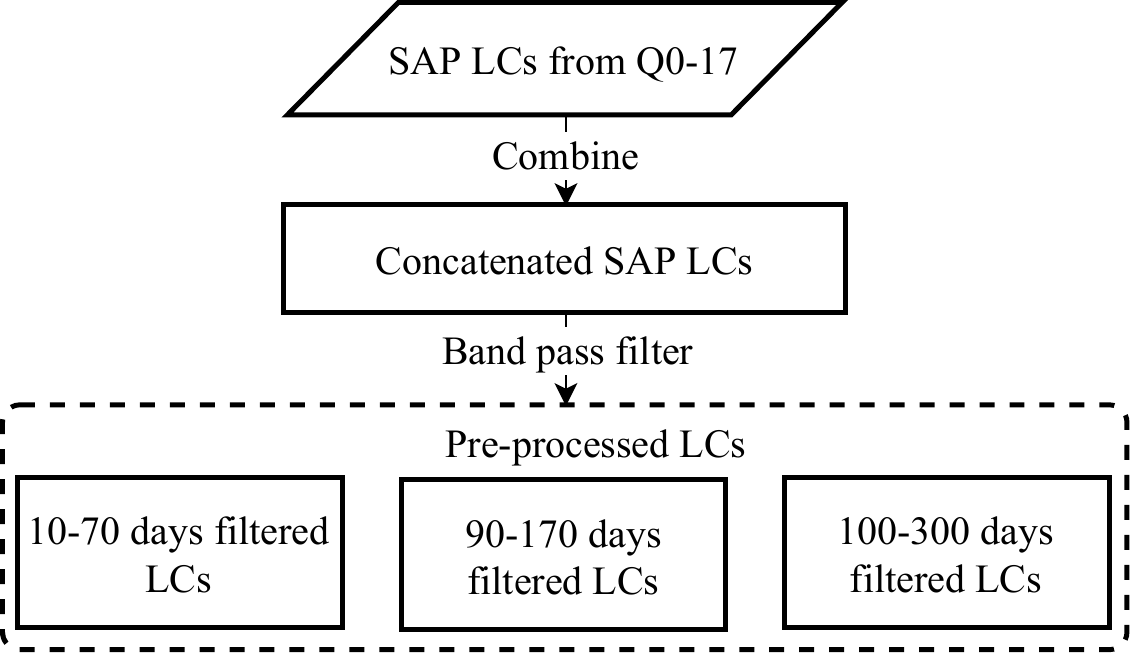}
	\caption{Detailed flowchart of the LC preparation and pre-processing. \label{fig: LC_pre}}
\end{figure}

We develop a pre-processing work-flow for every SAP LC, and the flowchart is shown in Figure \ref{fig: LC_pre}. Firstly, to increase the baseline (time span between start and end) to improve the frequency resolution, we use the whole database from quarter 0--17 (Q0--Q17). This includes Q0, Q1, and Q17, which are short and usually excluded from some studies \citep[e.g.,][]{Nielsen2013, McQuillan2014}. 
In order to obtain a complete LC with long baseline, we need a combination for different quarters. However, as shown in Figure \ref{fig: sap-examples}, the median flux of the SAP vary quarterly. Therefore, we directly combine the SAP LCs of different quarters from end to end in chronological order and shift the flux of every quarter to make sure there is no flux jump between neighbouring quarters. 

Although the thermal variations at the beginning of a quarter often cause some ramps of the flux, leading to some flux jumps or trends. However, if a flux jump appears regularly, it would cause a 90-day period, which could be dropped by our filters (discussed later) and similar period selection process (see Section \ref{sec:can_sele}). Even if it passes the filter and our selection process, we would remove it in our visually inspection because the turn points always at the start or end of the quarter. 
We also adjust the median flux of the concatenated SAP LCs so that it has the same median flux as the raw SAP LCs. Some examples of the concatenated SAP LCs are shown in Figure \ref{fig: sap-examples}. 

\begin{figure}
	\includegraphics[width=\columnwidth]{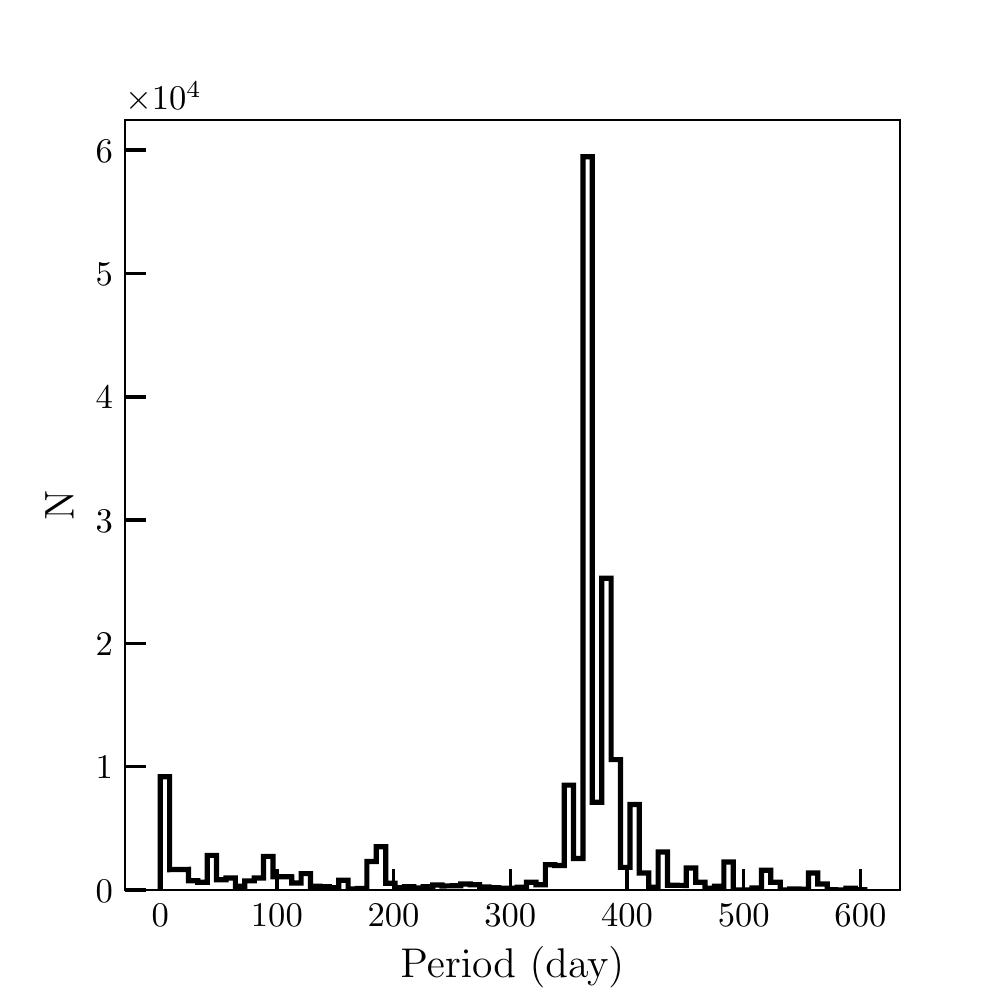}
	\caption{Distribution of period with maximum power. The period is cut off at 600 days. \label{fig: whole-distribution}}
\end{figure}

Since our LC combination is based on the raw SAP LCs, some systematic trends would affect the long-term period, the most serious one is the quarterly-rolling signal, often leading to a one-year-period and its harmonics, which could be clearly seen from Figure \ref{fig: sap-examples}. We also check those periods statistically, Figure \ref{fig: whole-distribution} shows the distribution of period with the maximum power of the whole \kepler concatenated SAP LCs through calculating the fast Lomb-Scargle directly. We could find some long period clusters on the histogram, especially around 90 days, 180 days and 360 days.

Therefore, in order to avoid these long-term systematic trends, we apply an iterative band-pass Butterworth filter \citep{butterworth1930theory} to pre-process the LCs. 
The Butterworth filter is a maximally flat magnitude filter, and it attenuates the specific frequency ranges without inducing ripples and other artefacts. This filter was often used to suppress noise \citep[e.g.,][]{Aasi2015, Collaboration2014} or obtain features \citep{Inceoglu2017} in the frequency domain. 

Naturally, according to period distribution in Figure \ref{fig: whole-distribution}, these three period clusters lead to three reasonable partitions in the periodogram. The first filter with cut-off frequency from 1/80 to 1/10 is used to pre-process the LCs for detecting period less than 90 days but higher than 30 days. A slightly higher frequency than 1/90 is chosen because of the wide range of the systematic effect. 1/10 cycle per day is chosen as the higher cut-off frequency because the high frequency signals like flares, transits or spikes caused by instrument should be ignored (an example is shown in the bottom right panel of Figure \ref{fig: example-lc}). The filter order is chosen as 30, using a higher or lower order does not vary much, but the cut-off frequency should also change so that the response is low enough to avoid systematic frequencies. We also compute the filter iteratively for 4 times to suppress the response around the cut-off frequency, because the systematic signals could be spread to a higher frequency. A larger iterative number would decline the response, but cost more computation time. 

\begin{figure*}
	\includegraphics[width=0.32\textwidth]{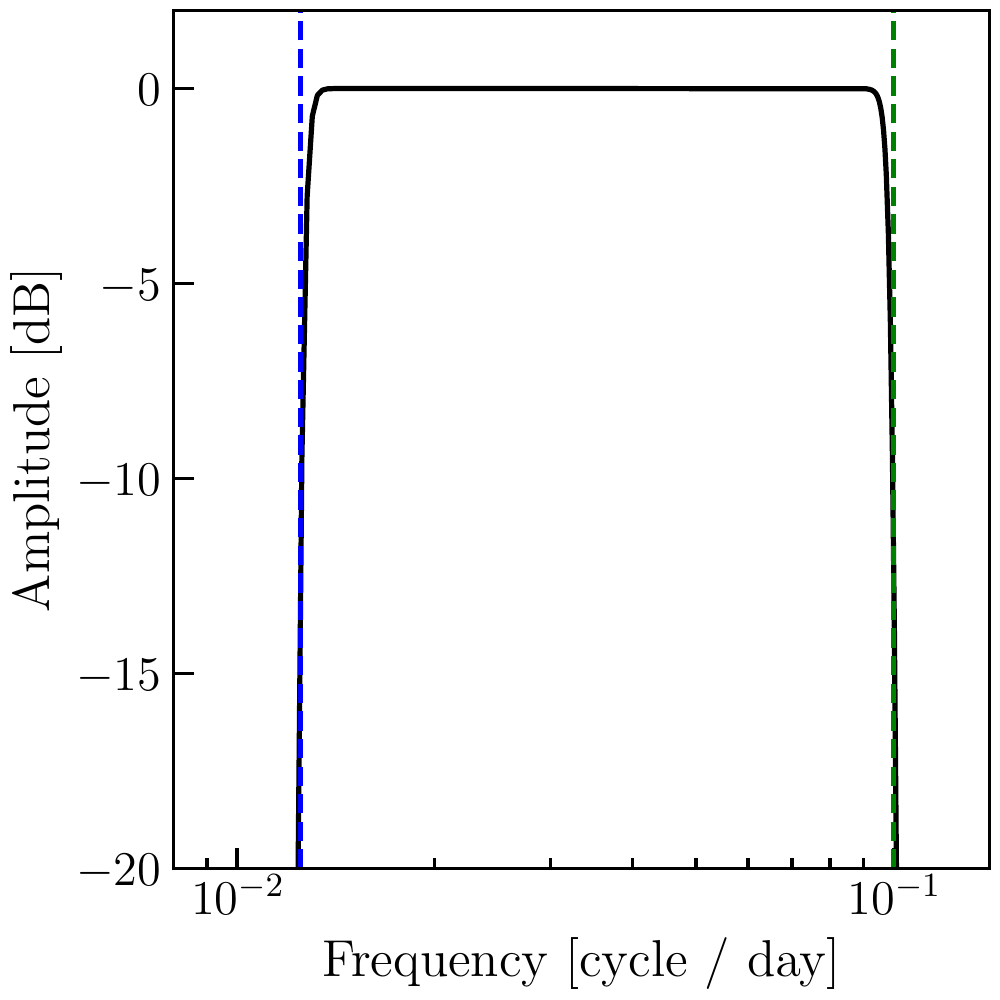}
	\includegraphics[width=0.32\textwidth]{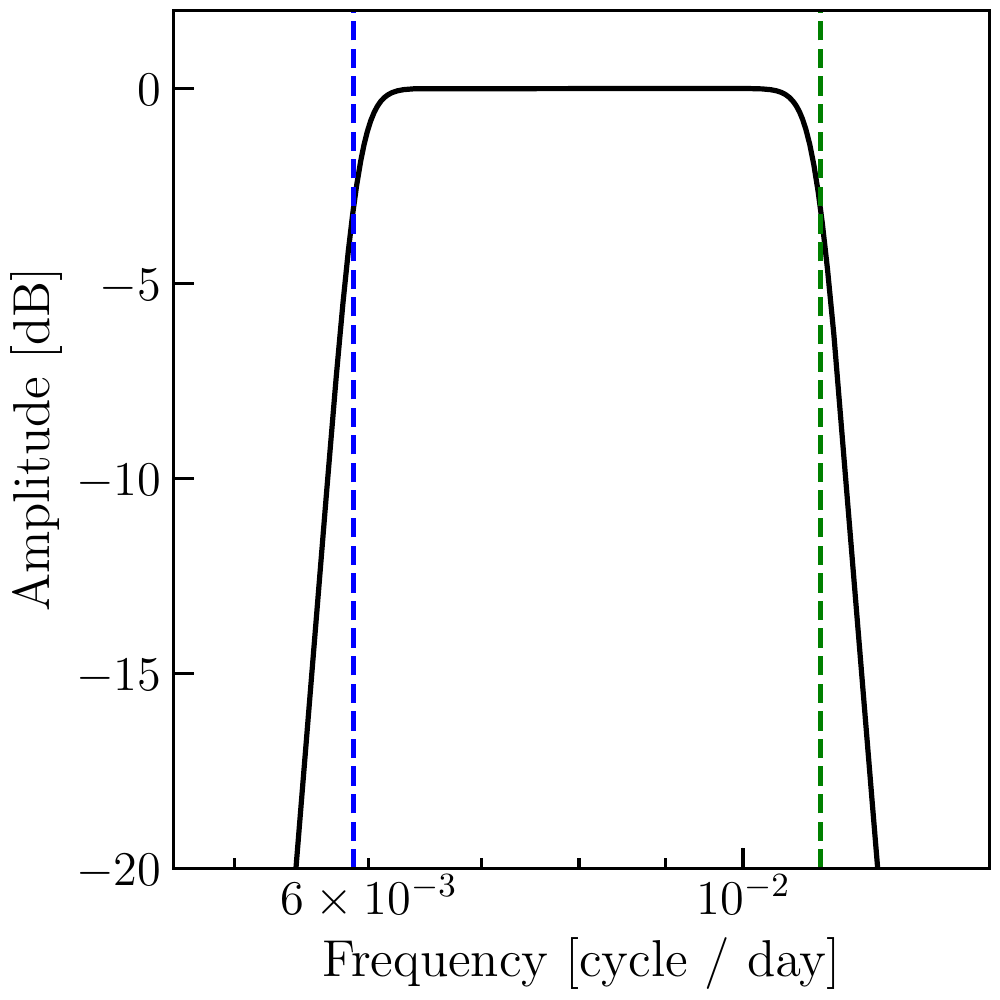}
	\includegraphics[width=0.32\textwidth]{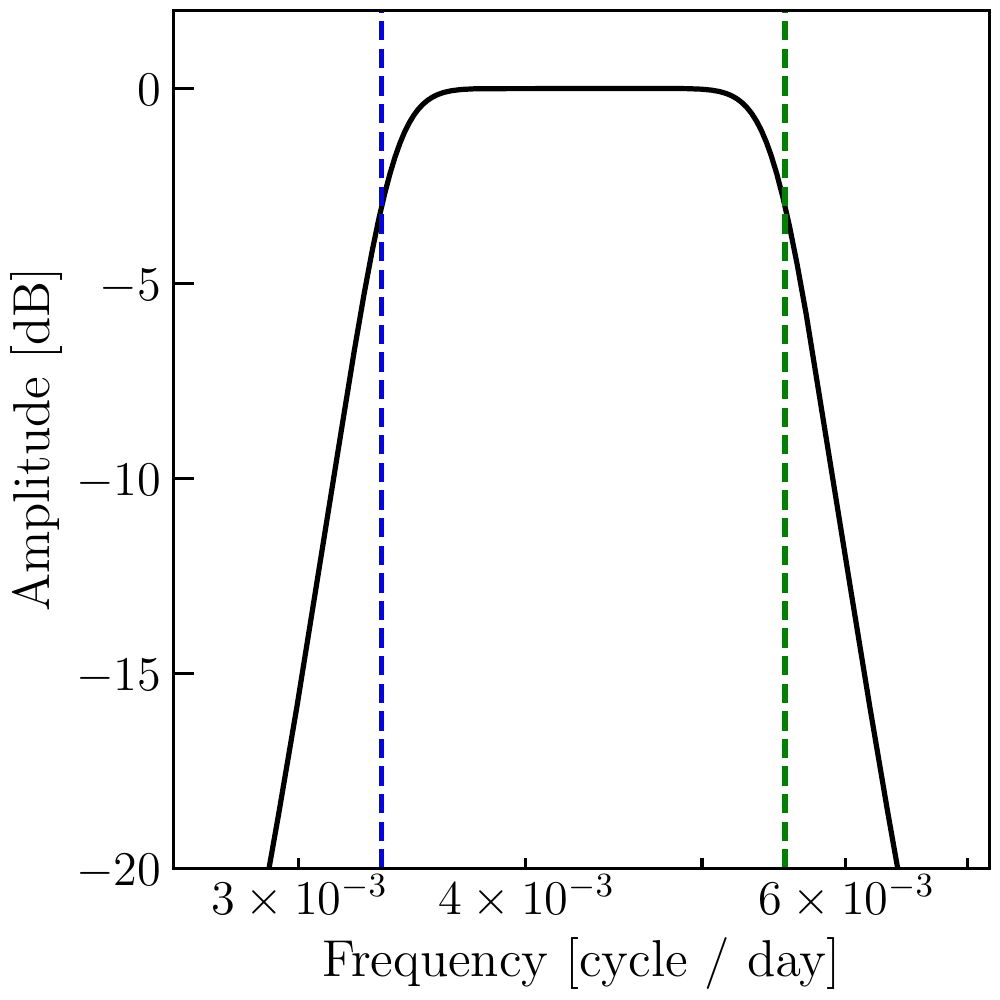}
	
	\caption{Three response curves of the Butterworth filter. From left to right, the band-pass cut-off frequency ranges are: 1/80 to 1/10 \textrm{c/d}, 1/170 to 1/90 \textrm{c/d} and 1/300 to 1/180 \textrm{c/d}. The lower cut-off frequencies are plotted in dashed green lines, and the higher cut-off frequencies are plotted in dashed blue lines. \label{fig: butter_filters}}
\end{figure*}

Then, we create another two filters with the cut-off frequency range from 1/170--1/90 and 1/300--1/180 cycle per day for longer period detection. These cut-off frequencies are chosen for the same reason as the first filter. Besides, the orders should be smaller because the high frequency signals are important for keeping the shape of a LC and the long-term trends dominate the LC, so we decrease the order to 10 and 5 for these two filters respectively. Figure \ref{fig: butter_filters} shows the response curves of the three band-pass filters, -20 dB means a signal is suppressed to its 1 per cent.

Thus, for every concatenated SAP LC, we apply those three band-pass filters to generate three types of filtered LCs. Some examples of the filtered LCs are given in Figure \ref{fig: example-lc}. We can see that the SAP LCs keep more long-term signals than PDC LCs, and most systematic trends are removed in our filtered LCs. This evidence could also be seen in the Lomb-Scargle periodogram, the filtered LCs have much cleaner periodogram. 

\section{Period Detection}\label{sec:detection}
Before we detect the periodic sources, another important factor is the effective time of observation, especially for slow rotators we are aiming at in this study. In principle, we believe that only when a periodic sinusoidal shape appears in the LC at least three times, can we confirm the periodicity, so it requires a minimum threshold of effective observation time when we select our sample. In our detection process, we set 270 (i.e., $ 90 \times 3 $), 540 (i.e., $ 180 \times 3 $) and 900 (i.e., $ 300 \times 3 $) days as the minimum effective observation time to be appropriate for those three types of the filtered LCs.
\begin{figure}
	\centering
	\includegraphics[width=0.9\columnwidth]{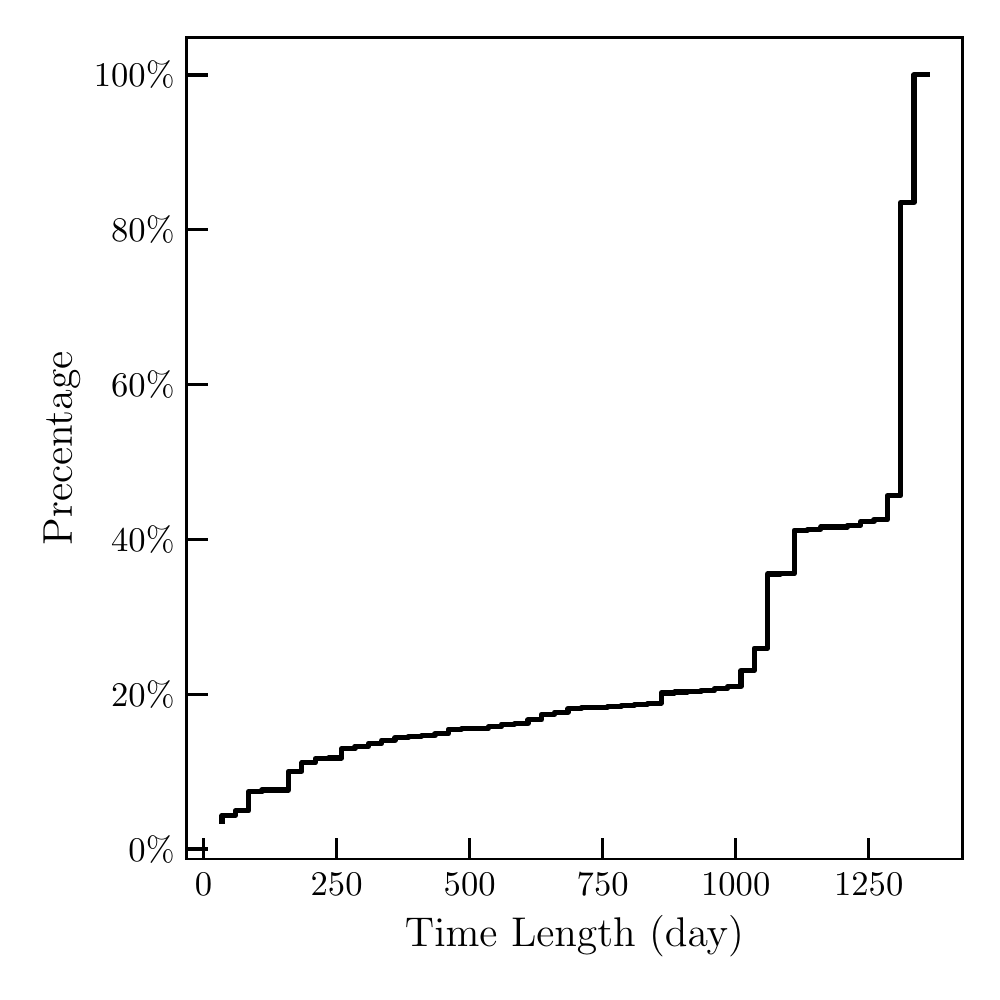}
	\caption{Cumulative distribution of effective observation time. \label{fig: time-distribution}}
\end{figure}
Figure \ref{fig: time-distribution} shows the cumulative distribution of the effective time of observations in the whole \kepler LC dataset. Roughly 80 per cent objects are observed for more than 1000 days; so, over 80 per cent \kepler SAP LCs are included in our sample. 

Then, we choose the fast Lomb-Scargle periodogram \citep{Press1989}, taking advantage of the fast calculation and rich information in its periodogram, providing a convenient result for the follow-up selection process. 
We also consider the false alarm probability (FAP) as a significant level and choose the probability threshold as $ 10^{-4} $. 
This threshold gives the probability that a signal with no periodic component would lead to a peak of this height \citep{VanderPlas2018}. 
Some examples of the SAP LCs and their periodograms with the significant levels are illustrated in Figure \ref{fig: example-lc}. 

After obtaining the periodogram of every filtered LC, we treat them slightly different. 
For every filtered LC with the cut-off frequency larger than 1/80, we extract two periods with prewhitening, which is most commonly used in asteroseismology.
The prewhitening method means iteratively subtract the fitted LCs from the filtered LCs and then calculate the Lomb-Scargle periodogram again.
However, for a filtered LC with cut-off frequency less than 1/90,  a weak rotational signal might mix with quarterly trends, only the most obvious periodic signal could be trusted. Therefore, we simply choose the most significant period as the period candidate.
The period detection error is set to 5 per cent, which is determined by our simulated LC test (described in Section \ref{subsec: threshold-determination}). 
In addition, we also fit those period candidates with the cosinoidal functions to estimate their amplitudes and phases.

\section{Simulated Data Test}\label{sec:mock_test}
Considering the difficulty of recovering astrophysical signals from the long-term artefacts dominated LCs, we need to be very cautious. 
Therefore, before we apply our detection approach to the pre-processed LCs, we should quantitatively justify our method.
Since our pre-processing and detection methods are designed for recovering periodical signals from the SAP-like long-term trends, we need a lot of trend-dominated mock LCs with known periods to test the ability of recovering.
To do so, we inject some periodic signals into the CBV LCs to produce our simulated LCs. 
After that, we detect the periodic signals in our simulated data by following the means described in Section \ref{sec:detection}, and then evaluate the completeness and reliability of our detection method. 

\begin{figure*}
    \includegraphics[width=0.32\textwidth]{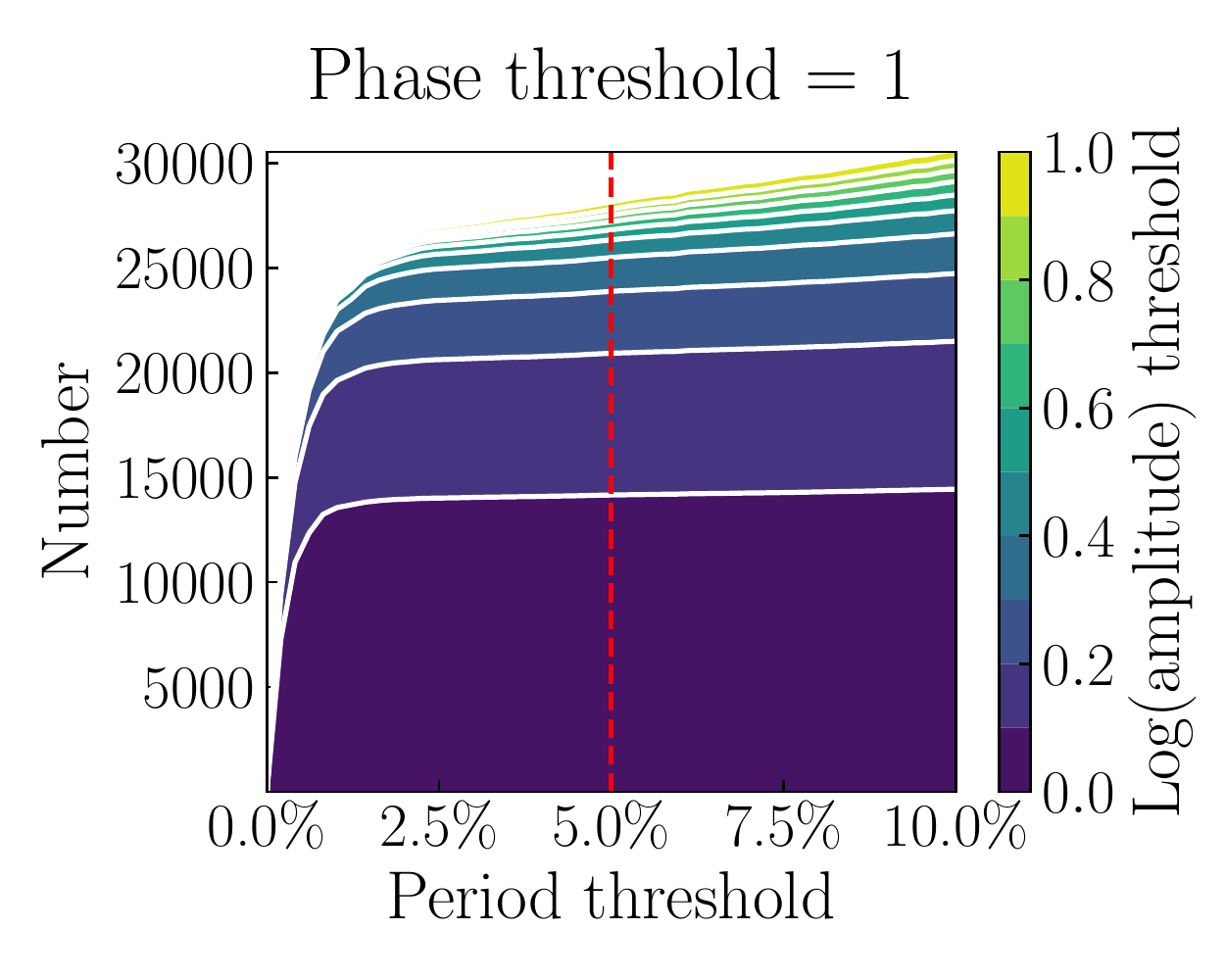}
    \includegraphics[width=0.32\textwidth]{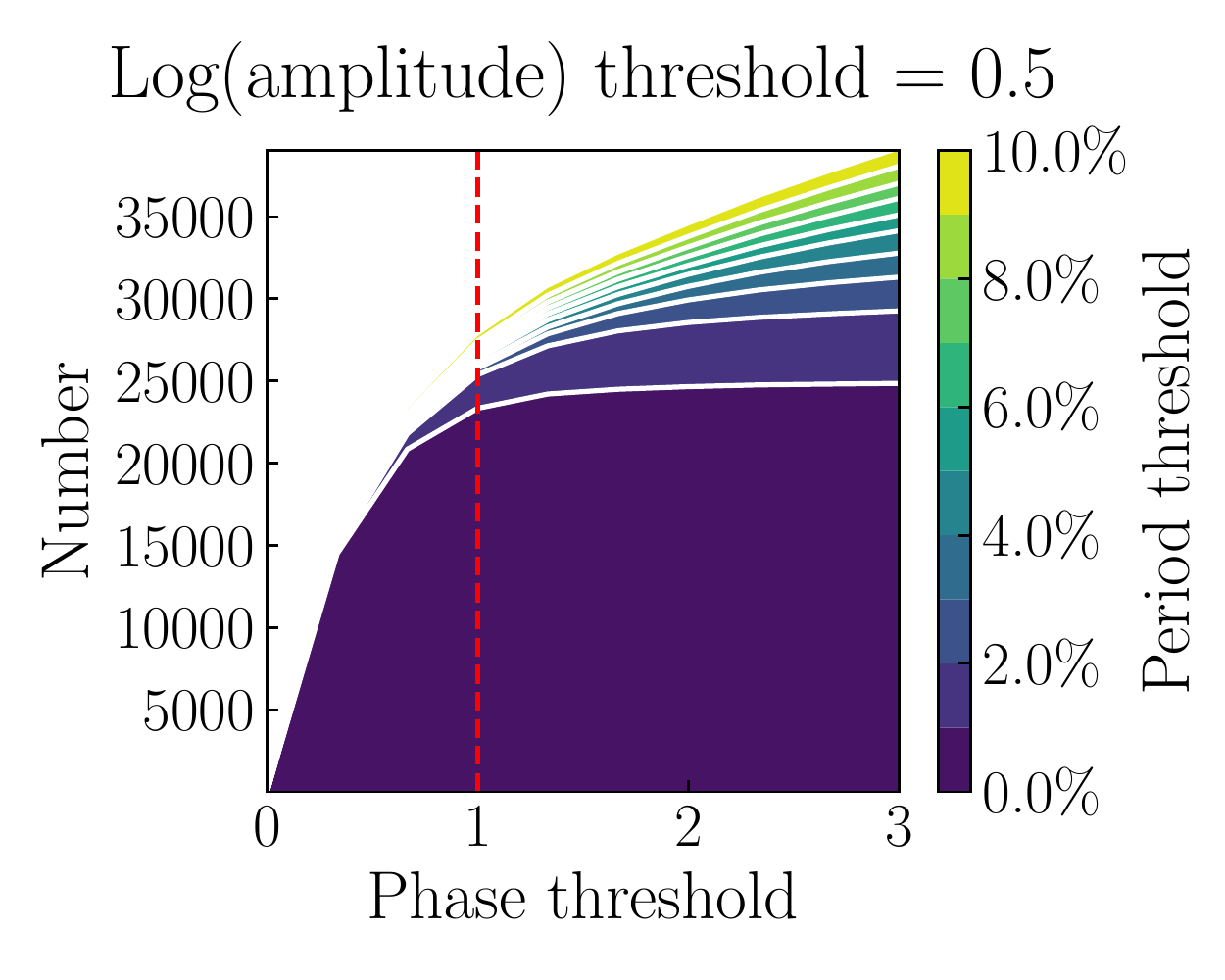}
    \includegraphics[width=0.32\textwidth]{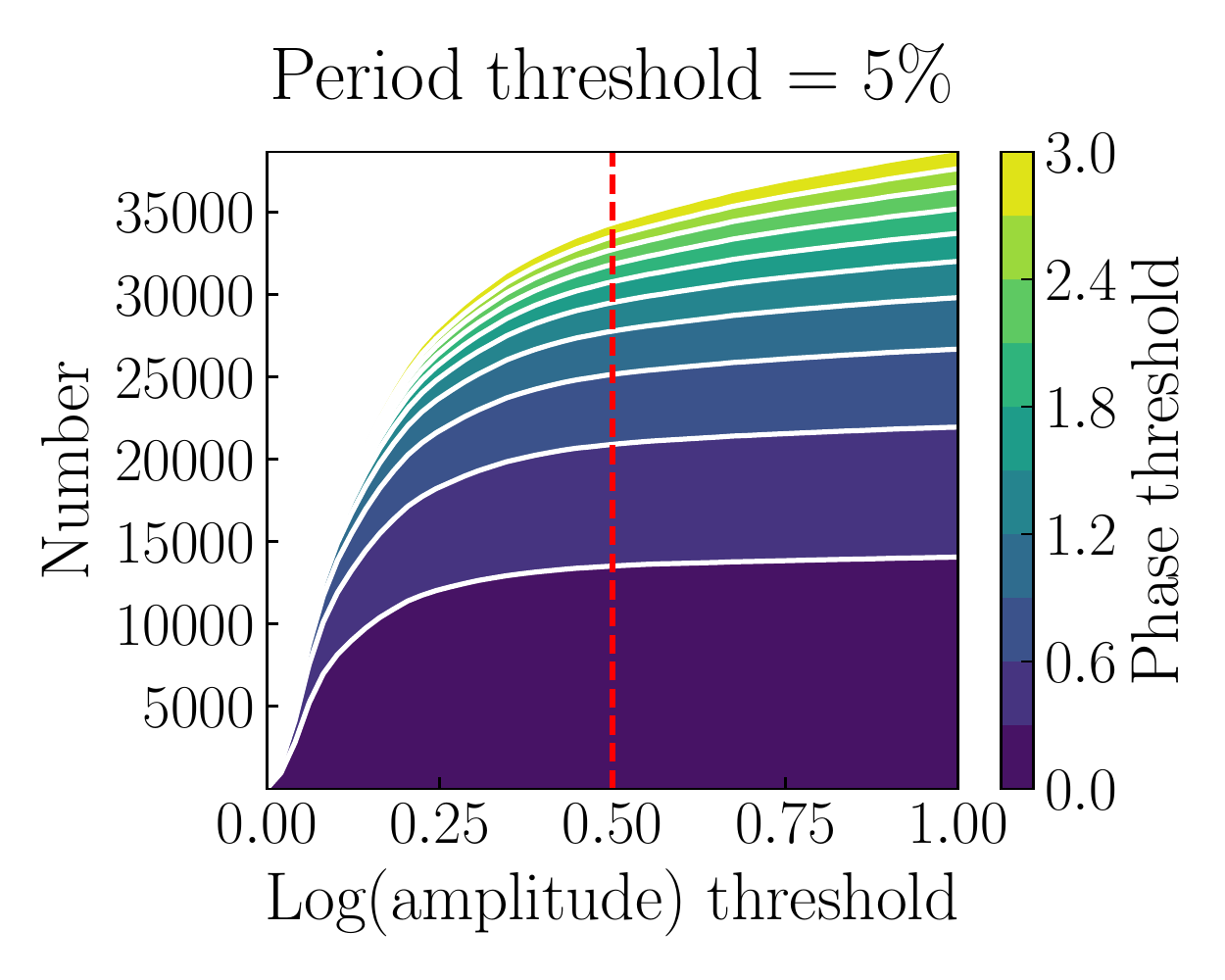}
	\caption{
	    Relationship between the detected number and the three thresholds for the period less than 90 days. Each one of the three sub-figures has a different fixed threshold (indicated by the title of every sub-figure), then shows how the detected number varies with the other two thresholds (one is plotted as the x-axis, the other is plotted as the contours with colourbar) .The dashed red line indicates the value of the x-axis.
	    \label{fig: le90_3d_phase}}
\end{figure*}

\begin{figure*}
    \includegraphics[width=0.32\textwidth]{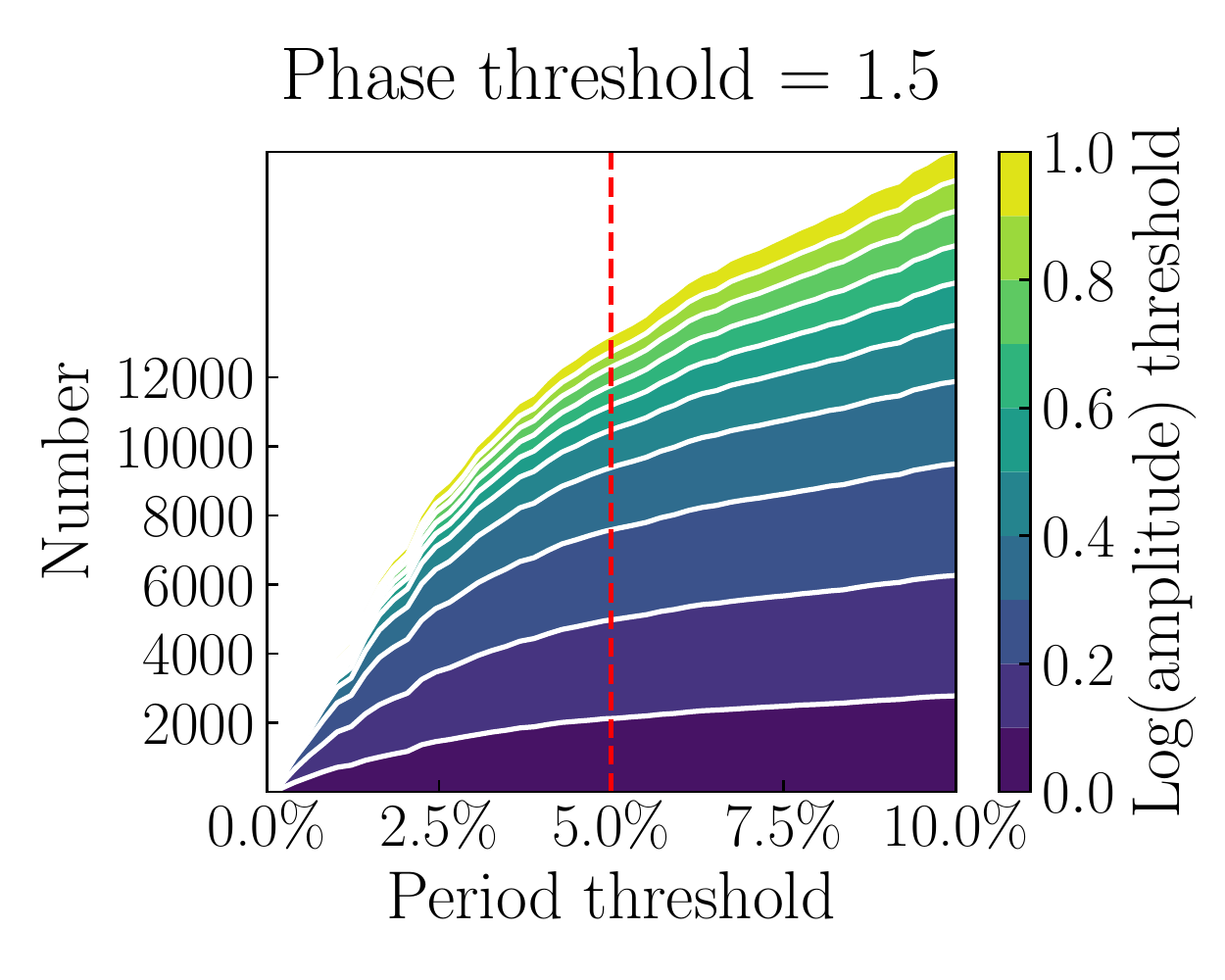}
    \includegraphics[width=0.32\textwidth]{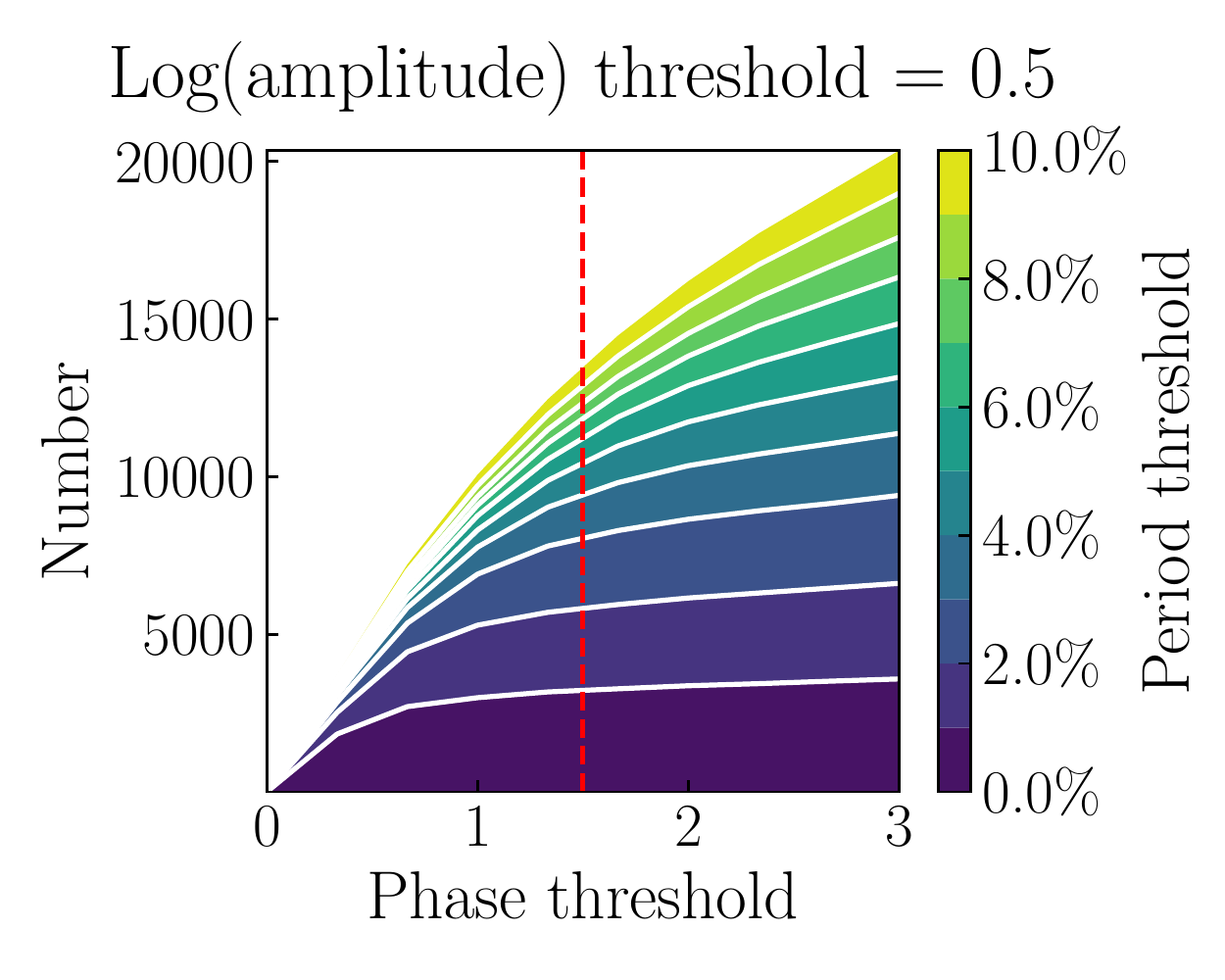}
    \includegraphics[width=0.32\textwidth]{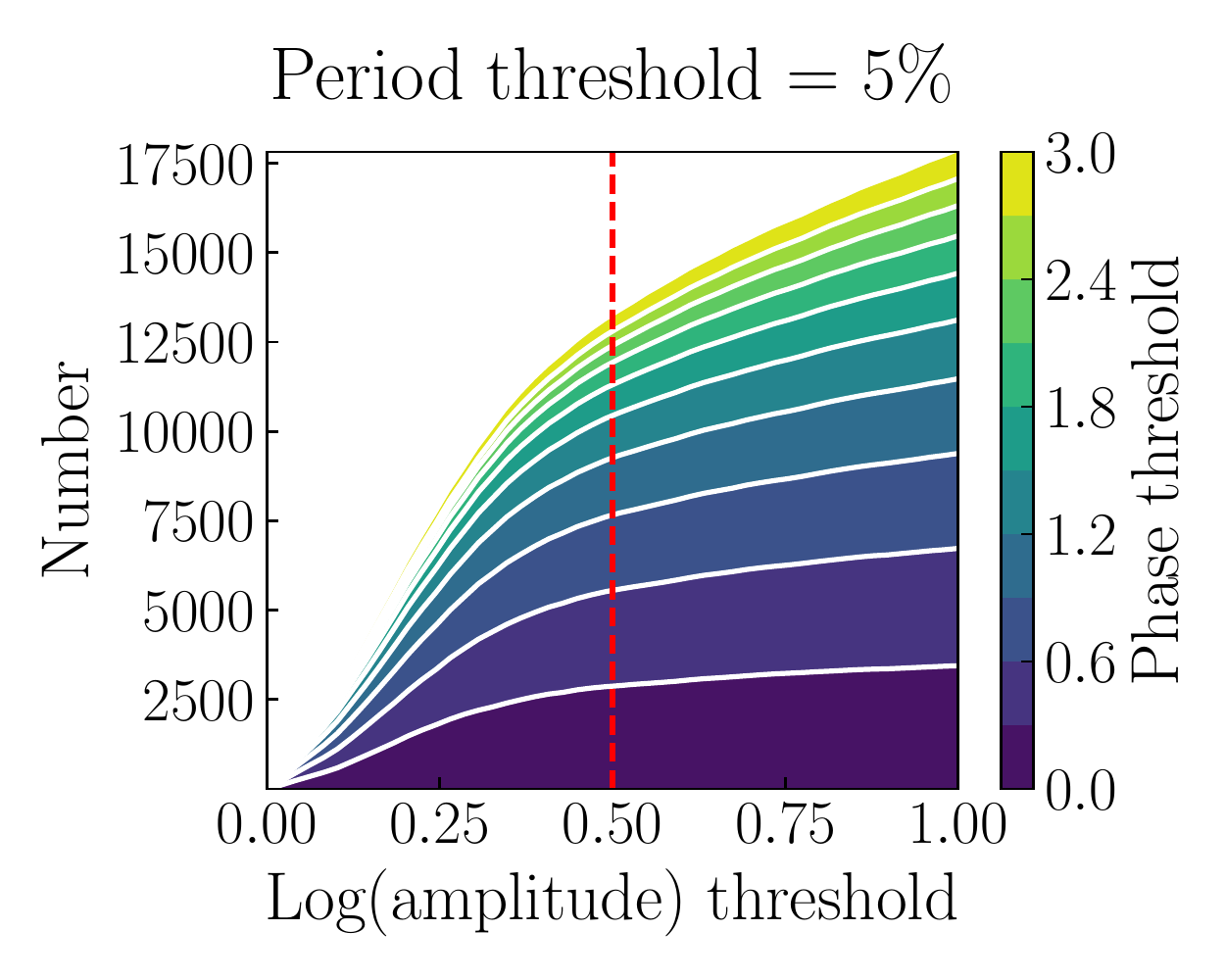}
	\caption{
	    Relationship between the detected number and the three thresholds for the period longer than 90 days but less than 180 days. Each one of the three sub-figures has a different fixed threshold (indicated by the title of every sub-figure), then shows how the detected number varies with the other two thresholds (one is plotted as the x-axis, the other is plotted as the contours with colourbar). The dashed red line indicates the value of the x-axis.
		\label{fig: la90le180_3d_phase}}
\end{figure*}

\subsection{Simulated LCs}\label{subsec:art_lc}
A simulated LC should be made by two components, one is the artificial LC, which shares the same time sampling and systematic trends as the real SAP LC. The other is the injected signal with the known period, amplitude and phase.
Similar works have been done by \citet{Aigrain2015} and \citet{Esselstein2018}, but they used models or known stellar samples to generate the artificial LCs, either lack the instrumental properties or contain some astrophysical variabilities. 
As mentioned in Section \ref{sec:lc_pre}, the first order of the CBV has the highest importance, and keep the long-term trends without astrophysical signals. It could be seen from Figure \ref{fig: results_figures}.
Thus, we use the CBV data to produce the artificial LCs. 
However, since the CBV keep the common trends of the 50 per cent most correlated objects, our simulated LCs only represent the average level of the long-term trends in the SAP. For a certain SAP LC, our results may be overestimated. Nonetheless, the CBV is still a pure sample only with the SAP-like long-term trends, which is useful to evaluate the performance of our methods to some extent.

\begin{figure}
	\includegraphics[width=\columnwidth]{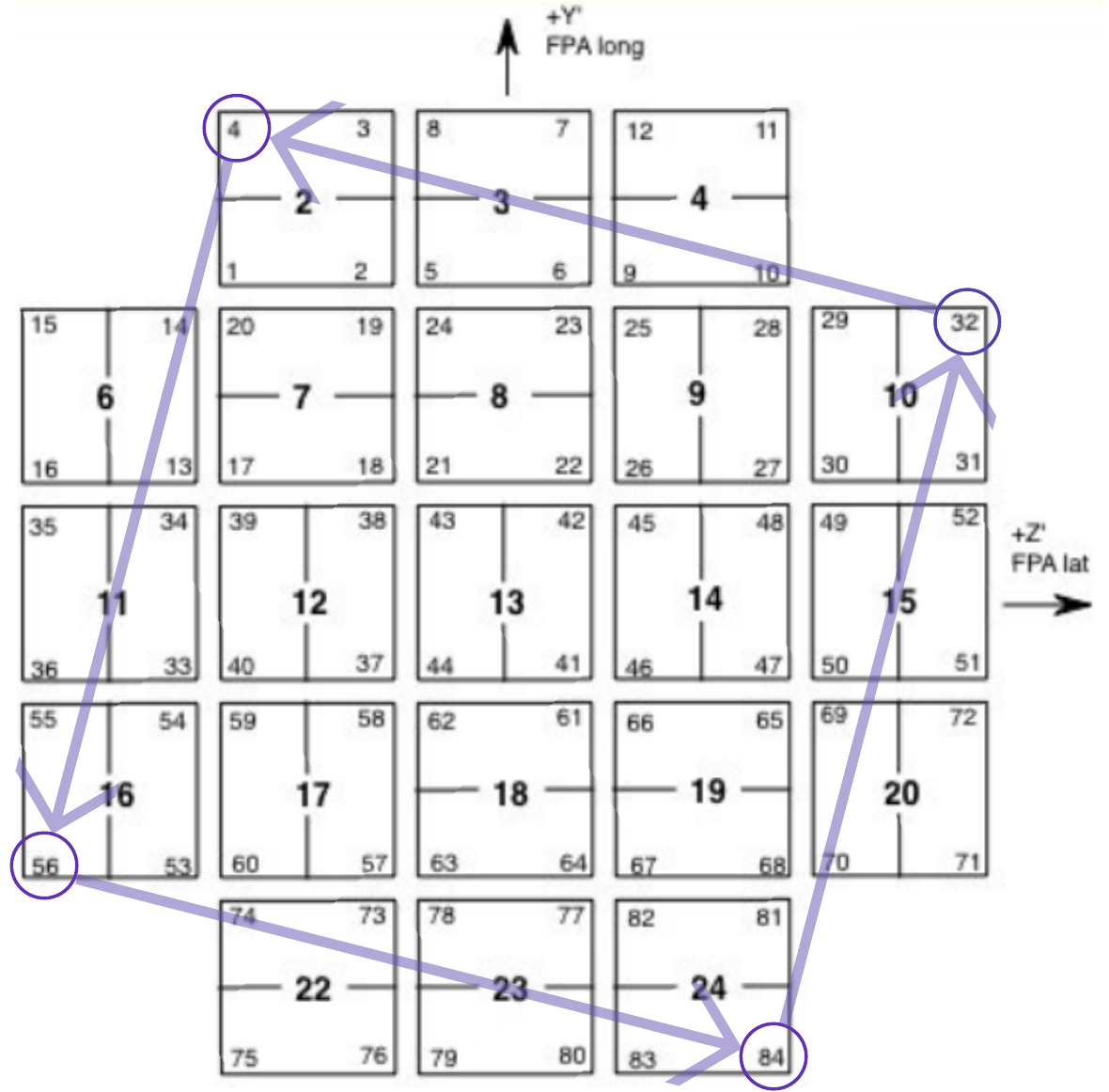}
	\caption{Modules and channels of \kepler focal plane from \citet{Thompson2016}. Each square shows 2 CCDs. The bold number in each square is the module number, and these smaller numbers at the corner are the channel numbers. The purple circles and arrows indicate the channel transformation with quarter changes. \label{fig: channel-figure}}
\end{figure}

According to the instrument handbook \citep{VanCleve2016}, the Detector Array Assembly consists of 21 modules, and each of them has 4 output channels with different readout noise, as shown in Figure \ref{fig: channel-figure}. 
Therefore, the channel numbers could roughly indicate the position of an object. Meanwhile, as the position of an object varies quarterly, the channel number changes accordingly. 
Based on this fact, we create a channel sequence by joining the channel numbers with a hyphen in time order, and then we fix the number to 0 if a quarter has no observation data.
For example, as denoted by the purple circles and arrows in Figure \ref{fig: channel-figure}, the channel sequence of KIC 000893676 is `0-32-4-56-84-32-4-56-84-32-4-56-84-32-4-56-84-32', the first number is zero because the Q0 data are empty.
In the \kepler SAP dataset, there are more than 7000 different channel sequences. 
It is different with a time invariant position ID --- sky group, even if two stars have the same sky group ID, they may have different channel sequences because of the lack of observations at some quarters.

\begin{figure}
	\includegraphics[width=\columnwidth]{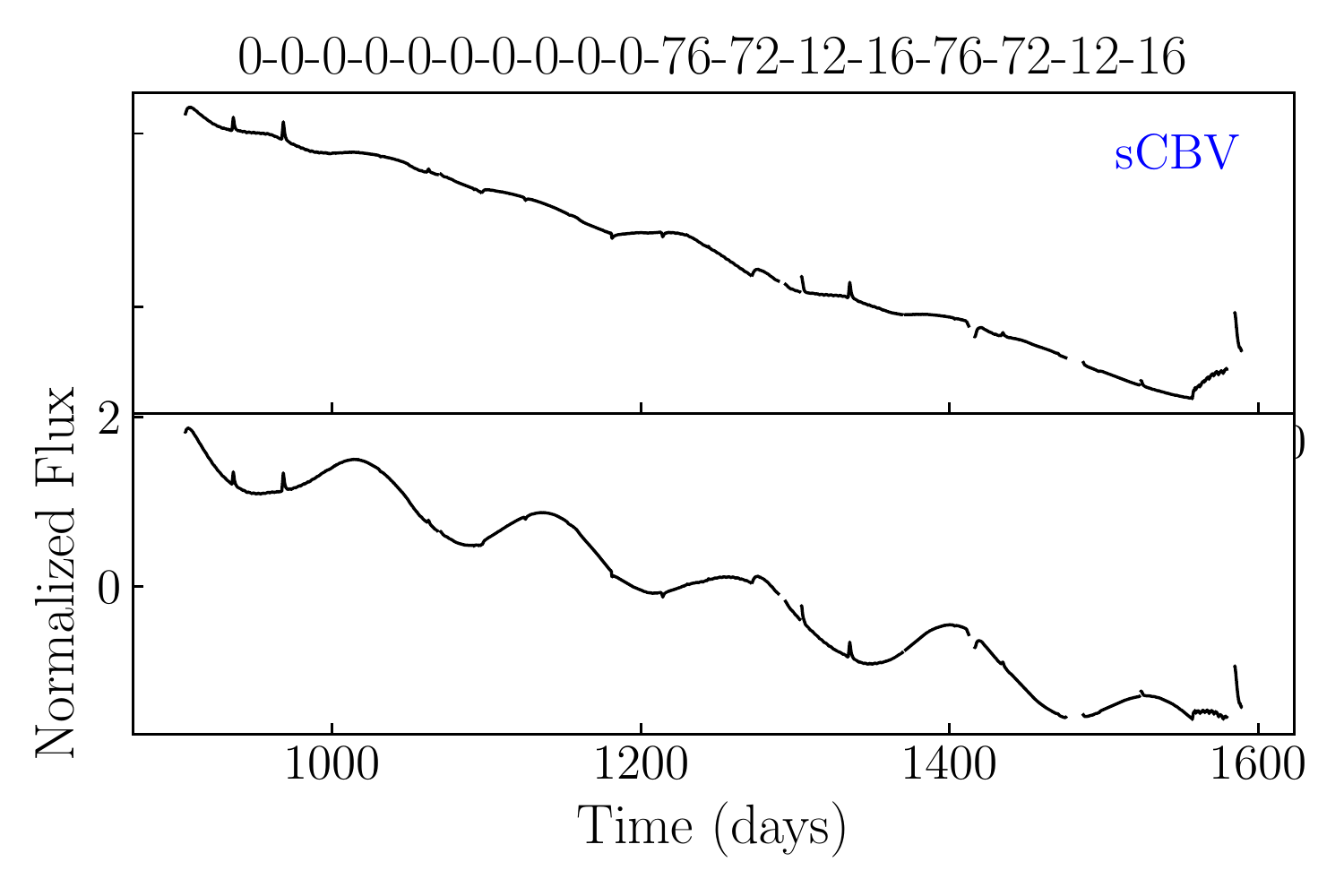}\\
	\includegraphics[width=\columnwidth]{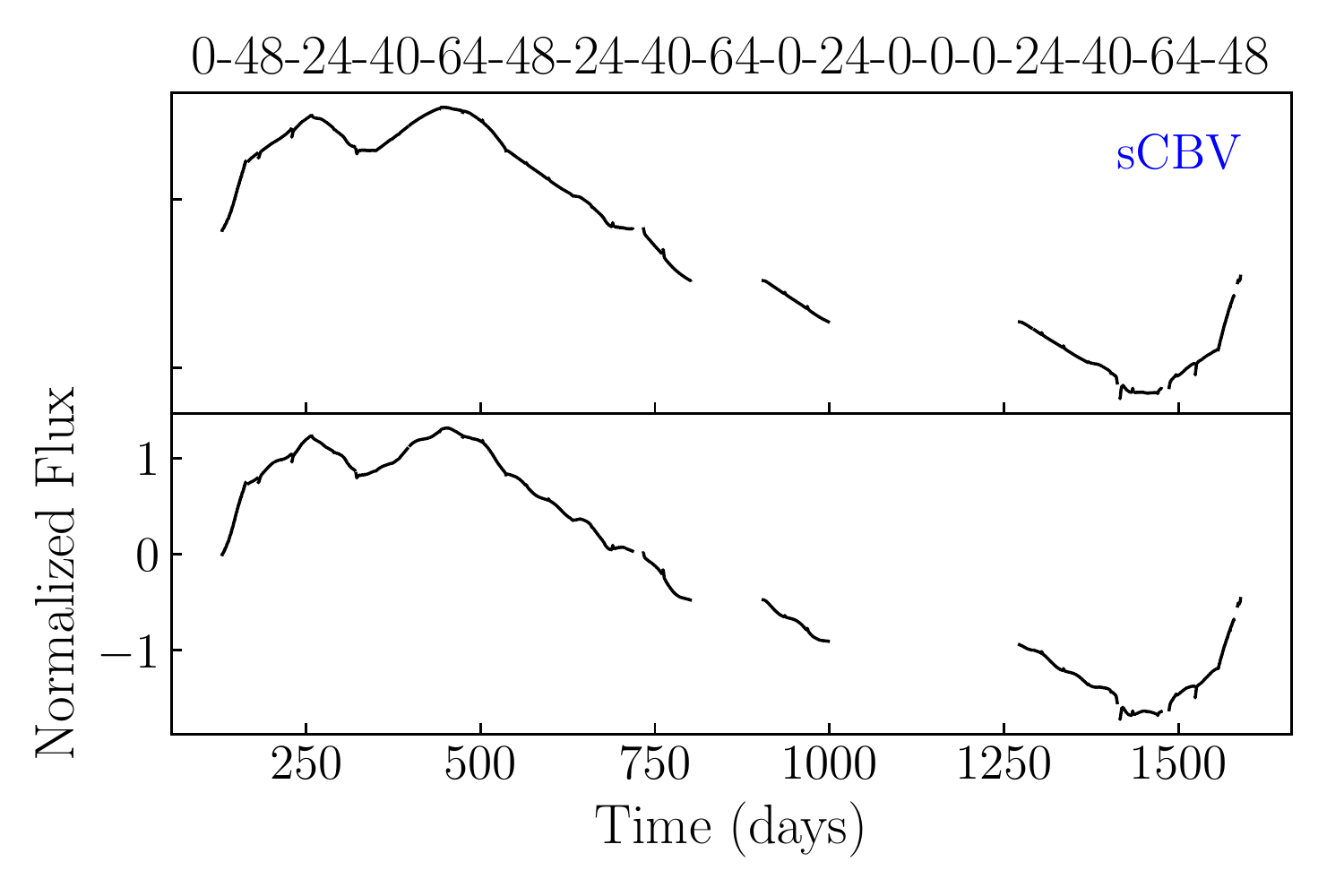}
	\caption{Two examples of sCBV LCs (entitled by their channel sequences) and corresponding simulated LCs. The simulated LCs are generated by injecting a cosinoidal signal. In the top figure, the injected period, amplitude and phase of the simulated LC are 130 days, 31.6 per cent and 4.34 respectively; in the bottom figure, the injected period, amplitude and phase of the simulated LC are 50 days, 2.5 per cent and 2.63 respectively. \label{fig: scbv-and-mock}}
\end{figure}

The CBV data are archived by the CCD channel number, and each vector is generated from one specific channel. 
Considering the volume of the data, we choose 1000 channel sequences randomly from the channel sequence list of the SAP data. 
Then, we combine the CBV data to produce a synthetic CBV (sCBV) LC for a certain channel sequence; this method is the same as the combination of SAP LCs (described in Section \ref{sec:lc_pre}).
Since the vector value of the CBV is not a physical quantity, a normalization approach is useful for the subsequent injection process. 
A z-score normalization is applied to the sCBV LCs, and the formula is given by 
\begin{equation}
Z = \frac{V - \mu}{\sigma},
\end{equation}
where $ Z $ is the normalized value, $ V $ is the raw vector value; $\mu \text{ and } \sigma $ are the mean and standard deviation of the raw vector values. Some normalized sCBV LCs are shown in Figure \ref{fig: scbv-and-mock}.

Then, we construct a series of cosinoidal signals in the form of 
\begin{equation}
S = A \cos(2 \pi / P\  t + \varphi),
\end{equation}
where $ S $ indicates the injected signal, $ A, P \text{ and } \varphi $ are the amplitude, period and phase of the injected signal respectively. We choose 20 samples evenly from 2.5 to 5.5 dex ppm as the injected amplitudes, and the periods vary from 30 to 300 days within 10 days interval. 
We also randomly sample the phase from a uniform distribution over $ [0, 2\pi) $. 
Finally, after add these different combinations of $ S $ into every artificial LC, $ 540,000 $ simulated LCs are generated in total. 

\subsection{Threshold determination}\label{subsec: threshold-determination}
Then, we pre-process the simulated LCs and detect the period candidates. We choose the closest period when there are more than one period candidates.
However, to evaluate whether the detected signals are successfully recovered from the simulated LCs, we need to determine the detection error thresholds for the period, amplitude and phase. 
If we increase these thresholds, more objects would appear in our results, but the ratio of false positive also becomes higher, vice versa. 
Therefore, we determine these thresholds through checking the effect when a threshold varies. 

As shown in Figure \ref{fig: le90_3d_phase}, for the filtered LCs with cut-off frequency from 1/80 to 1/10, the detected numbers vary with the thresholds change.
We could clearly find that the recovered number increase when the thresholds get larger, but the increasing rates decrease as the thresholds increase. That is because the valid objects should share the similar parameters (including period, amplitude and phase) to the injected signals. 
When the increasing rates drop to a constant value, the increment is mainly caused by errors. In our work, we choose three strict thresholds because of the effect of the long-term trends. As indicated by the dashed red lines of Figure \ref{fig: le90_3d_phase}, our thresholds are close to where the increasing rate is constant. The period threshold looks slightly larger but still less than previous works (10 per cent from \citet{Aigrain2015}, 20 per cent from \citet{Esselstein2018}).
Consequently, we choose 5 per cent as the relative period threshold, 0.5 as the $ \log(\text{amplitude}) $ threshold and 1.0 as the phase threshold.

A similar approach is applied to the filtered LC with the cut-off frequency less than 1/90 but larger than 1/170. 
However, as shown in Figure \ref{fig: la90le180_3d_phase}, the valid number increases slower than in Figure \ref{fig: le90_3d_phase}, the thresholds are not as accurate as the shorter period detection. 
That also means the longer period detection might have a higher false positive rate.
We also choose the 5 per cent as the relative period error and 0.5 as the $ \log(\text{amplitude}) $ error, but the phase error should be 1.5 since the valid number rises slower when the phase larger than 1.5. 
We choose the same thresholds for the third type of filtered LCs, although the valid number is too little to check its variation.


\subsection{Test results}\label{subsec: test-results}

\begin{figure*}
	\includegraphics[width=0.306\textwidth]{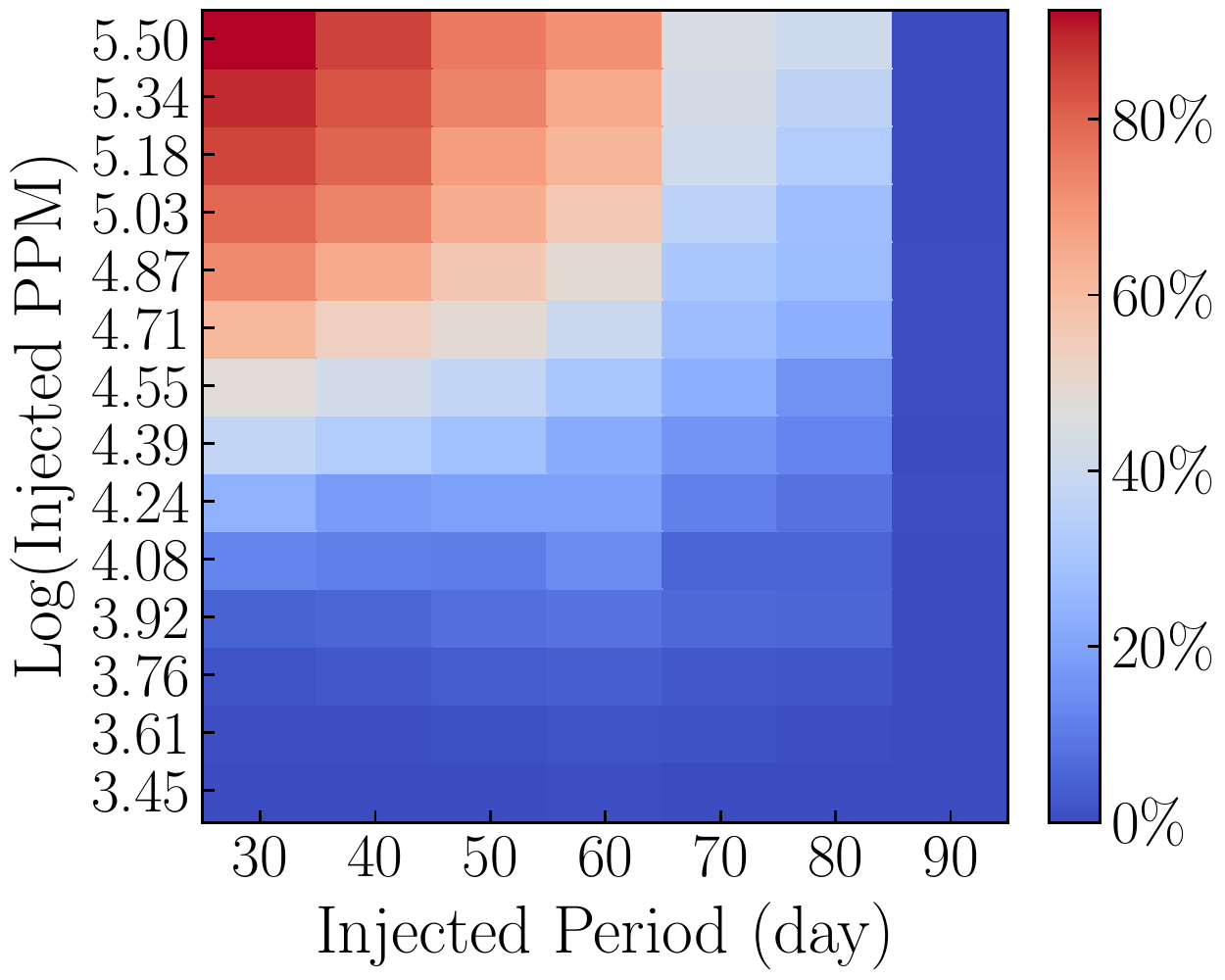}
	\includegraphics[width=0.31\textwidth]{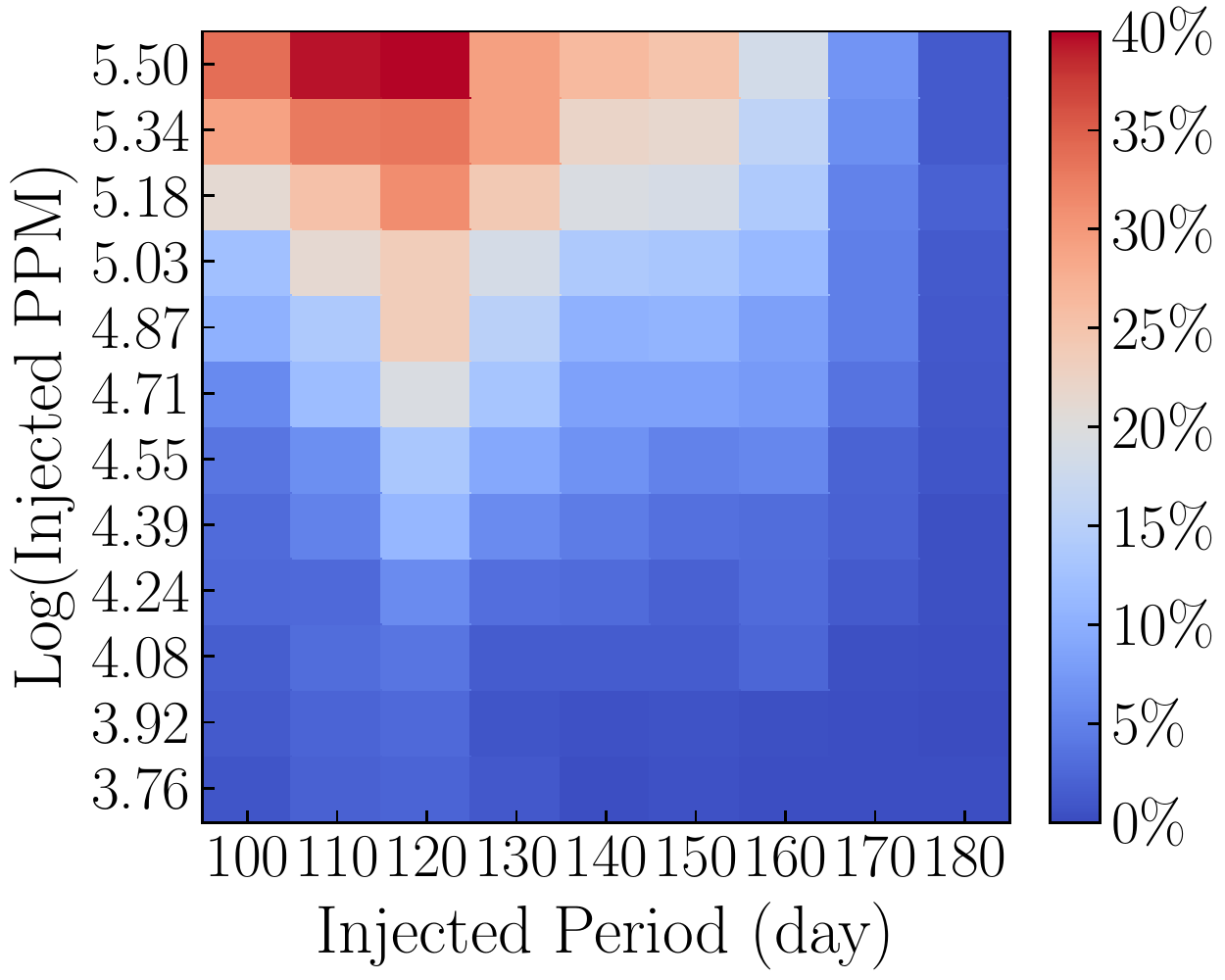}
	\includegraphics[width=0.368\textwidth]{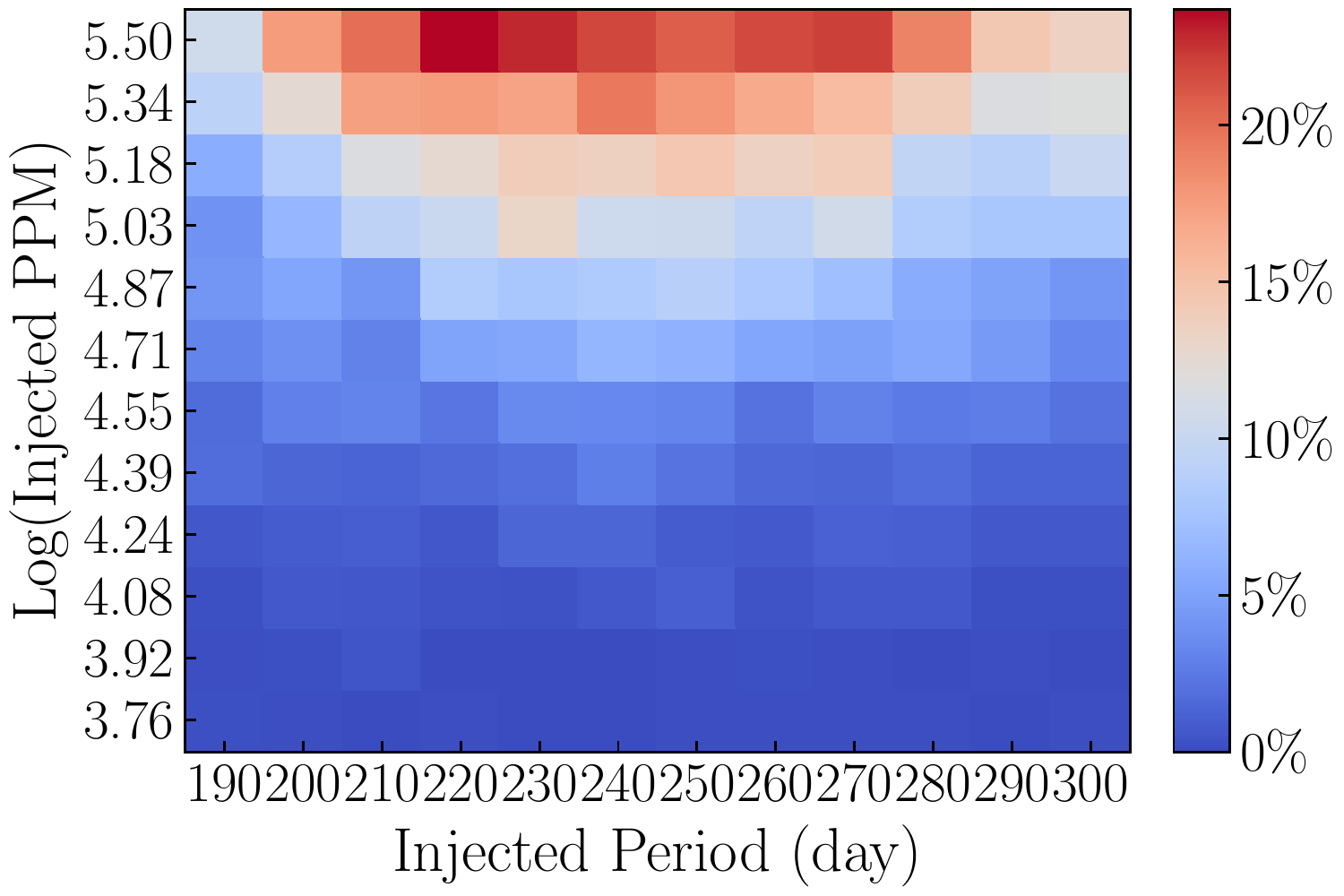}
	
	\caption{Completeness of simulated LC test for different injected period ranges.  \label{fig: mock_results-comp}}
\end{figure*}

The completeness is defined as the ratio of the recovered number and the injected number for every injected period-amplitude bin.
As in Figure \ref{fig: mock_results-comp}, we analyse the completeness of our detection method for different injected period ranges and amplitude ranges. 
We can obviously find that the completeness is sensitive to the injected period and amplitude, the longer injected period and lower amplitude are more difficult to be detected.
This evidence presented in Figure \ref{fig: mock_results-comp} confirms our previous analysis for the SAP LCs in Section \ref{sec:lc_pre}. 

\begin{figure*}
	\includegraphics[width=0.306\textwidth]{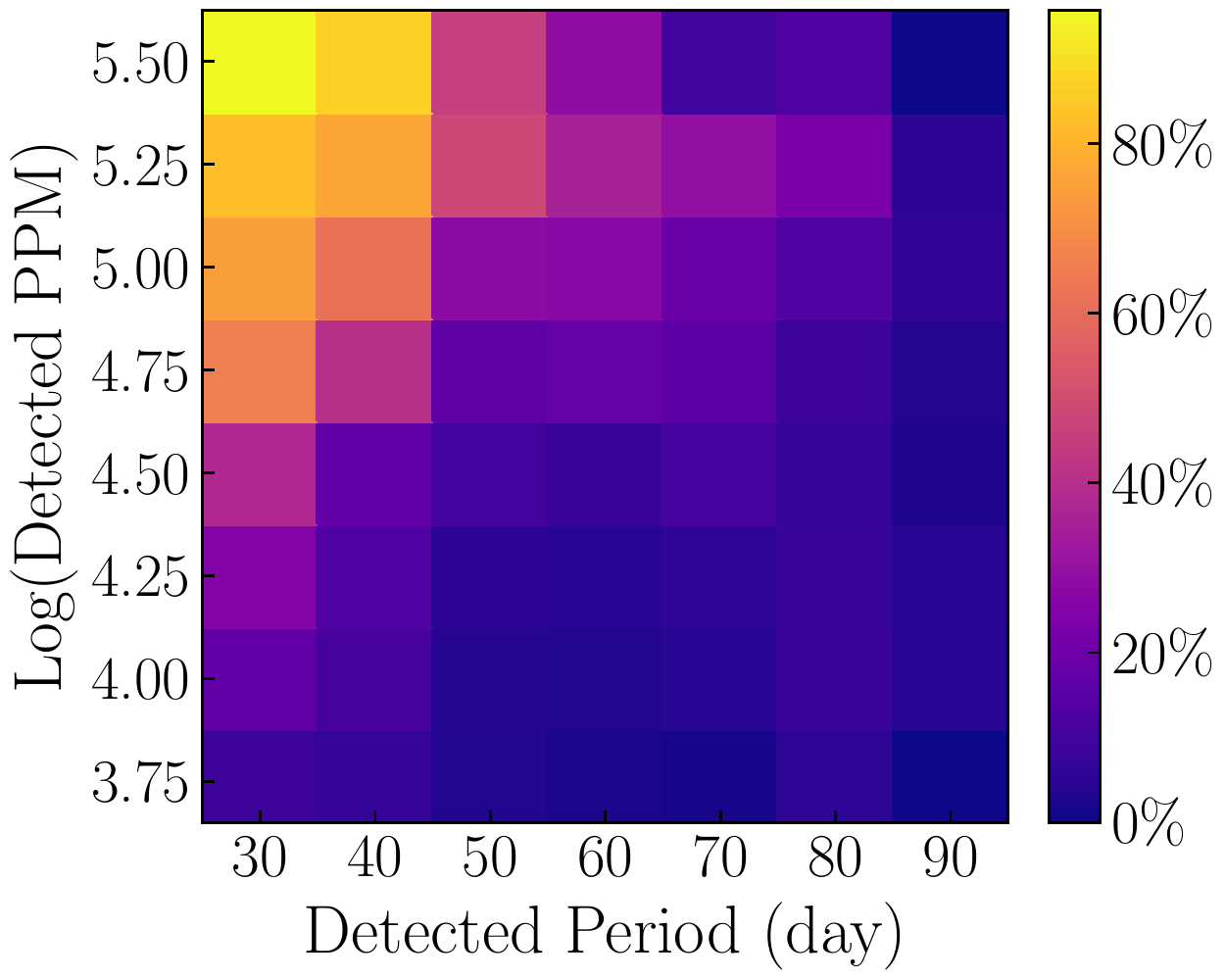}
	\includegraphics[width=0.31\textwidth]{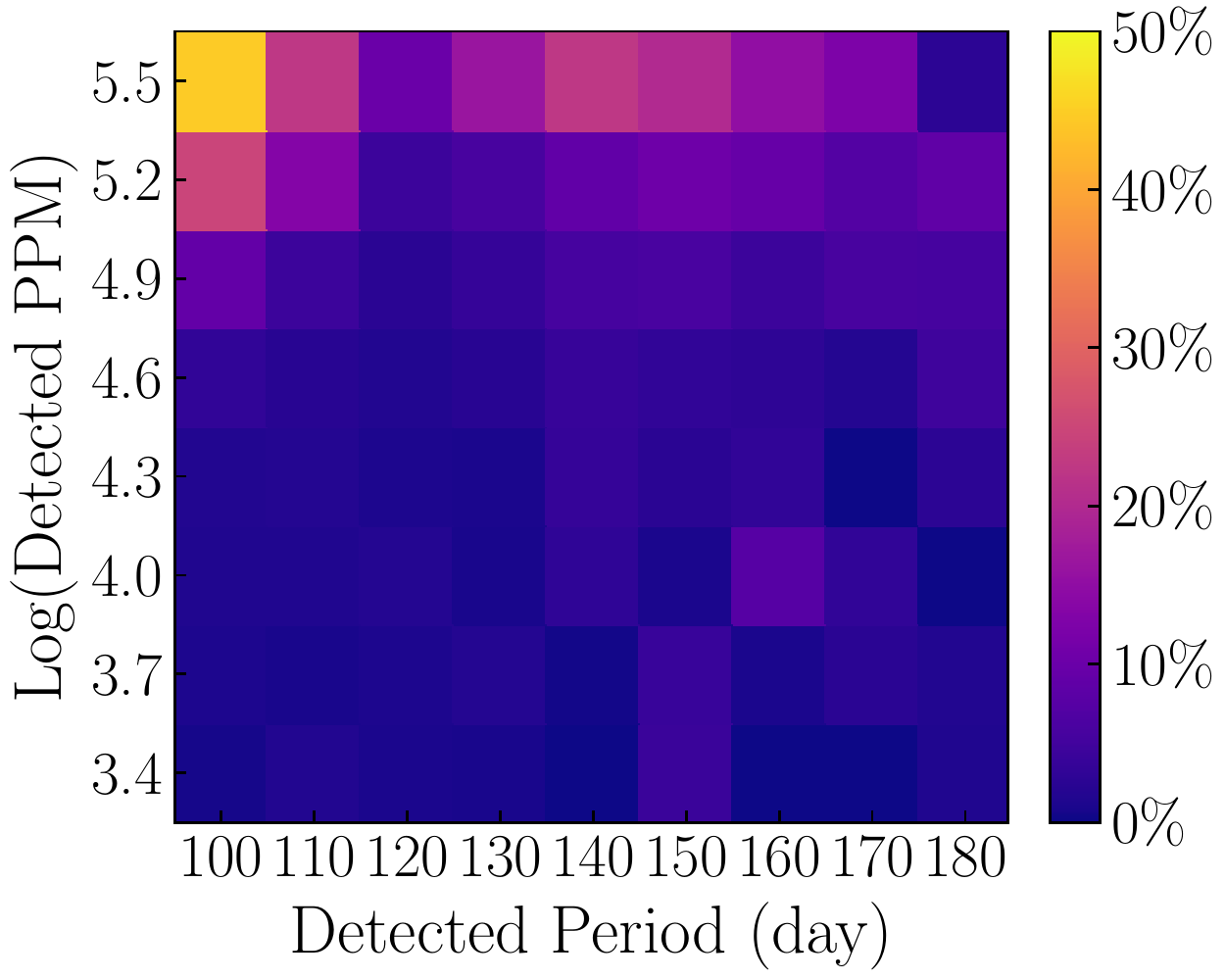}
	\includegraphics[width=0.368\textwidth]{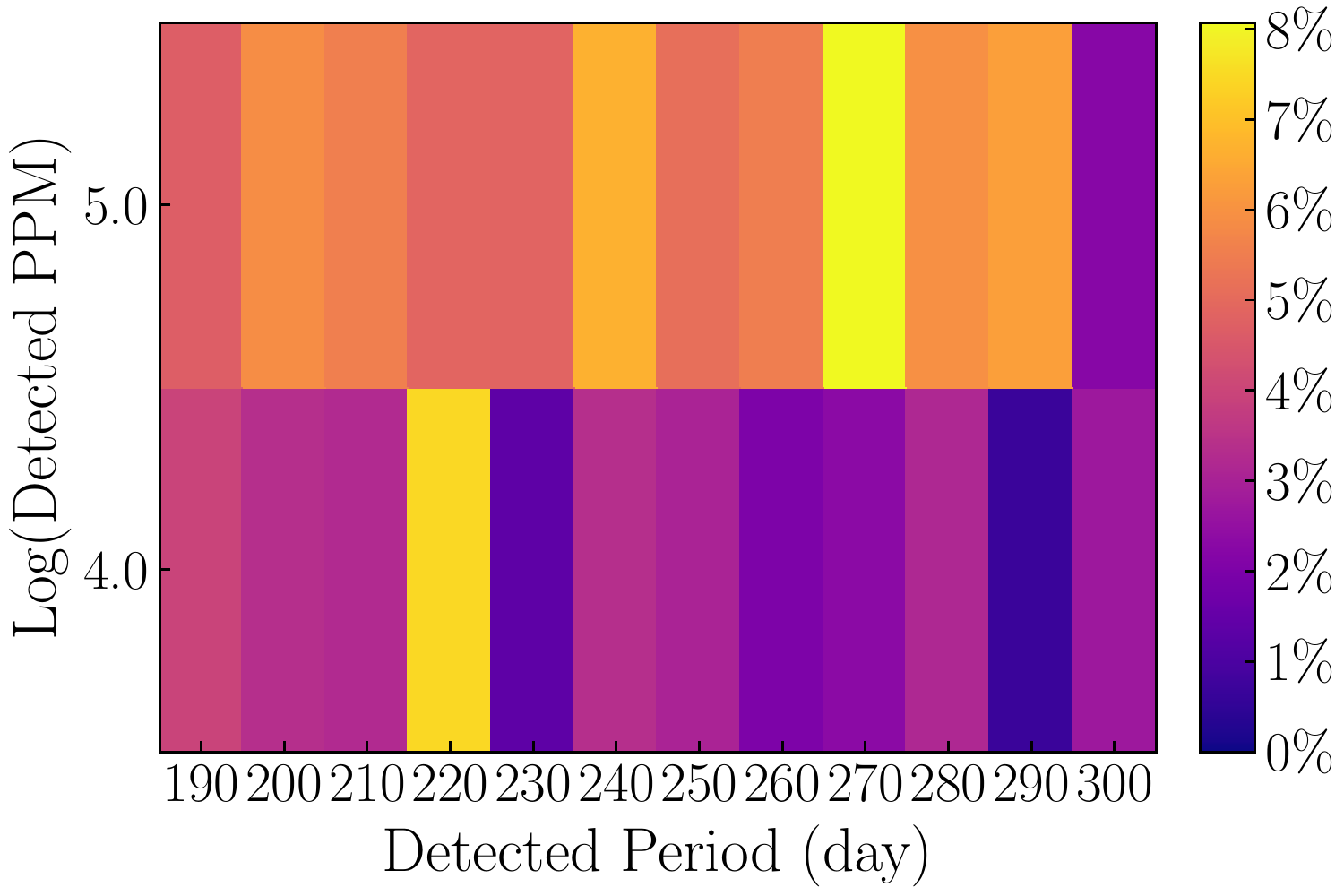}
	
	\caption{Reliability of simulated LC test for different injected period ranges. Since the detected results are different with the injected values, we change the edges of some bins. \label{fig: mock_results-reli}}
\end{figure*}

From the left panel in Figure \ref{fig: mock_results-comp}, it can be seen that our detection limit is extended to a longer period but highly constrained by the amplitude. 
Generally, the completeness is positively related to the amplitude but negatively related to the period.
Similar behaviours are shown in the middle panel and the right panel. 
In the middle panel, the recovered signals tend to appear around the 120 days with an obvious lower amplitude; this freak increments may be polluted by a 3-order harmonic of some one-year systematic trends. The right panel shows that the period longer than 200 days could be detected with possibility higher than 10 per cent only when the amplitude larger than 5 dex ppm. 
Therefore, in consideration of the detection limits, we are more likely to find a long period with large amplitude. 

We also study the reliability of our simulated data test, and the results are shown in Figure \ref{fig: mock_results-reli}. 
Similar to the completeness, the reliability is calculated by dividing the number of valid objects by the total detected objects in every detected period-amplitude bin. 
The reliability also shows the similar trends with the completeness, but the quantity is lower. 
This result also reveals an obvious trade-off between completeness and reliability, which is also indicated in previous results \citep{Aigrain2015, Esselstein2018}. 
In our work, the reliability is not particularly important, because we have a strict selection process. 

\section{Candidate Selection}\label{sec:can_sele}
\begin{figure}
	\centering
	\includegraphics[width=0.9\columnwidth]{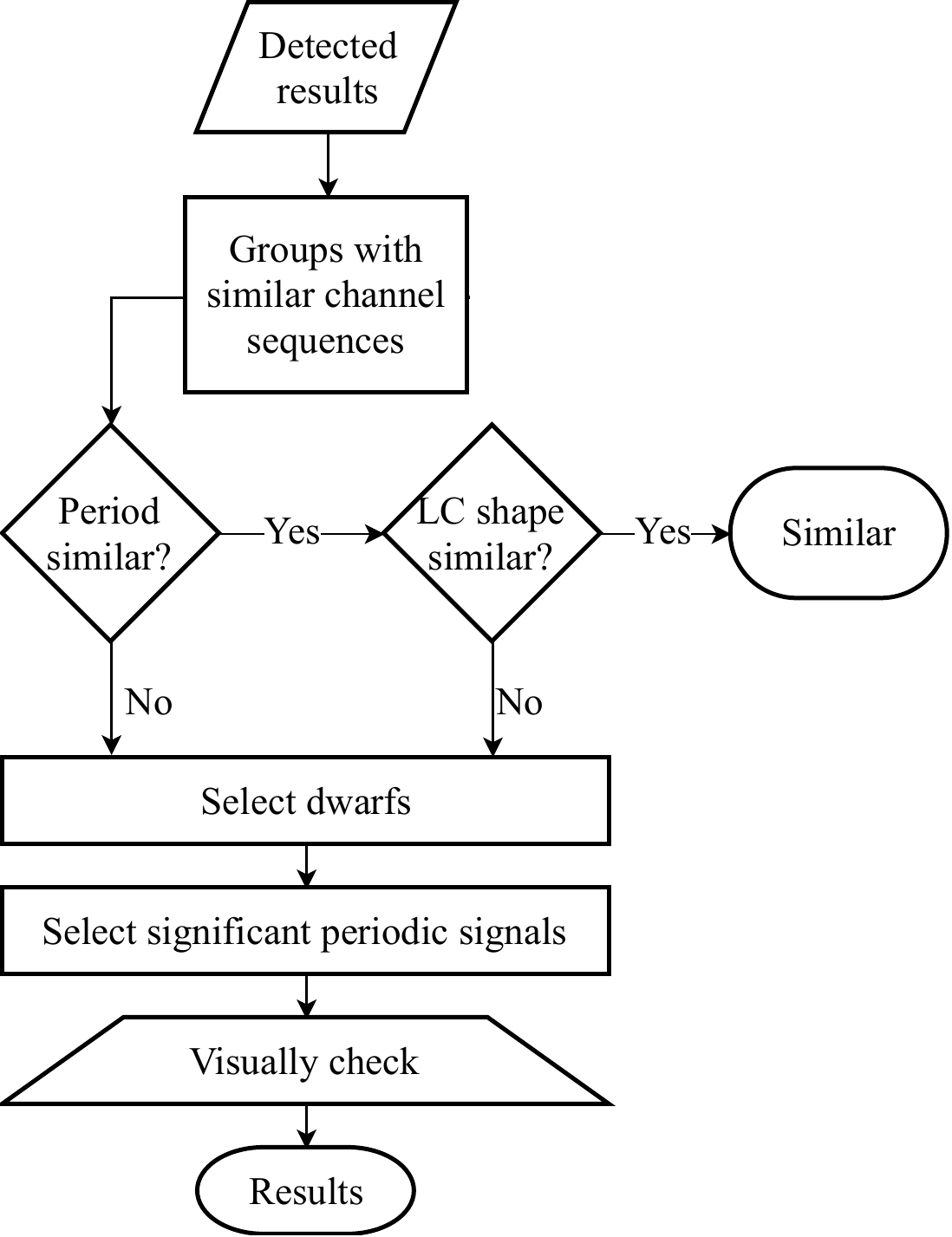}
	\caption{Detailed flowchart of the candidate selection. \label{fig: flow_can_sel}}
\end{figure}

Based on the results of the simulated data test, our detection method have the ability to detect more long period objects from \kepler SAP LCs. 
Thus, we apply our detection method to the pre-processed LCs in \kepler, for the 10--80 days filtered LCs, we detect periods for 181,766 objects; for the 90--170 days filtered LCs, we detect periods for 175,214 objects; for the 180--300 days filtered LCs, we detect periods for 165,500 objects. 
However, our reliability is not high enough to provide the long rotation period candidates directly, it is necessary to create a strict selection process after the detection. 
Generally, our selection method is simply dropping the objects with similar periods and LC shapes, because we believe that the stellar variation is unique but the instrumental effect is in common.
Figure \ref{fig: flow_can_sel} shows the detailed flowchart of our selection process.

%
According to the previous studies on planet candidates validation \citep[e.g.,][]{Twicken2016} and flare verification \citep{Yang2017}, systematic errors always cause the same variations on nearby sources; so those studies often used a pixel level inspection approach to remove the background noise and centroid offsets, etc.
Besides the connection of spatial distribution, the time correlation should also be considered. 
In the procedures of planet candidates vetting, \citet{2016ApJS..224...12C} used an `Ephemeris Match Indicates Contamination' method to vet the false positive threshold crossing events. 
Therefore, we should also remove the characteristics when they are displayed at the same time.

However, transits and flares are usually short-term signals, for the long-term variations, we only need to focus on the artefacts that the time scale exceeds the rotation period. 
As mentioned in Section \ref{subsec:art_lc}, we create a channel sequence for every star. 
The channel sequences are determined by the information of an object in both position and time, so the similar channel sequence may introduce similar trends. 
Actually, by observing the characteristics of SAP LCs, we often find many objects with similar LC shapes, although they are completely different objects but with similar channel sequences. For example, the top three concatenated SAP LCs are similar in Figure \ref{fig: similar_lc_shape}. 
Therefore, comparing the similarity of the LC on a similar channel sequence can effectively identify the systematic trends.


Then, we group these objects by their channel sequences, and similar channel sequence (at most one channel number is different at the same quarter) would be classified as the same group. 
In each group, we analyse their similarity of the detected periods and LC shapes.
\begin{enumerate}
	\item Period similarity\\
	According to the conclusion of simulated data test in Section \ref{subsec: threshold-determination}, if the two detected periods are close within the 5 per cent tolerance, they are considered similar. \\

	\item LC shape similarity\\
	For each pair of normalized LCs, we calculate the Hausdorff distance $ H(\mathrm{LC_1}, \mathrm{LC_2}) $\citep{Hausdorff} to evaluate their similarity. Hausdorff distance is widely used to check the similarity of two datasets\citep[e.g.]{Huttenlocher1993, Montes2018, FilaliBoubrahimi2018}. A general form of Hausdorff distance is defined as 
	\begin{align}
	h(A, B) & = \max_{a\in A}{\{\min_{b\in B}{\{d(a, b)\}}\}}\\
	H(A, B) & = \max{\{h(A, B), h(B, A)\}},
	\end{align}
	where $ a $ and $ b $ are points of sets $ A $ and $ B $ respectively, $ d(a, b) $ is a metric between these points. In this work, we take $ d(a, b) $ as the Euclidean distance between $ a $ and $ b $. 
	
	In order to determine whether they are similar LCs, we build a binomial classification based on the Hausdorff distance. We randomly choose 100 channel sequences and then select 5 SAP LCs in each channel sequence, after removing 75 LCs because of their observation time are less than 270 days, 425 LCs survived as the training set. Then, according to the similarity of the LC shapes with each other, we visually tag these training LCs with `similar' or `dissimilar' in each channel sequence. For example, in Figure \ref{fig: similar_lc_shape}, the upper three LCs shapes look similar to each other, so we mark them as `similar'; while the shapes of the three LCs below look very different to each other, so we mark them as `dissimilar'.
	
	\begin{figure}
		\includegraphics[width=\columnwidth]{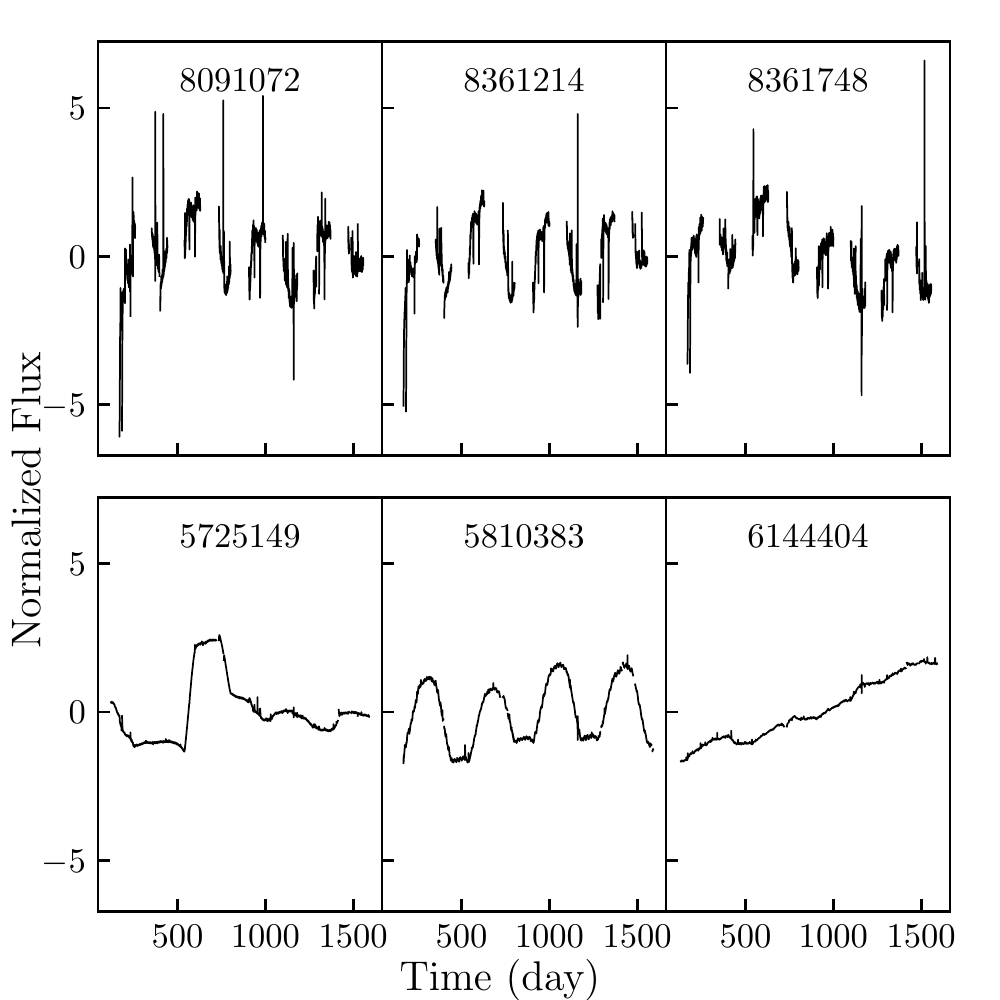}
		\caption{Six combined SAP LCs of our training set. 
			The number above in every sub-figure is the corresponding KIC ID. 
			The channel sequences of the upper three LCs and the lower three LCs are `0-0-44-0-42-0-44-0-42-0-44-0-42-0-44-0-42-0' and `69-69-9-13-73-69-9-13-73-69-9-13-73-69-9-13-73-69', respectively. \label{fig: similar_lc_shape}}
	\end{figure}
	
	After that, we calculate the Hausdorff distance for every LC combination with same channel sequence. Comparing with the manual tags, if we choose $ H_t $ as the distance threshold to separate similar and dissimilar, we could find the false positive rate and true positive rate vary as $ H_t $ changes. Therefore, the best distance threshold should have a larger true positive rate and lower false positive rate. We could plot a receiver operating characteristic (ROC) curve to find the best distance threshold. As shown in Figure \ref{fig: roc-curve}, the x-axis and y-axis of the ROC curve are the false positive rate (FPR) and true positive rate (TPR) respectively, which are defined as
	\begin{equation}
	\mathrm{FPR} = \frac{\mathrm{FP}}{\mathrm{FP + TN}}, 
	\end{equation}
	and
	\begin{equation}
	\mathrm{TPR} = \frac{\mathrm{TP}}{\mathrm{TP + FN}}.
	\end{equation}
	FP, TP, FN and TN represent the number of false positives, true positives, false negatives and true negatives respectively. The TPR and FPR are also known as `sensitivity' and `1 $ - $ specificity'.
	The ROC curve is a commonly used method for threshold determination and model performance assessment. Figure \ref{fig: roc-curve} gives the ROC curve with the Hausdorff distance threshold varies from 0 to 500 and the step is 1. Since the closer to the top-left corner the better performance of the classification, we choose the best distance threshold with calculating the distance to the top-left corner $ d_\mathrm{corner} $ given by
	\begin{equation}
	d_\mathrm{corner} = \sqrt{(1 - \mathrm{sensitivity})^2 + (1 - \mathrm{specificity})^2}
	\end{equation}
	When the Hausdorff distance is 90, $ d_\mathrm{corner} $ is the minimum, so the Hausdorff distance criterion is set to 90 in our work.
		
	\begin{figure}
		\includegraphics[width=\columnwidth]{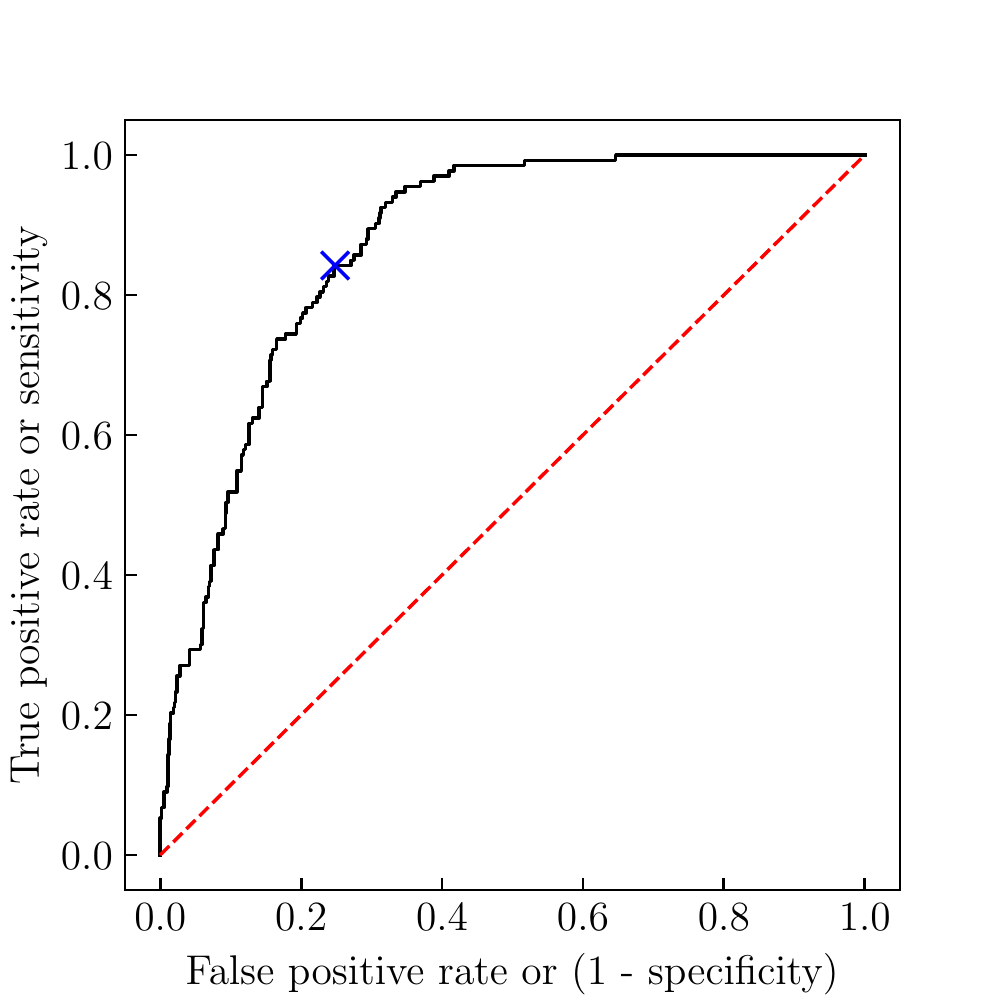}
		\caption{ROC curve for different Hausdorff distance threshold. 
			The black step line shows the FAP and TAP at different Hausdorff distance.
			The dashed red diagonal line means random guess, and the blue cross marker indicates the point with Hausdorff distance = 90, which has the lowest distance to the top-left corner on ROC curve. \label{fig: roc-curve}}
	\end{figure}
	
\end{enumerate}

\begin{figure*}
	\includegraphics[width=\columnwidth]{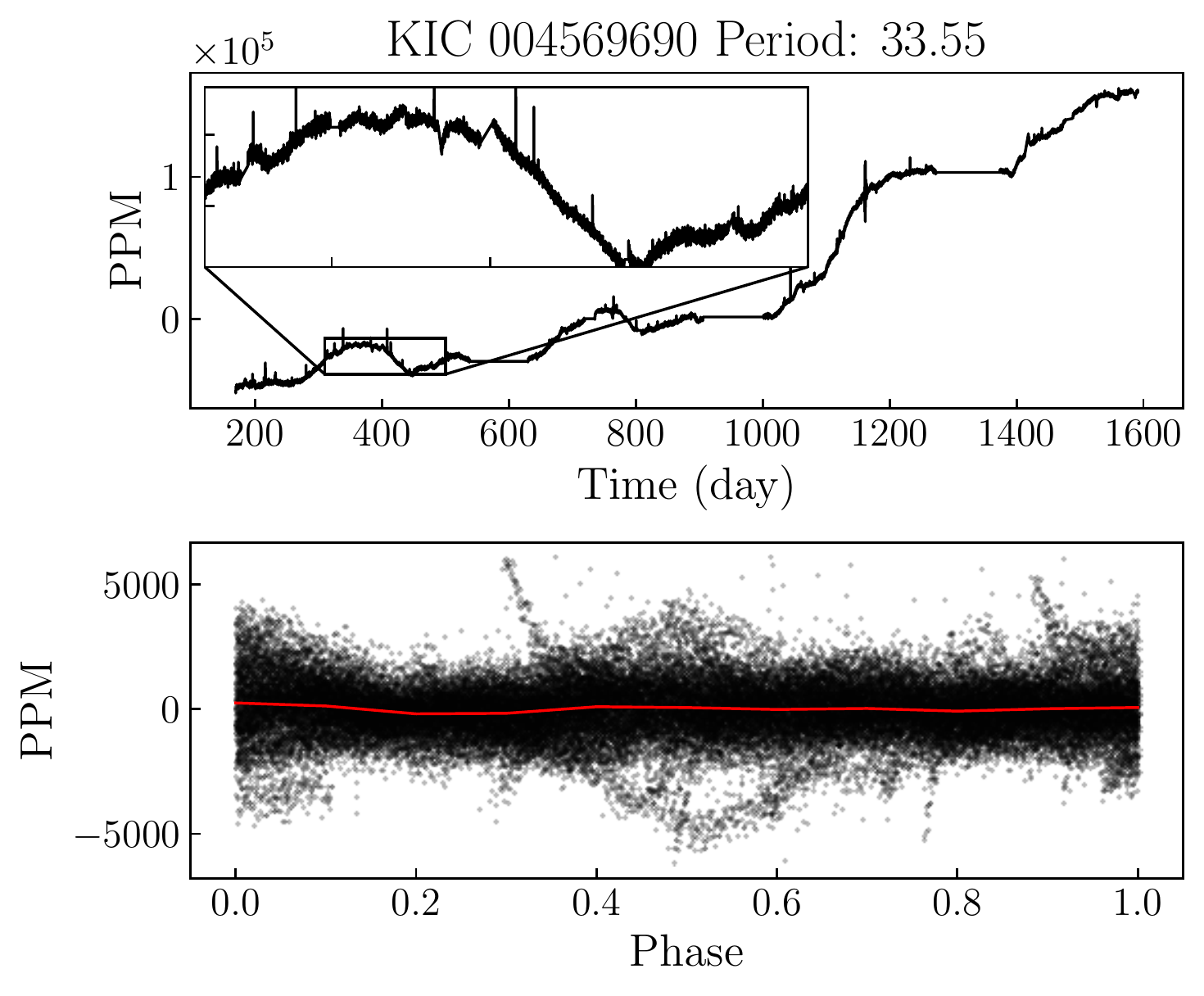}
	\includegraphics[width=\columnwidth]{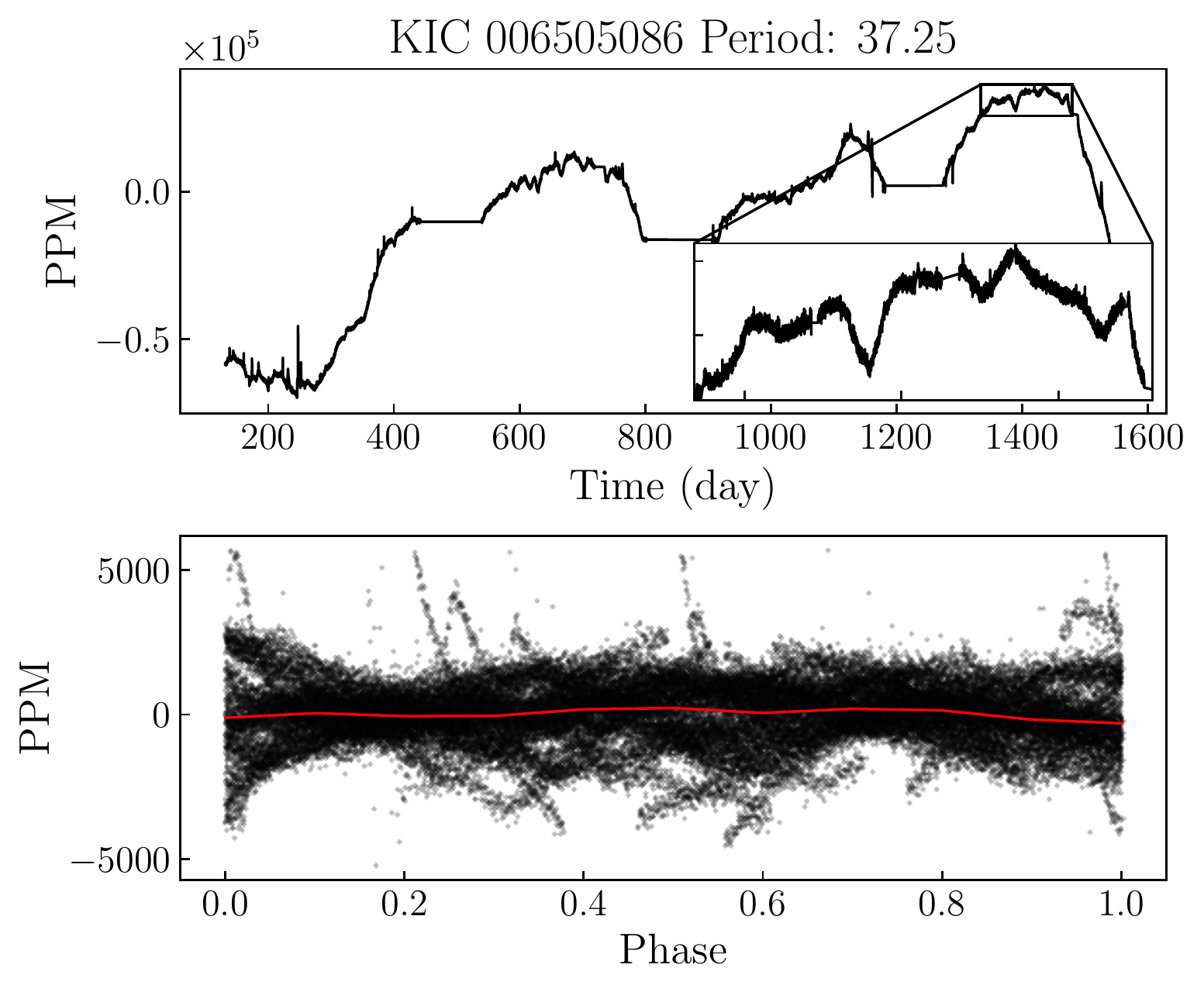}
	\caption{Two examples of the SAP LCs with periods from \citet{McQuillan2014} but being removed in our sample. In each sub-figure, the top panel is the concatenated SAP LC and its zoomed inset plot, the bottom panel is the folded LC. The title of each sub-figure indicate the KIC ID and the period from \citet{McQuillan2014}.}
	\label{fig: visually_inspect_bad}
\end{figure*}
\begin{figure*}
    \includegraphics[width=0.329\textwidth]{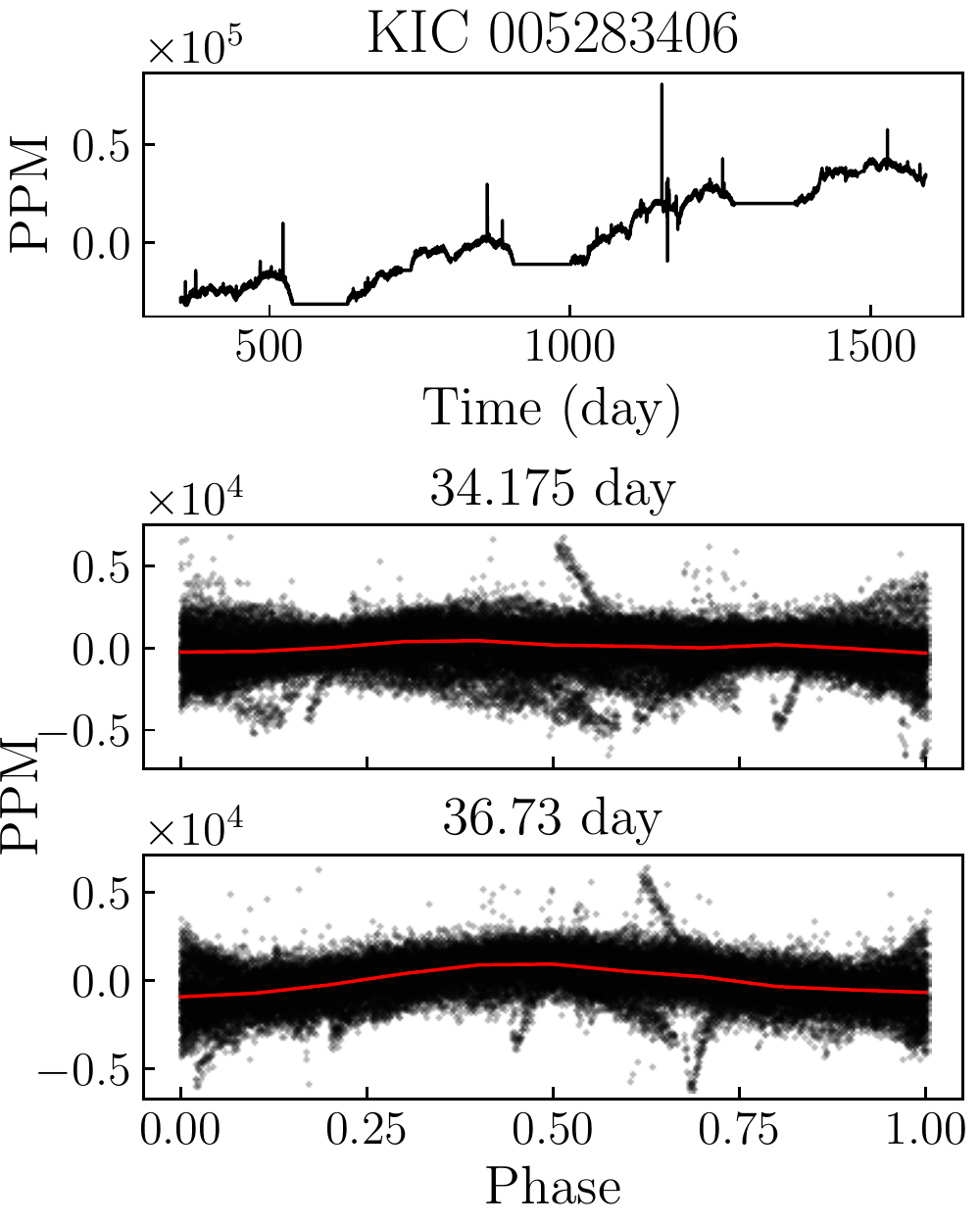}
	\includegraphics[width=0.329\textwidth]{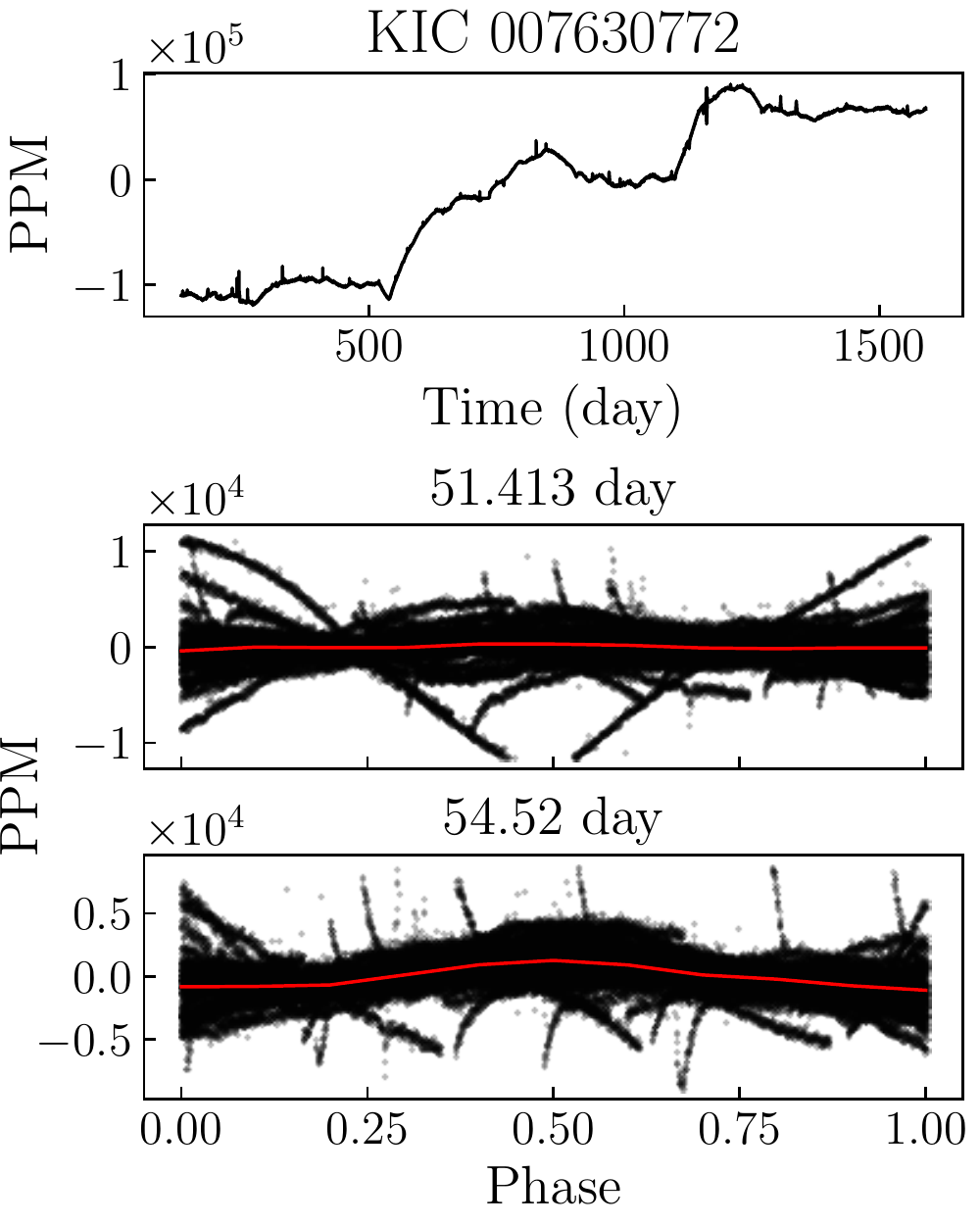}
	\includegraphics[width=0.329\textwidth]{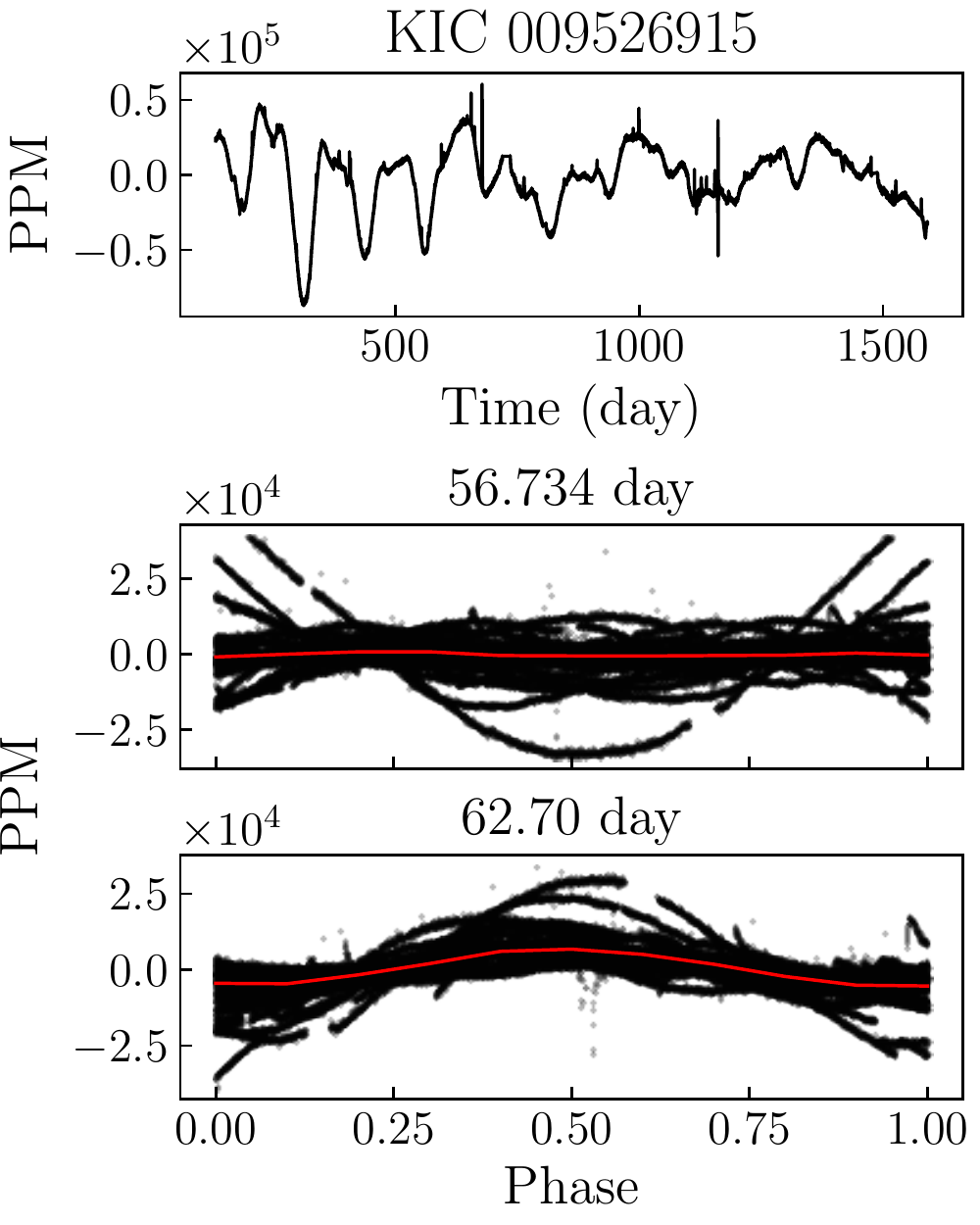}
	\caption{Two examples of the SAP LCs with better periods in our detection. In each sub-figure, the top panel is the concatenated SAP LC, the middle panel and the bottom panel are the folded LCs with period from \citet{McQuillan2014} and our results respectively. The KIC ID and periods are shown on the title of each panel.}
	\label{fig: visually_inspect_good}
\end{figure*}

\begin{figure*}
	\includegraphics[width=0.32\textwidth]{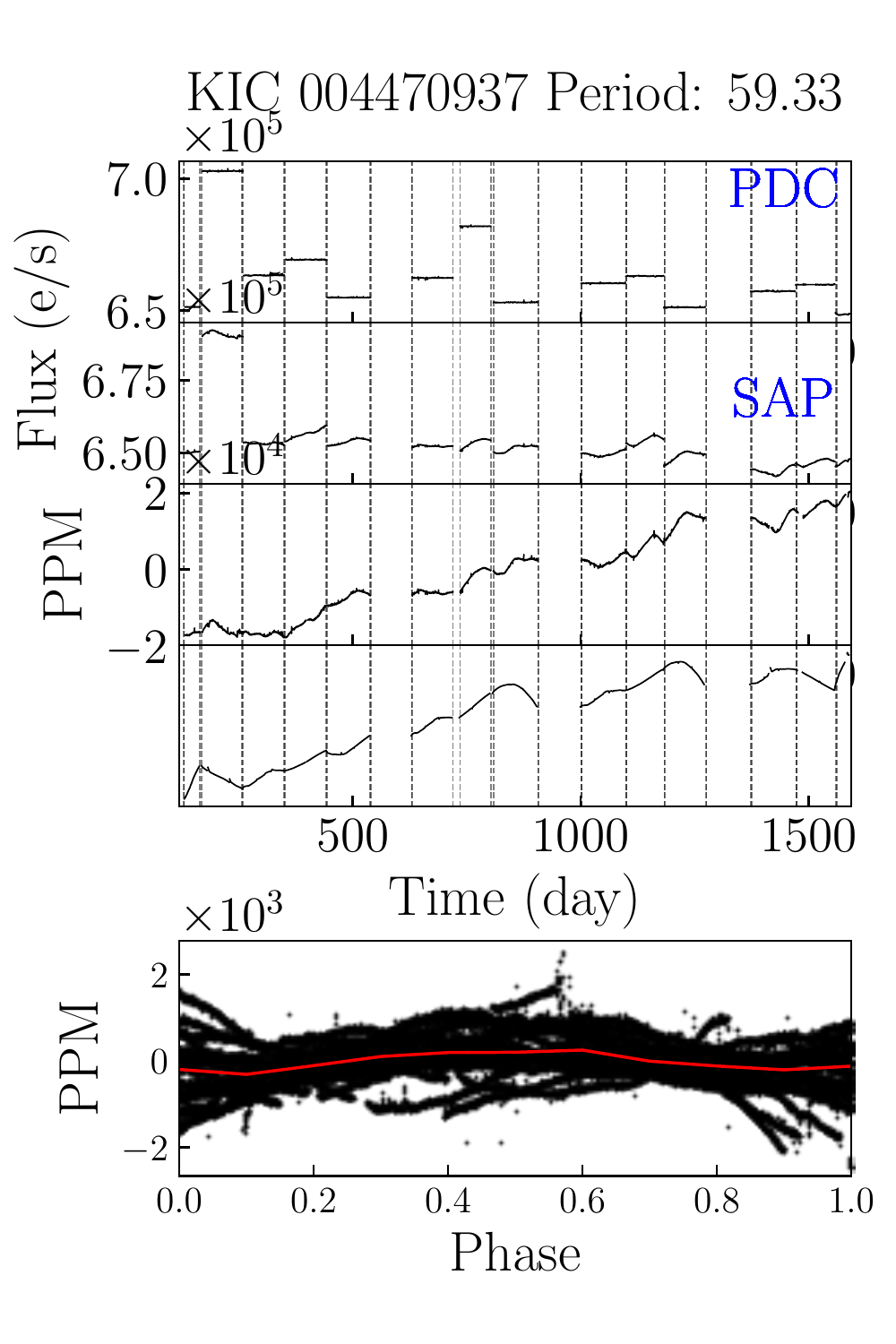}
	\includegraphics[width=0.32\textwidth]{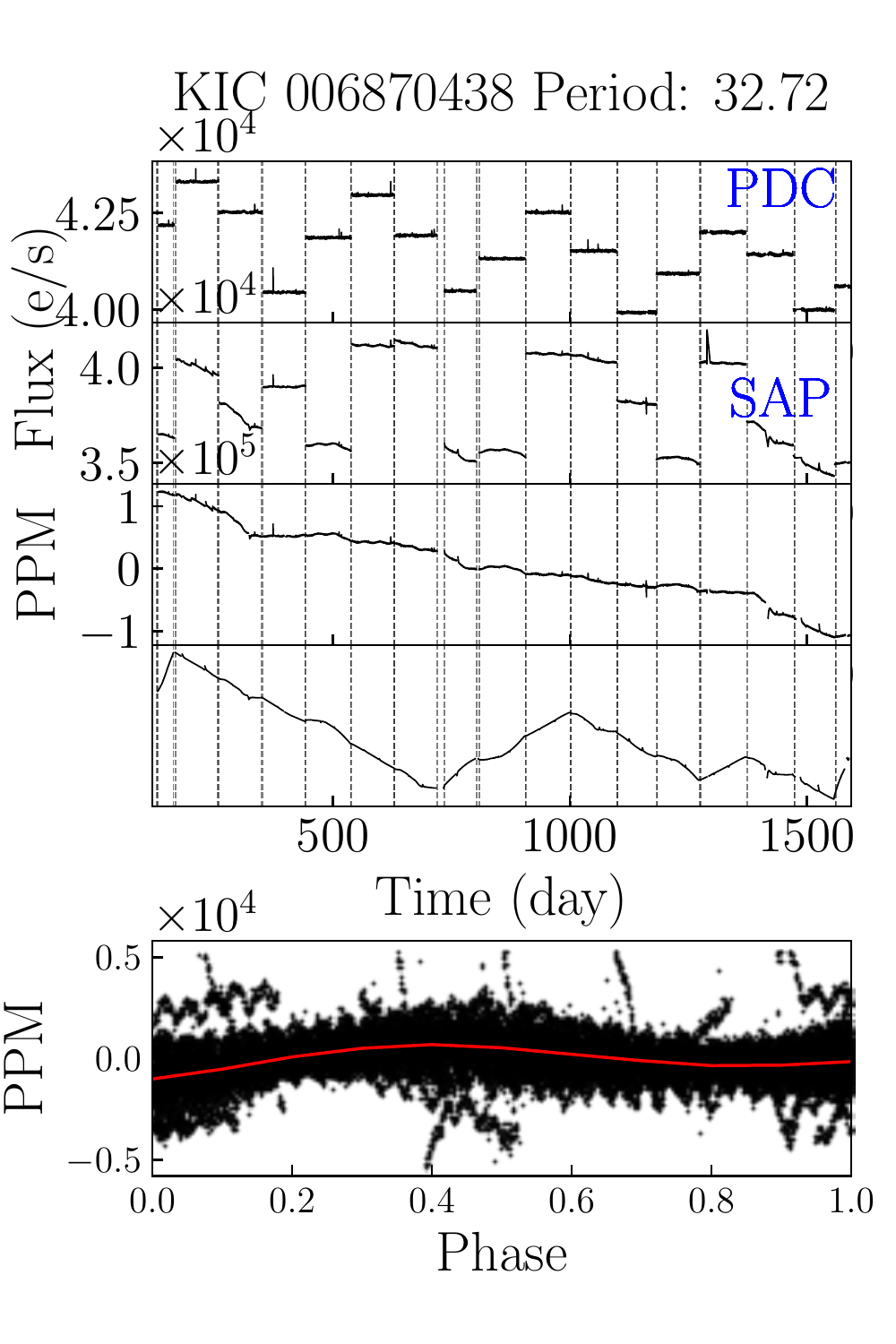}
	\includegraphics[width=0.32\textwidth]{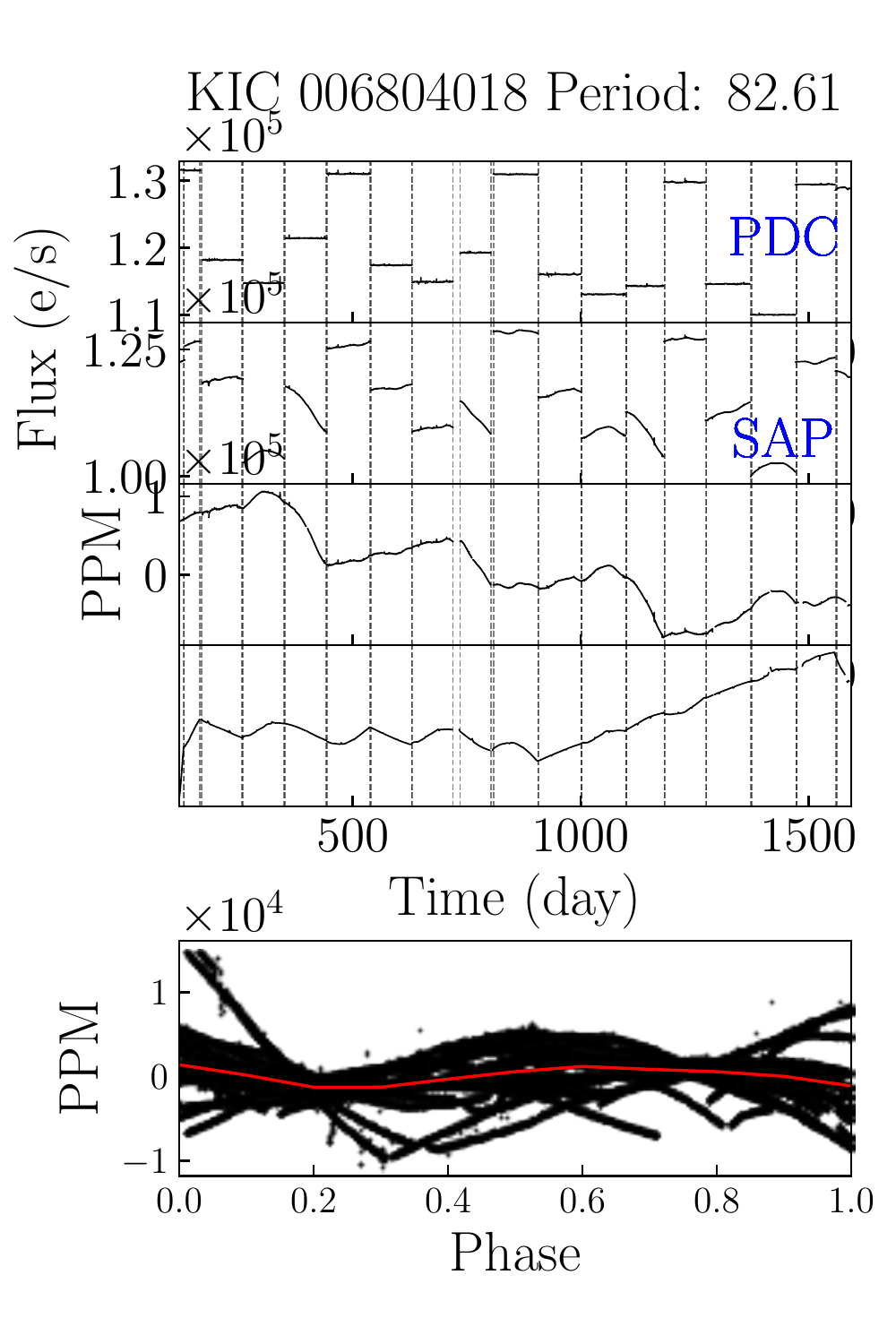}
	\includegraphics[width=0.32\textwidth]{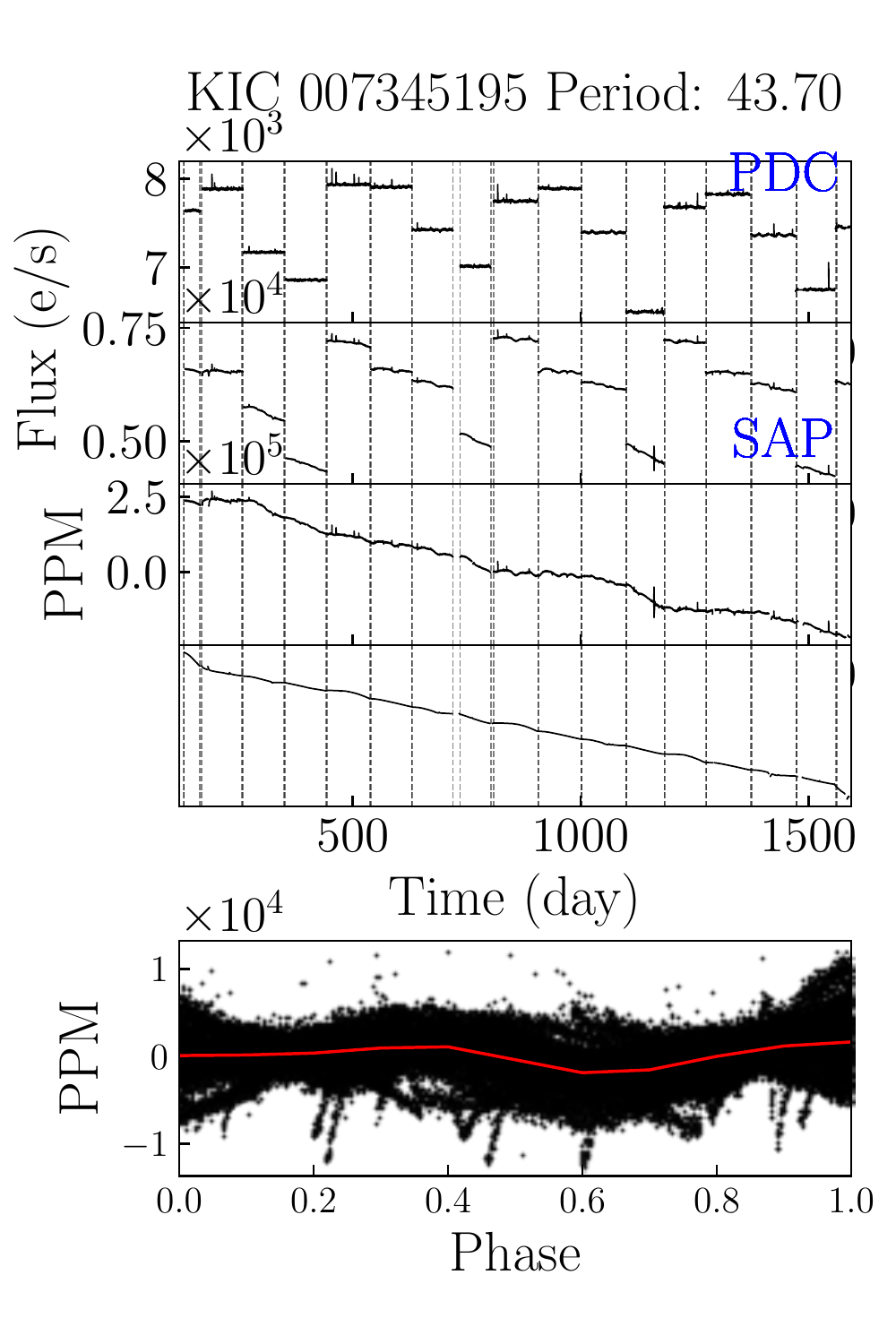}
	\includegraphics[width=0.32\textwidth]{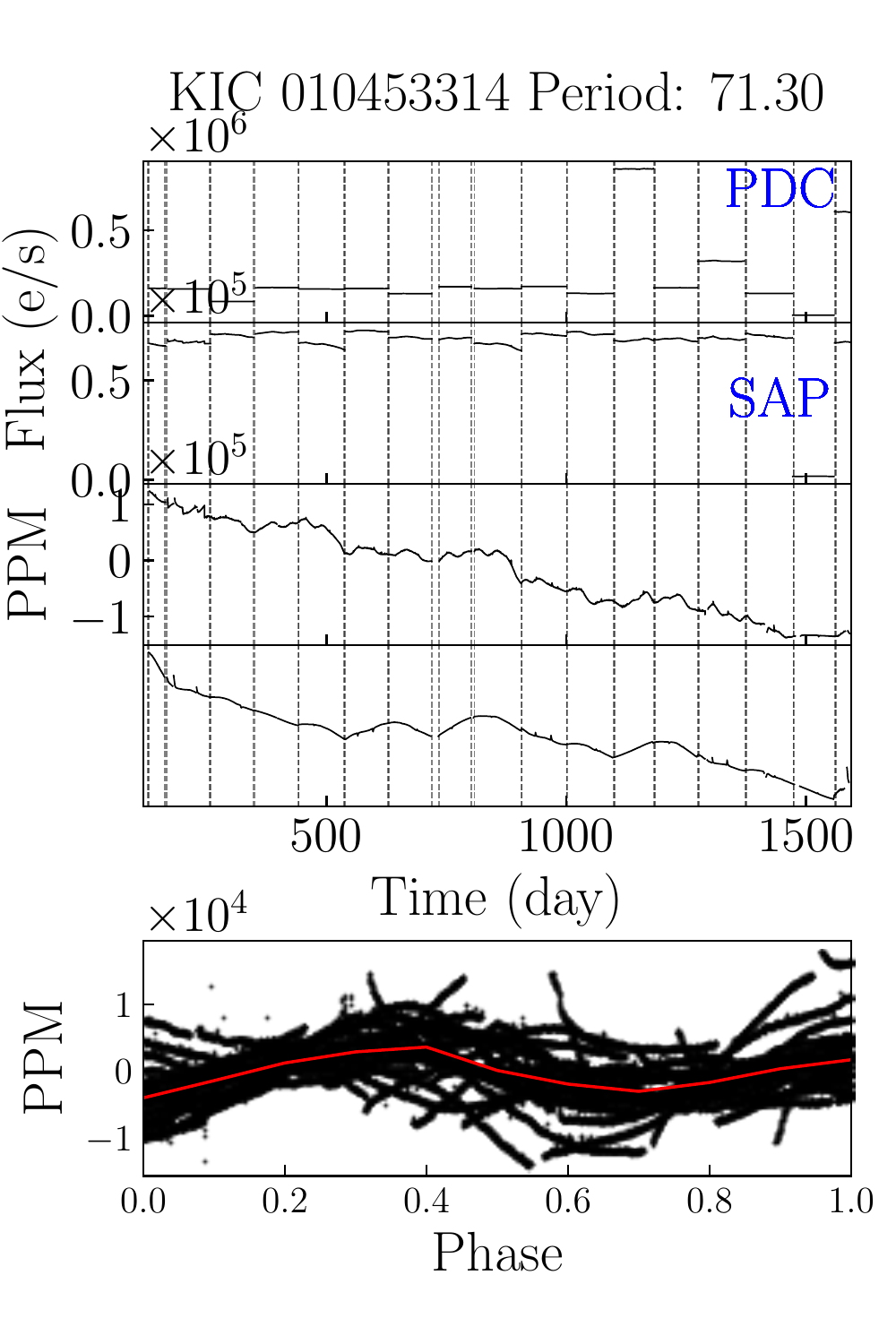}
	\includegraphics[width=0.32\textwidth]{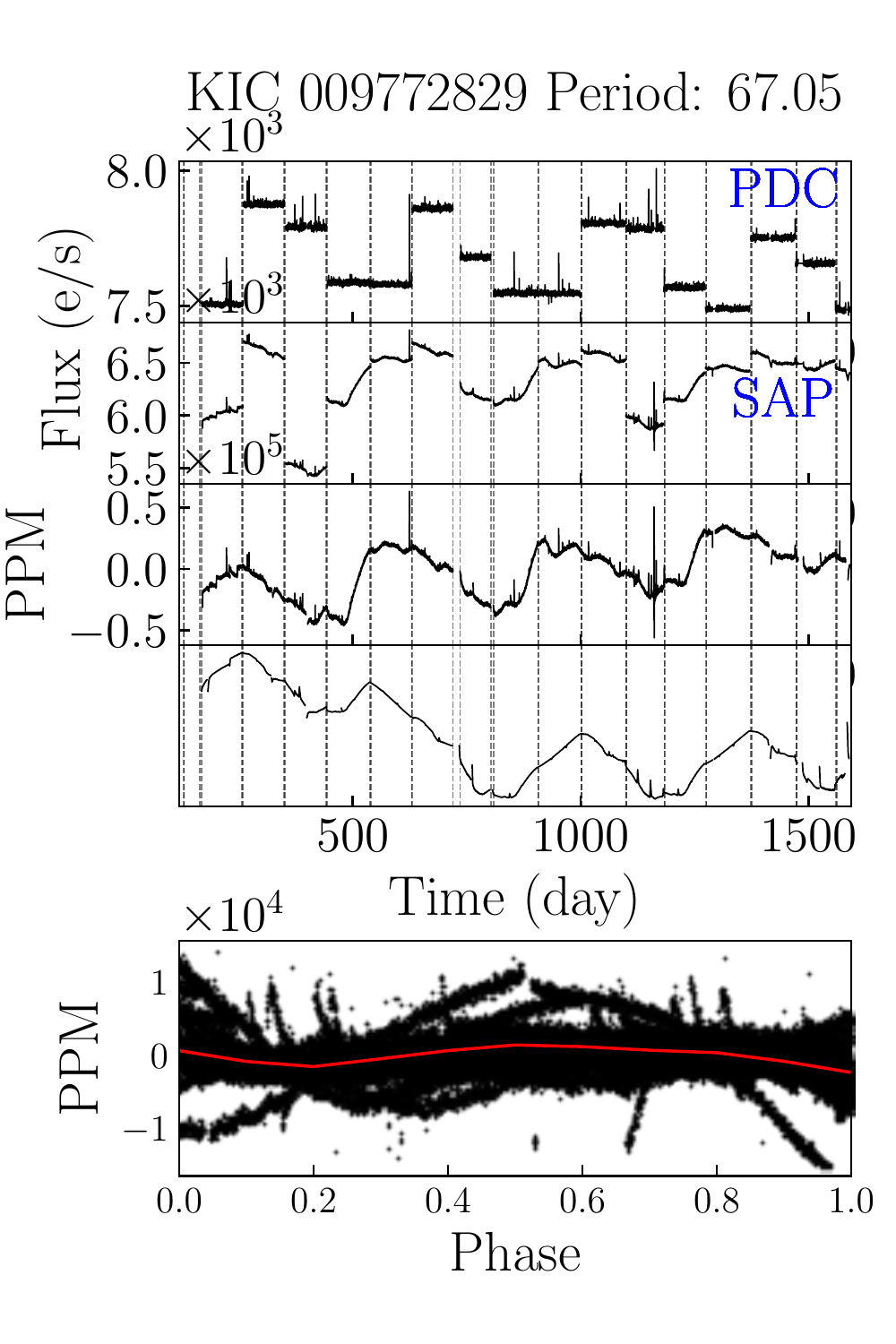}
	\caption{
		Six examples of results after our selection process. 
		In each sub-figure, the title shows the KIC ID and its period, the black line in the top two panels display the raw PDC and SAP LCs. 
		The next two panels show the concatenated SAP LCs and the sCBV LCs, the bottom panel contains the folded LC points and median value curve.}
	\label{fig: results_figures}
\end{figure*}

After clarifying the definition of those two similarities, we could select dissimilar candidates correspondingly. For each object in every channel sequence group, we check whether there are more than 3 stars that are similar to its period and LC shape (i.e., the Hausdorff distance is less than 90).
If this condition is true, we believe that the period is caused by artefact. 

Practically, we determine the similarity for every LC in our three types of filtered LCs respectively. Besides, because the first type of filtered LC has two candidate periods, for each one of the first type of filtered LC, only when both of the two detected periods meet the conditions of similar period and LC shape, can we mark this LC as `similar'. For every filtered LC, we also compare the similarity with its corresponding sCBV LC. Finally, we found 
72,831 similar objects for the first filtered LC, 126,376 similar objects for the second filtered LC and 126,242 similar objects for the third filtered LC.

Then, since we are aiming at finding long period rotators, we choose the $ \log g > 4.0 $ as the threshold to select main-sequence stars. The Kepler Input Catalog (KIC) provides basic parameters for most sources in the \kepler mission field of view \citep{Brown2011}, however, several studies have found the KIC parameters are inaccurate and have some systematic defects \citep[e.g.,][]{Hekker2011, Liu2014}. \citet{McQuillan2014} used the $ T_{\mathrm{eff}}-\log g $ cuts derived by \citet{Ciardi2011} with KIC parameters. Instead, we also find this threshold would introduce some sub-giants and contaminate our sample. 

Therefore, we need to use a corrected \kepler stellar parameter catalogue to make our sample as reliable as possible. \citet{Huber2014} derived the atmospheric properties from different observational techniques such as photometry, spectroscopy, asteroseismology and exoplanet transits. After DR25, \citet{Mathur2017} published a new version of the revised KIC parameters, which contains 197,096 targets from \kepler DR25 dataset, and they released the stellar properties catalogue, `stellar17', at the Mikulski Archive for Space Telescopes\footnote{\url{http://archive.stsci.edu/kepler/stellar17/search.php}} (MAST) and the NASA Exoplanet Archive\footnote{\url{http://exoplanetarchive.ipac.caltech.edu}}. Some unclassified stars with no reliable parameters in `stellar17' are removed.

Another vital work was done by the Large Sky Area Multi-Object Fiber Spectroscopic Telescope\footnote{Also named as the Guo Shoujing Telescope. \url{http://www.lamost.org}} (LAMOST), which produced a great number of low-resolution spectra \citep{Su1998, Cui2012}. \citet{DeCat2015} initiated the LAMOST-\kepler (LK-project) to acquire LAMOST observations for objects in the \kepler field. Subsequently, \citet{Ren2016} derived the atmospheric parameters by adopting the LAMOST Stellar Parameter pipeline (LASP). They indicated that the parameters derived from the LK-project are a useful tool to generate a corrected KIC parameters catalogue.
Therefore, we use the official LASP parameters from the A, F, G and K type star catalogue of LAMOST DR5 and DR4 to correct the stellar17 and KIC parameters. When we meet a replicated object in different catalogues, the priority of different parameters is LAMOST DR5 > LAMOST DR 4 > stellar17 > KIC. 

Removing Eclipsing Binaries (EBs) is also necessary for us, and finally, the known EBs recorded in the third version of the \kepler Eclipsing Binary Catalog \citep{Kirk2016} are removed from our sample. \citet{McQuillan2014} used the KOI catalog to remove planetary transit signals, while \citet{McQuillan2013a}, \cite{Mazeh2015} and \cite{Angus2018} found some rotation periods in the KOI dataset. Therefore, we only remove the EBs, for keeping the possibility of finding some long periods in the planetary systems. 

After the selection process of dwarf and EB, there are 79,744 dissimilar stars remaining in the first type of filtered LCs, 34,481 in the second type of filtered LCs and 28,776 in the third type of filtered LCs. 

As shown in Section \ref{subsec: test-results}, in order to improve the completeness of our detection as much as possible, many false positive periods were introduced into our detection results. Therefore, we choose the phase dispersion minimisation (PDM) \citep{Stellingwerf1978} method to select objects with significant periodic signals. We study the results of the PDM method (i.e., $ \theta $) to distinguish whether the period of candidate is significant. Considering the poor quality of SAP data, we choose 0.9 as the threshold of $ \theta $. Then, 9940 objects remaining in the first type of filtered LCs, 676 in the second type of filtered LCs and 457 in the third filtered LCs. 


\begin{figure*}
    \centering
	\includegraphics[width=0.329\textwidth]{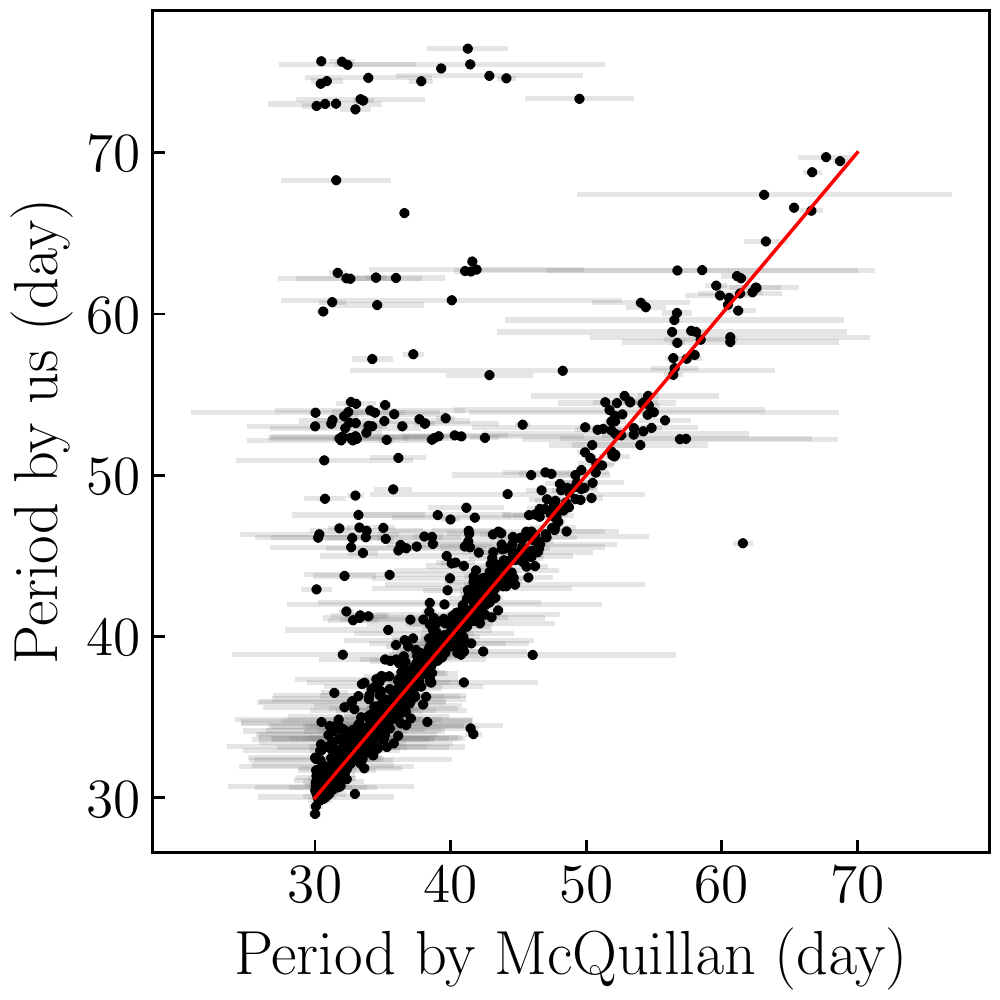}
    \includegraphics[width=0.329\textwidth]{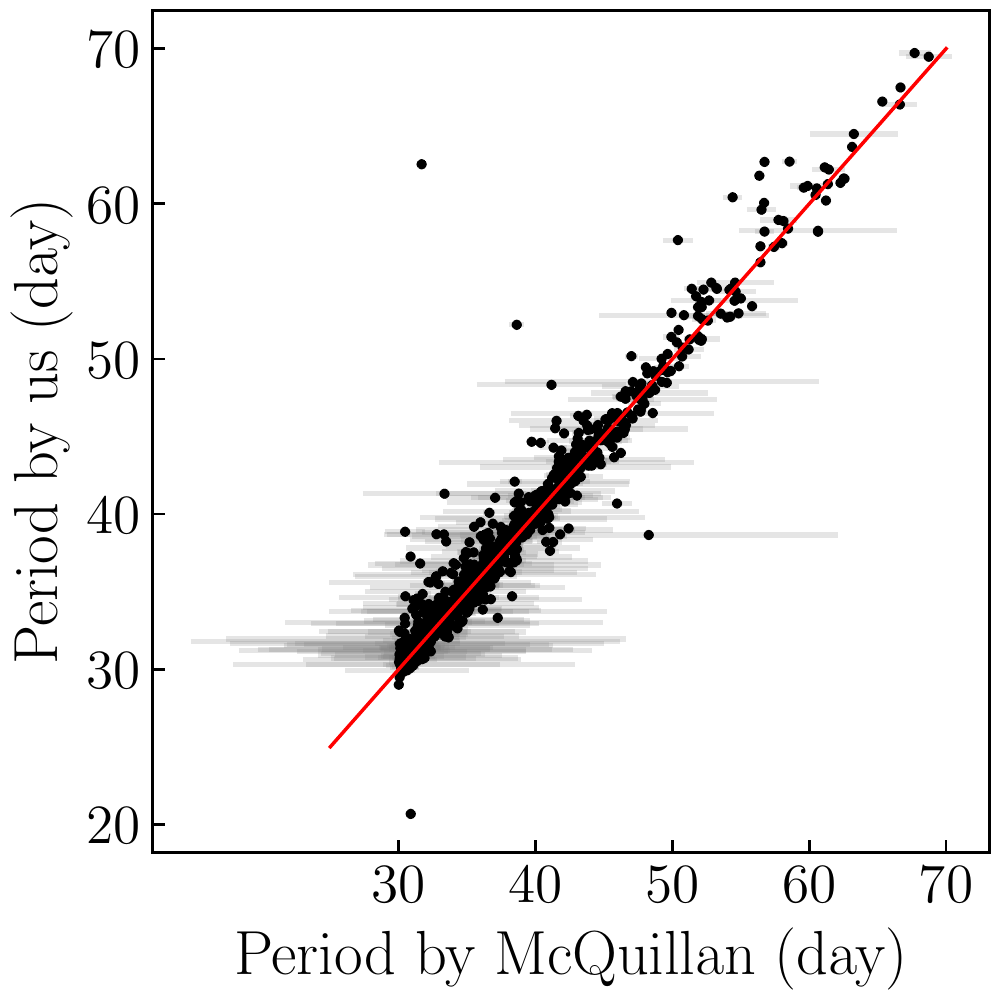}
	\includegraphics[width=0.329\textwidth]{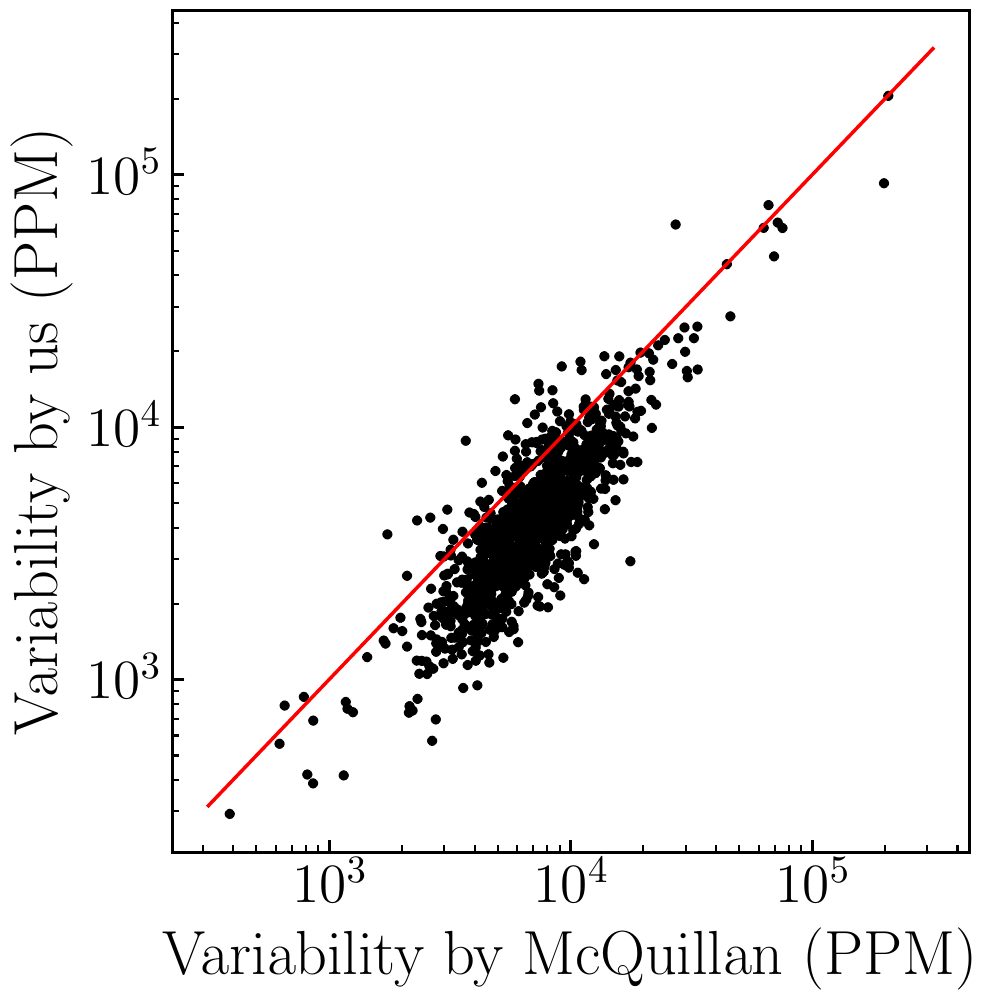}
	
	\caption{Comparisons of our detected results and those of \citet{McQuillan2014}. The left panel shows the period comparison between our results and those of \citet{McQuillan2014}. The middle panel shows the period comparison between our inspected results and those of \citet{McQuillan2014}. The right panel is the comparison of variability from our inspected results and those of \citet{McQuillan2014}. The red line illustrates the 1:1 line.}
	\label{fig: mcq-comparison}
\end{figure*}

\section{Comparison and Results}\label{sec:results}
After our period detection and selection procedure, 1224 objects are cross-matched with the results of \citet{McQuillan2014}.
As introduced in Section \ref{sec:intro}, the periods of \citet{McQuillan2014} are relatively longer than other previous studies. Therefore, we show a direct comparison between the periods measured in our work and those detected by \citet{McQuillan2014} (left panel of Figure \ref{fig: mcq-comparison}). 
If we trust the results from the \citet{McQuillan2014}, 953 of 1224 periods are within 5 per cent errors, which means we have nearly 80 per cent reliability of period after selection process, apparently higher than the simulated test results. Although the improvement is significant, we still could not trust the detected periods directly. 
Thus, to obtain a reliable result, as the number of candidates decrease after the selection process, a visually check could be applied. 
In our visually inspection, we recalculate the Lomb-Scargle and fold the LCs to check whether there are better periods, and we also zoom in the LCs to check their variations. If the a periodical variation always turn at the start or end of quarters, we would not trust the period; if the periodical signals always follow the discontinuities (might caused by the sudden pixel sensitivity dropouts (SPSD)), and if the period is apparently unstable, we would not trust them either.

Then we visually check the 1224 objects, most of the periods are consistent with the results of \citet{McQuillan2014}, showing the similar folded SAP LCs.
Some of the objects are removed because the periodical signals are so weak (low signal over noise ratio in the SAP) that we could not find the best folded LCs, even if we tried the periods from the \citet{McQuillan2014}. Some of the periodical variations often follow the discontinuities, as shown in Figure \ref{fig: visually_inspect_bad}; for those objects, even though we could find the better periods, we still remove them because of their serious instrumental effects. 
We also find few objects have obviously better periods of our results than those of \citet{McQuillan2014}. Three examples are shown in \ref{fig: visually_inspect_good}, it is clear that our periods yield better folded LCs. Some periods are updated in our results by manually adjusting. Finally, after the visually inspection, 1107 objects are preserved. Table \ref{tab:mcq_results} list their parameters and periods from \citet{McQuillan2014} and us. 

\begin{table}
	\caption{A portion of inspected results cross-matched with \citet{McQuillan2014}, the full table is available online.\label{tab:mcq_results}}
	\begin{threeparttable}
		\begin{tabular}{cccccc}
			\hline
            KIC ID    & Teff    & $\log g$  & $[\mathrm{Fe}/\mathrm{H}]$ & PRot & Period \\
                     & (K)     & (dex) & (dex) & (day) & (day) \\ \hline
            4750609  & 5042.09 & 4.64  & 0.01 & 35.39 & 35.71 \\
            4644300  & 4817.15 & 4.50  & 0.17 & 36.14 & 34.83 \\
            2986505  & 4488.08 & 4.51  & 0.04 & 40.73 & 40.18 \\
            1865663  & 4155.31 & 4.49  & 0.04  & 31.32 & 31.97 \\
            4254223  & 3879.62 & 4.51  & -0.32 & 31.15 & 31.31 \\
            4553344  & 5135.66 & 4.73  & 0.14 & 33.95 & 34.41 \\
            8431125  & 4588.25 & 4.57  & -0.12 & 31.14 & 31.61 \\
            10753394 & 4261.92 & 4.62  & 0.02 & 32.14 & 32.48 \\
            8935997  & 5228.16 & 4.54  & -0.43 & 32.73 & 32.57 \\
			\hline
		\end{tabular}
		\begin{tablenotes}
			\item PRot are taken from \citet{McQuillan2014}.
		\end{tablenotes}
	\end{threeparttable}
\end{table}

After the inspection of these cross-matched objects, we also compare our revised periods with those from \citet{McQuillan2014}. As shown in the middle sub-figure of Figure \ref{fig: mcq-comparison}, only a few objects have significant different periods with \citet{McQuillan2014}, and we believe in our periods because of the folded results.

We also show a similar plot that compares our variability and $ R_{\mathrm{per}} $ (amplitude of periodic variability, taken from \citet{McQuillan2014}) in the right panel of Figure \ref{fig: mcq-comparison}. Most variabilities are similar but our amplitudes are slightly smaller than those of \citet{McQuillan2014}. 
The reason is that our folded LCs are aligned at the median value. 
In our process of folding, we align the median of the points in each period to zero, and then calculate the variability range from the median of the folded LC (illustrated by the red lines in Figure \ref{fig: results_figures}).
This process makes a LC less likely to be influenced by some occasional variations. 

Then, according to the results of comparison and the improvement after our visually inspection, we visually check all the selected candidates of Section \ref{sec:can_sele}, and 165 objects are newly found.
A portion of our new results are illustrated in Table \ref{tab:results}, some examples are shown in Figure \ref{fig: results_figures}.


\begin{table*}
	\caption{A portion of our new results, the full table is available online.\label{tab:results}}
	\begin{threeparttable}
		\begin{tabular}{cccccccc}
			\hline
			KIC ID   & Teff    & $\log g$ & $[\mathrm{Fe}/\mathrm{H}]$ & Period & Period Error & Variability\\
			& (K)     & (dex)    & (dex)  & (days) & (days)       & (ppm)\\ \hline
			8302593  & 4897.99 & 4.70     & -0.13 & 34.89  & 1.74         & 759.45\\
			8915957  & 7001.38 & 4.25     & 0.39  & 46.94  & 2.35         & 42988.00\\
			9974087  & 6960.90 & 4.26     & 0.34  & 41.54  & 2.08         & 2947.22\\
			6192231  & 5397.96 & 4.07     & 0.32  & 33.26  & 1.66         & 14871.34\\
			7673194  & 5247.48 & 4.56     & 0.12  & 35.35  & 1.77         & 590.01\\
			9458505  & 4795.23 & 4.64     & 0.05  & 33.57  & 1.68         & 1712.66\\
			7773978  & 5300.39 & 4.24     & -0.03 & 30.85  & 1.54         & 1907.20\\
			7134214  & 5756.06 & 4.20     & 0.55  & 59.77  & 2.99         & 910.34\\
			6880781  & 4465.40 & 4.72     & -0.27 & 35.96  & 1.80         & 967.43\\
			10665957 & 4842.81 & 4.61     & 0.11  & 35.65  & 1.78         & 1314.89\\
			9639369  & 4742.12 & 4.60     & 0.11  & 36.81  & 1.84         & 1113.66\\
			11074541 & 4763.66 & 4.68     & 0.00  & 35.07  & 1.75         & 2872.73\\
			11650280 & 5878.22 & 4.14     & -0.27 & 59.76  & 2.99         & 42244.16\\
			9941066  & 4013.57 & 4.58     & -0.27 & 32.44  & 1.62         & 7251.10\\
			8242434  & 4565.28 & 4.62     & -0.03 & 46.47  & 2.32         & 1692.52\\
			11197878 & 5386.17 & 4.55     & 0.16  & 31.32  & 1.57         & 3954.30\\
			11363367 & 6595.13 & 4.17     & -0.35 & 48.68  & 2.43         & 10134.28\\
			8674074  & 4050.34 & 4.53     & -0.06 & 58.50  & 2.93         & 1648.42\\
			8674005  & 4796.85 & 4.66     & -0.30 & 31.15  & 1.56         & 1144.26\\
			7466500  & 6077.76 & 4.28     & 0.05  & 48.80  & 2.44         & 1556.85\\
			\hline
		\end{tabular}
	\end{threeparttable}
\end{table*}

\begin{figure*}
    \includegraphics[width=0.32\textwidth]{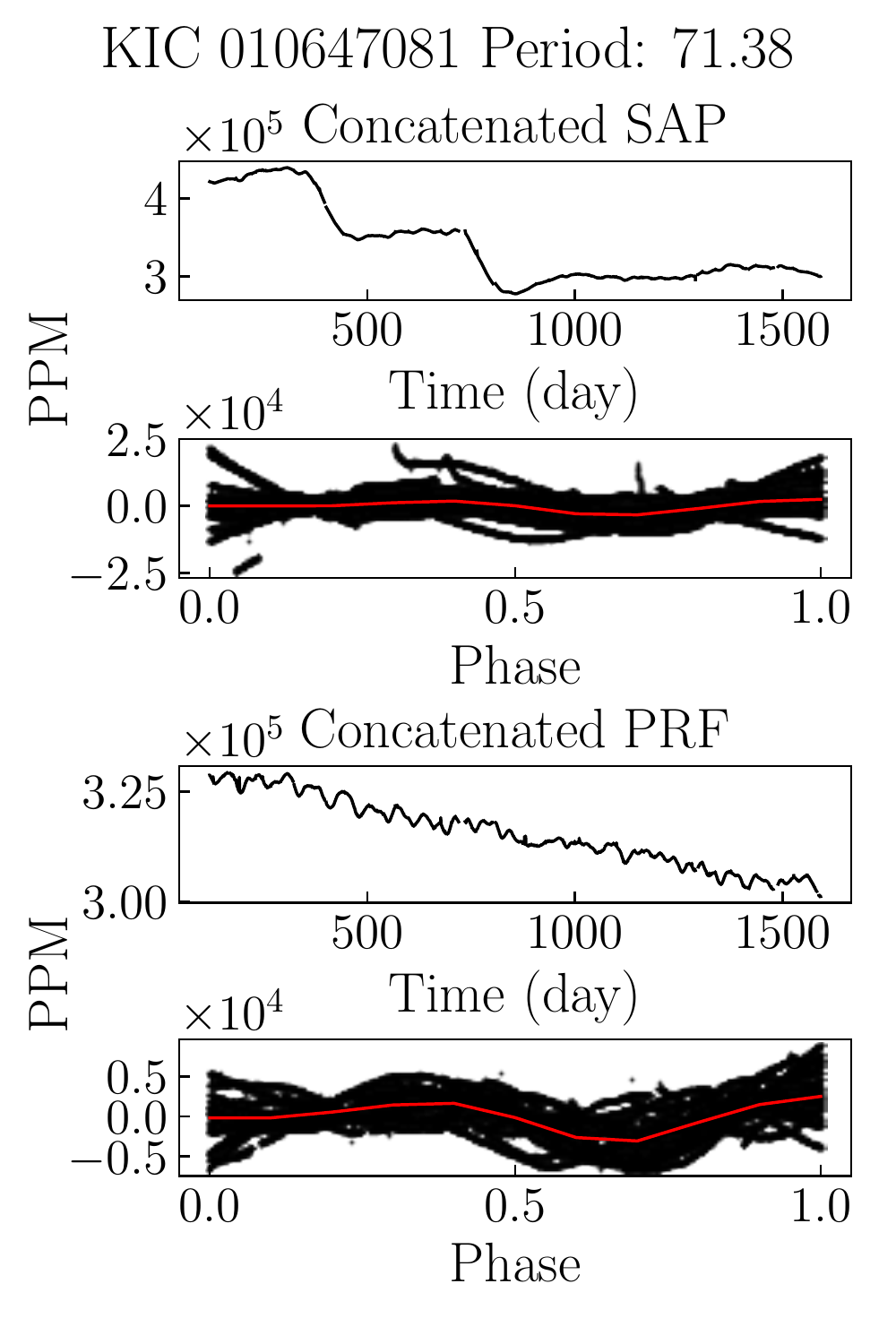}
    \includegraphics[width=0.32\textwidth]{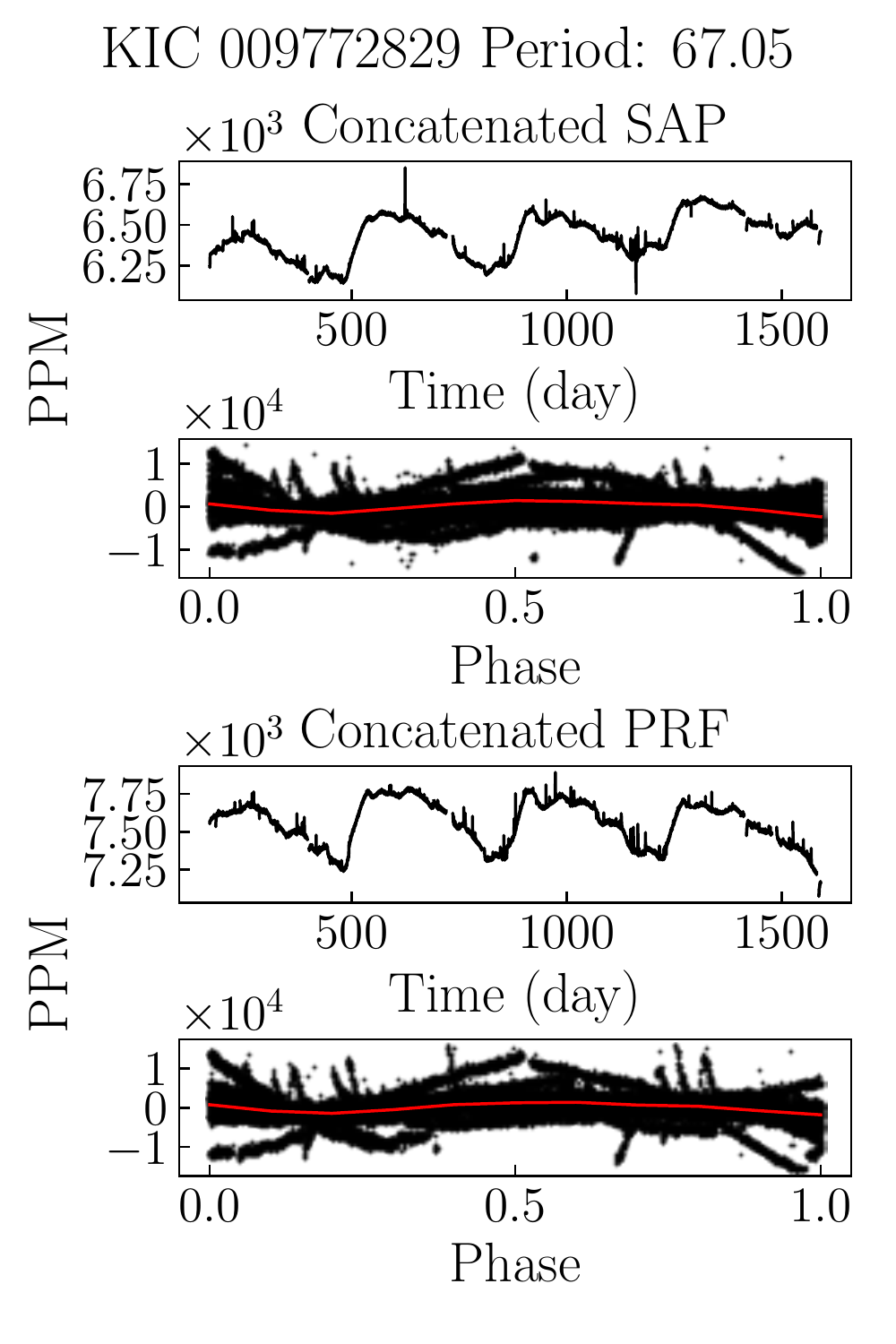}
    \includegraphics[width=0.32\textwidth]{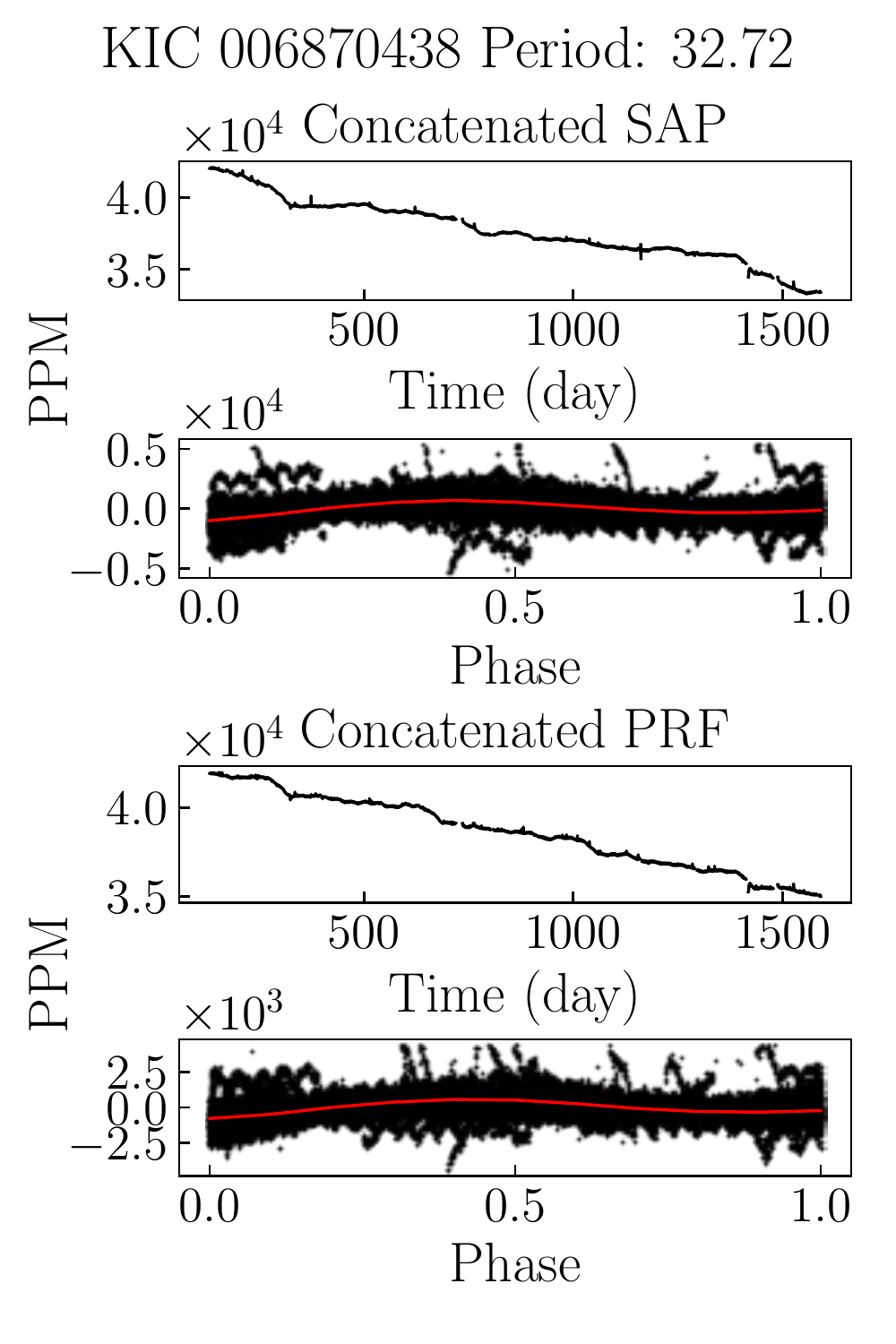}
	\caption{
		Three examples of the comparisons of the PRF fitted LCs and the SAP LCs. In each sub-figure, the title shows the KIC ID and its period, the upper two panels show the SAP LC and folded LC correspondingly; the bottom two panels show the PRF fitted LC and folded LC correspondingly. The red lines indicate the median value of the folded points.}
	\label{fig: prf_results}
\end{figure*}

\section{Photometry inspections}\label{sec:phot_insp}



In our work, we choose the SAP because of its preservation of long term signals, but since the SAP reduction used the fixed apertures, it may includes more systematics or not keep all the target flux \citep{Montet2017}. So, there should be some photometry inspections for our long rotation period objects before further discussion of the activity. Here we choose the Target Pixel File (TPF) and Full Frame Image (FFI) to investigate our SAP LCs.

Firstly, we tried the Pixel Response Function (PRF) model \citep{Bryson2010} fitting photometry (also known as Point Spread Function (PSF) fitting photometry) for our newly found objects by using the lightkurve package. 
\citep{2018ascl.soft12013L}\footnote{\url{https://github.com/KeplerGO/lightkurve}} The PRF fitting is successfully used in planet investigations\citep[e.g.][]{Batalha2013, Torres2011}, and crowded K2 fields for faint stars\citep{Vanderburg2014, Libralato2016}.
In our fitting process, we choose the default Gaussian prior of star position, background, focus and motion, and the Poisson posterior for the maximum likelihood estimator. Then, we remove the 5-sigma outliers with sigma-clipping method and combine the fitted LCs of every quarter (following the same method described in Section \ref{sec:lc_pre}).

After we visually inspect the fitted LCs with folding and compare them with the SAP LCs, we find that many of the fitted LCs preserve the long-term trends, but some shapes of the trends are different with the SAP; and some fitted LCs show clearer periodical signals than the SAP. Most of the fitted LCs have the same periodical signals, which yield similar folded LCs. Besides, few objects show apparently different LCs with the SAP, for these objects, we apply a manually adjustment for their apertures in the next aperture inspection. In general, most of our periods are confirmed with the PRF fitting photometry.
Here we plot three examples of the concatenated PRF fitted LCs and the concatenated SAP LCs correspondingly in Figure \ref{fig: prf_results}.

\begin{figure*}
    \includegraphics[width=0.495\textwidth]{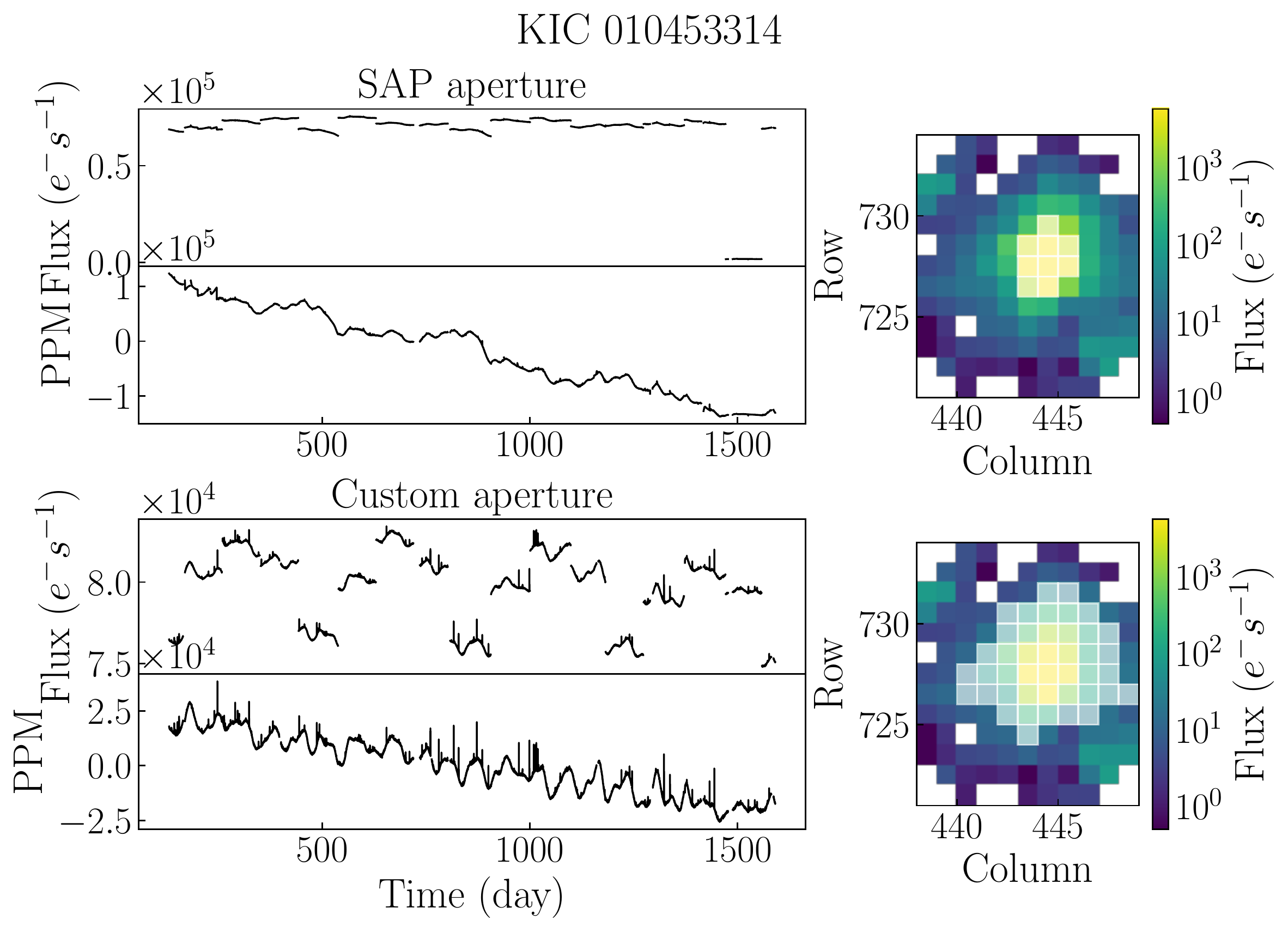}
    \includegraphics[width=0.495\textwidth]{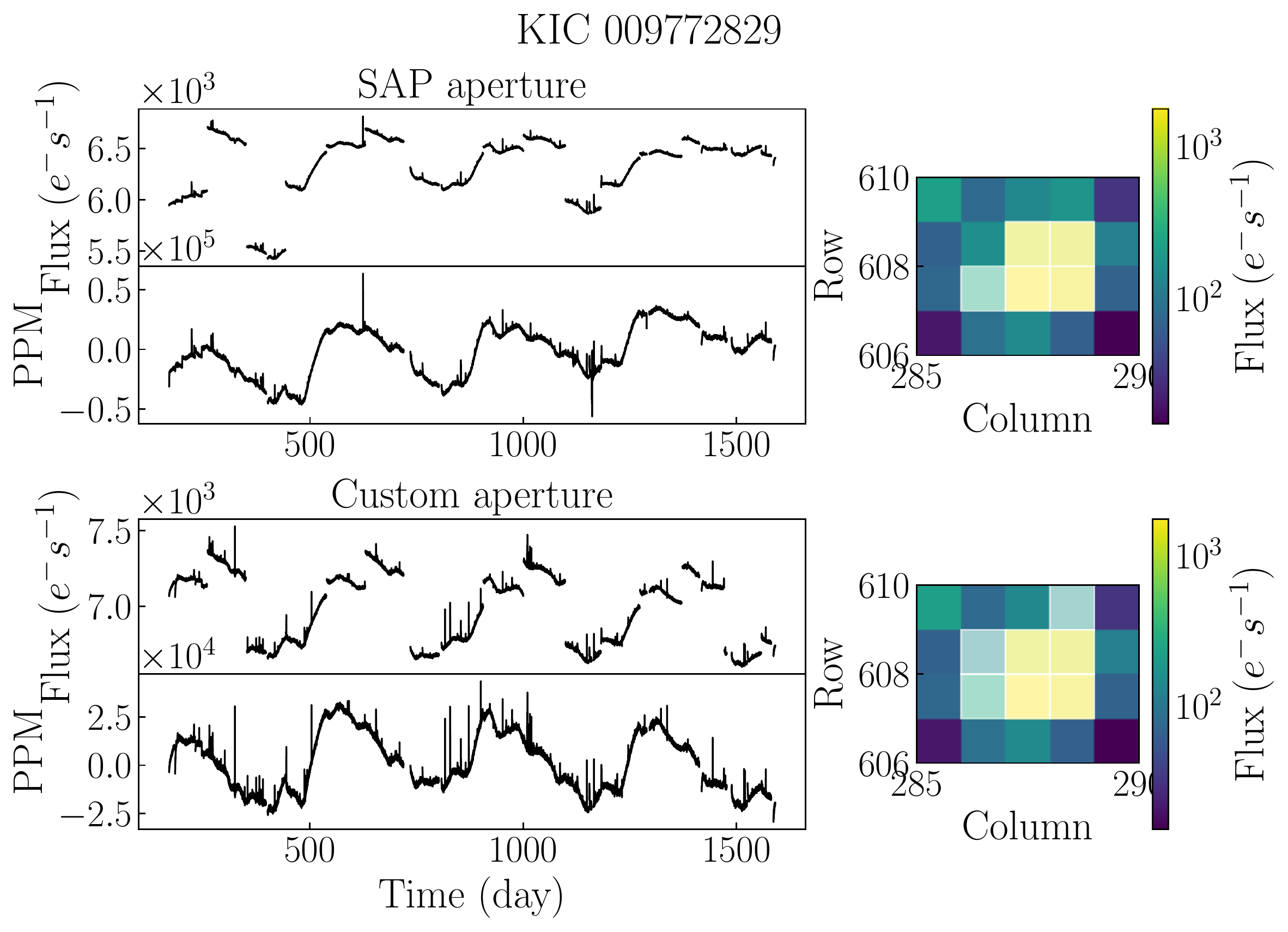}\\
    \includegraphics[width=0.495\textwidth]{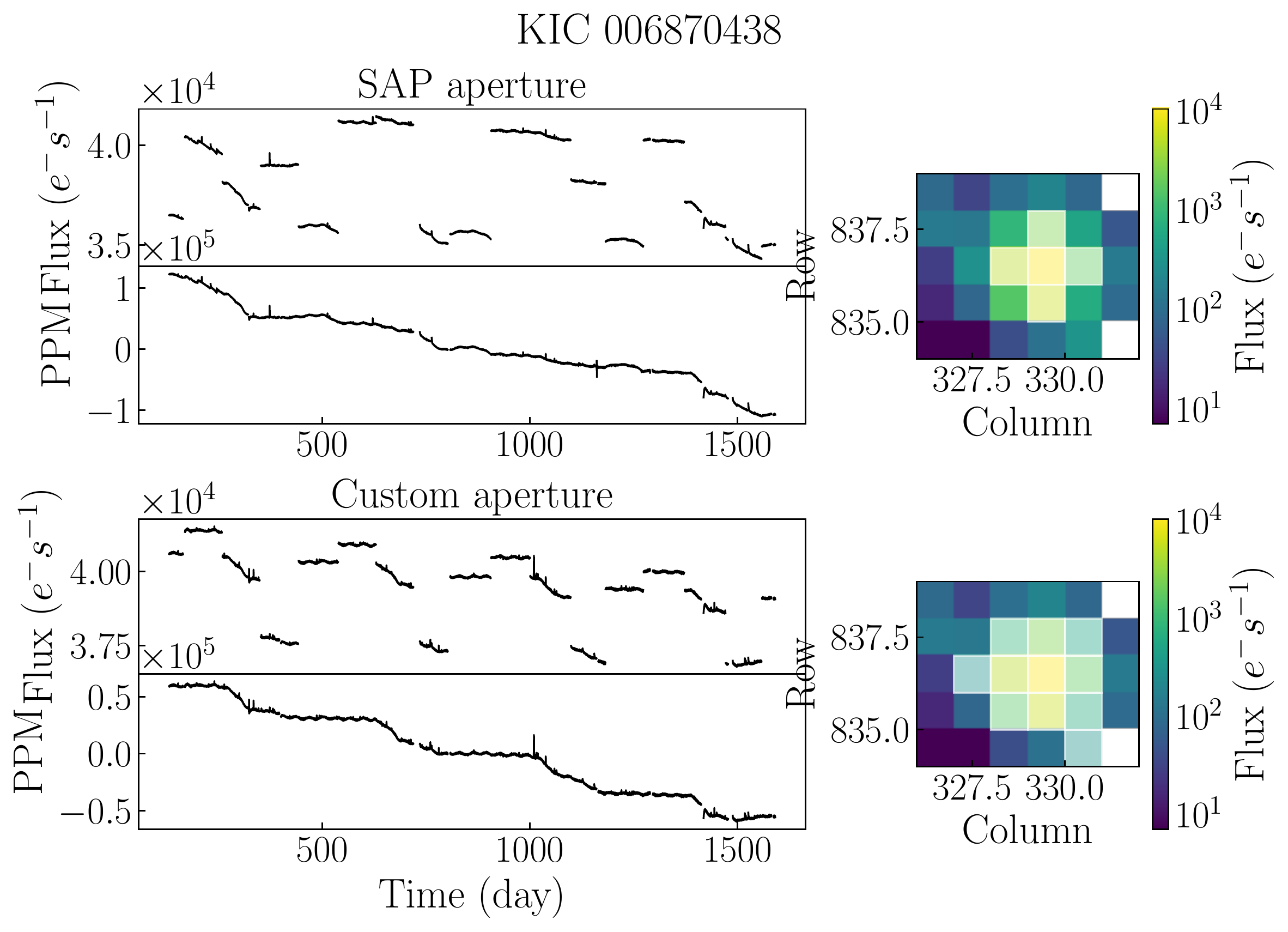}
    \includegraphics[width=0.495\textwidth]{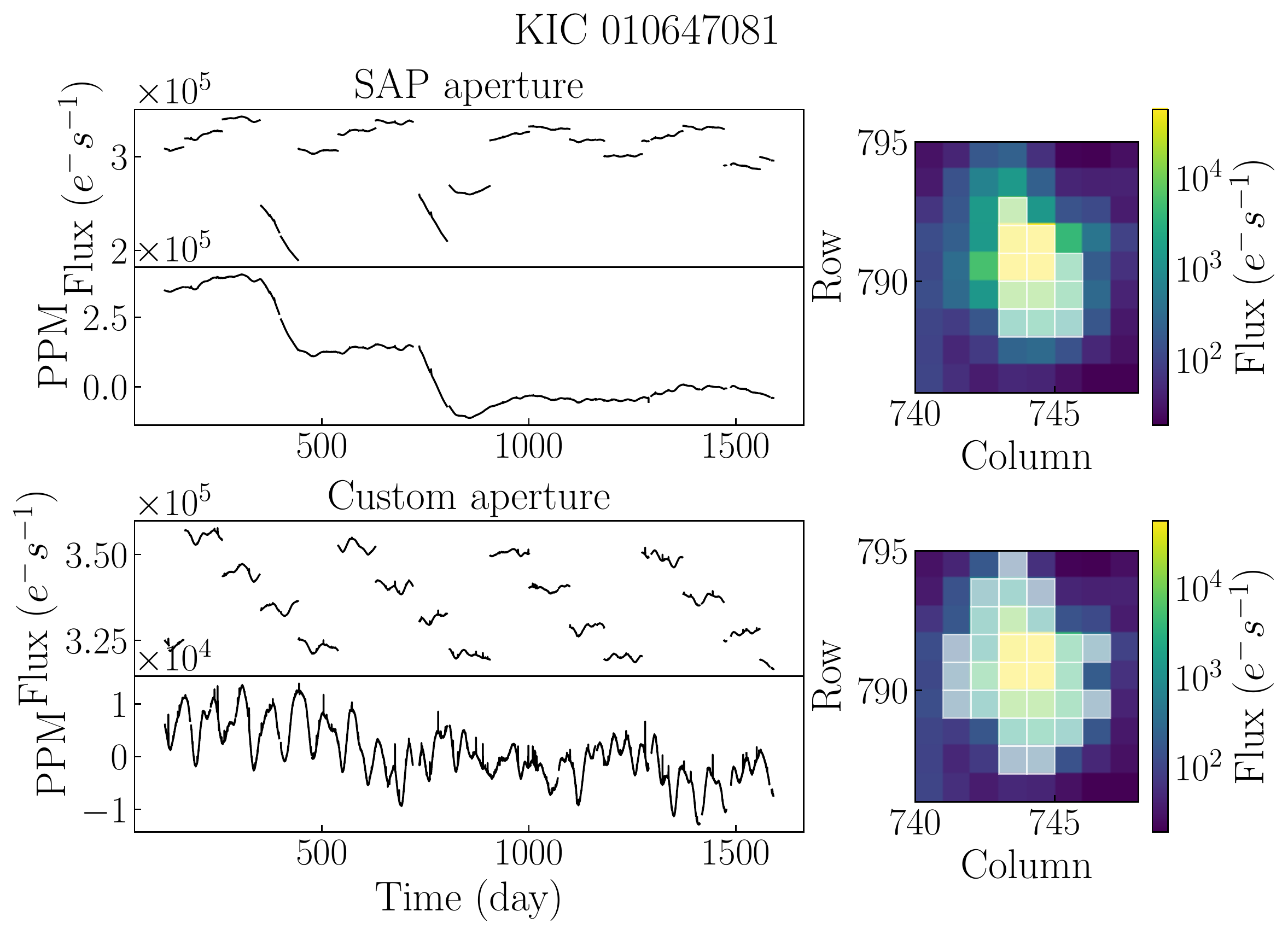}
	\caption{
		Four examples of the comparisons of the larger apertures and the SAP. In each sub-figure, the LCs are shown in the left side, the pixel images are shown in the right side and the apertures are masked by the white semitransparent pixels. The top two LCs are the SAP LCs, the upper one is the raw SAP, the lower one is the concatenated SAP LC. The bottom two LCs have the larger apertures.}
	\label{fig: larger_aper_results}
\end{figure*}

Then, we perform aperture photometry with a larger aperture. If the aperture of the SAP is too small, increasing the aperture size would be helpful to include more of the stellar flux, and the systematics may change so that it could be distinguished. Therefore, we choose an aperture to mask the connected pixels with flux greater than the 60th percentile, and we use the same procedure as the PRF fitting photometry to combine the LCs and remove outliers.

We also inspect these larger aperture LCs and compare them with the SAP LCs. Most of objects show a more apparent periodical signals, because a larger aperture includes more flux; although many of them still keep the long-term trends but the variations are different. 
So, most of our results are confirmed by the larger aperture photometry. However, we also find few objects show totally different LCs with the SAP, for these objects and the uncertain objects in the PRF fitting photometry, we have a inspection with manually changing their aperture masks. 
After the manually inspection, we find that three of the larger apertures include some pixels of the nearby stars, after we exclude the contamination, they have the similar signals to the SAP; Four objects are sensitive to the size of apertures, so we add the asterisk signs to their KIC ID in Table \ref{tab:results}. 

\begin{figure*}
    \includegraphics[width=0.9\columnwidth]{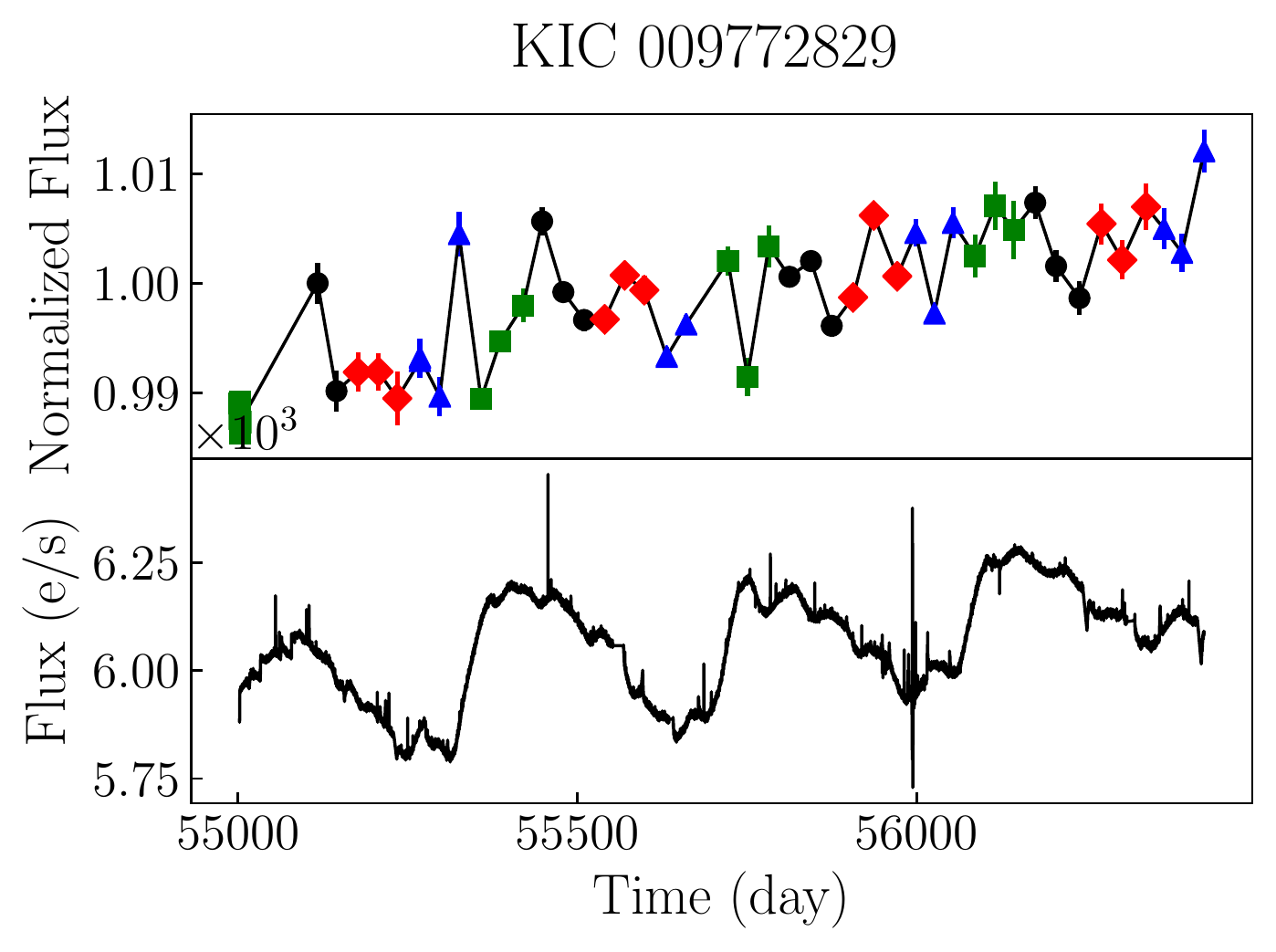}
    \includegraphics[width=0.9\columnwidth]{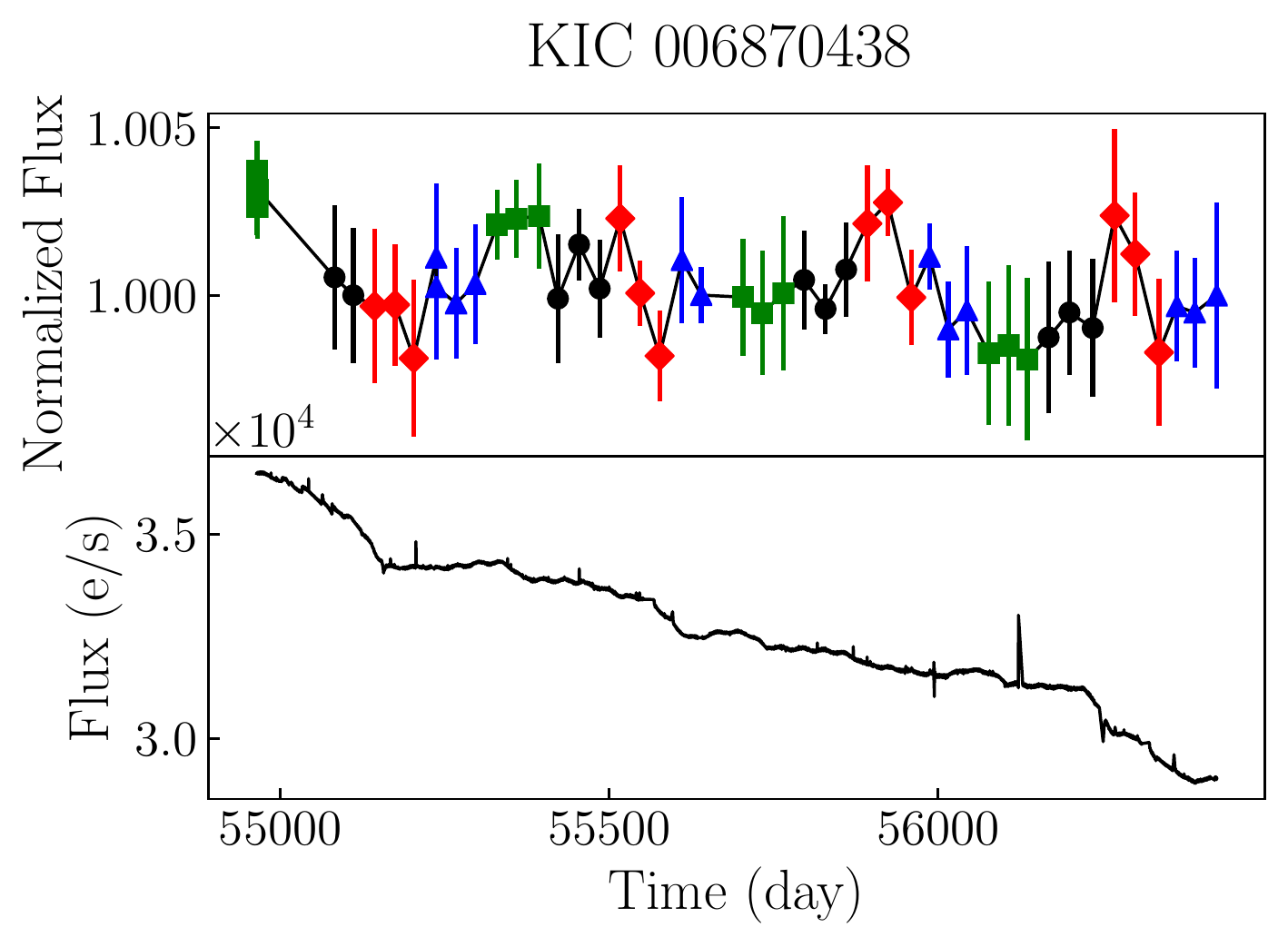}\\
    \includegraphics[width=0.9\columnwidth]{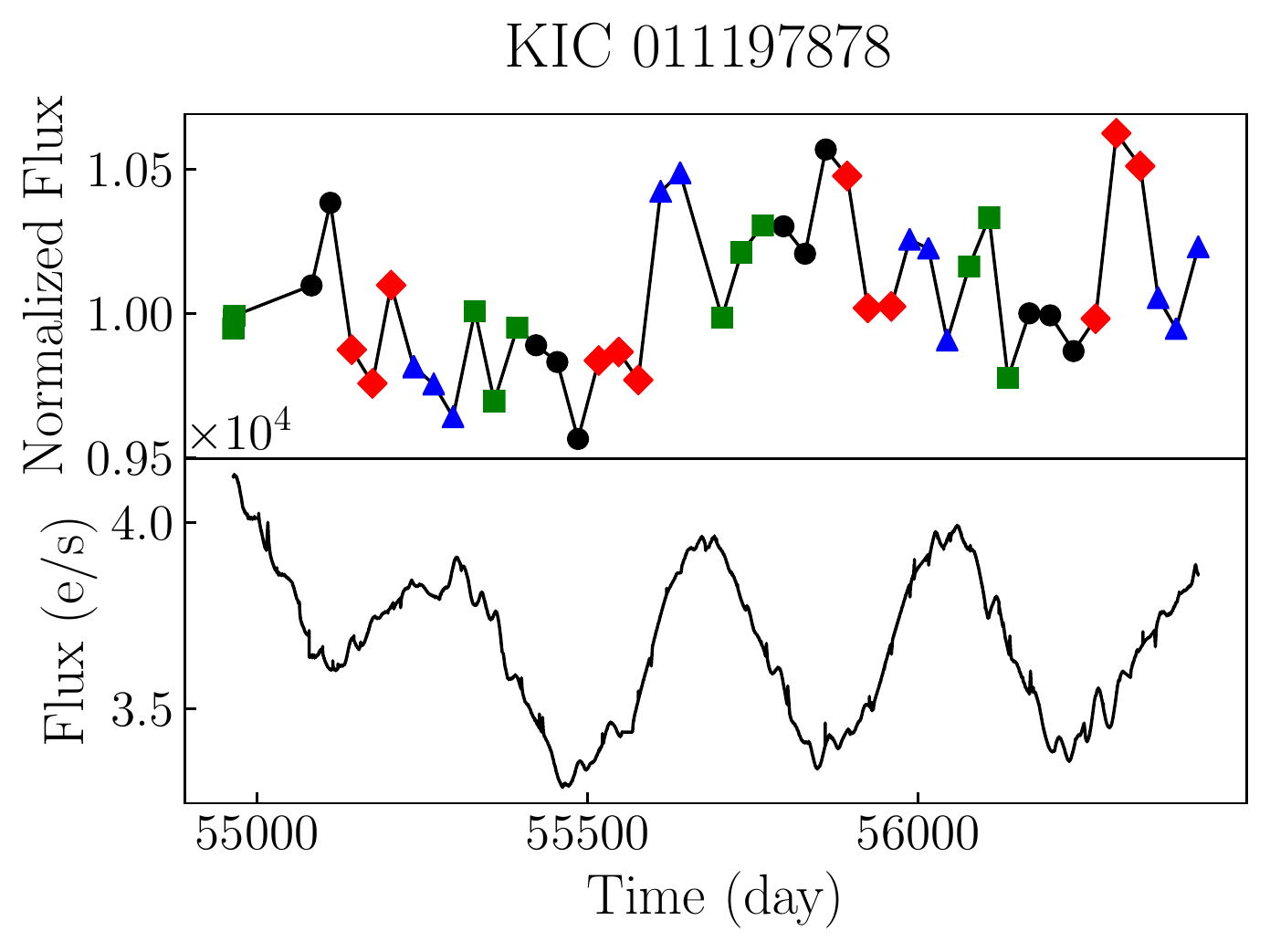}
    \includegraphics[width=0.9\columnwidth]{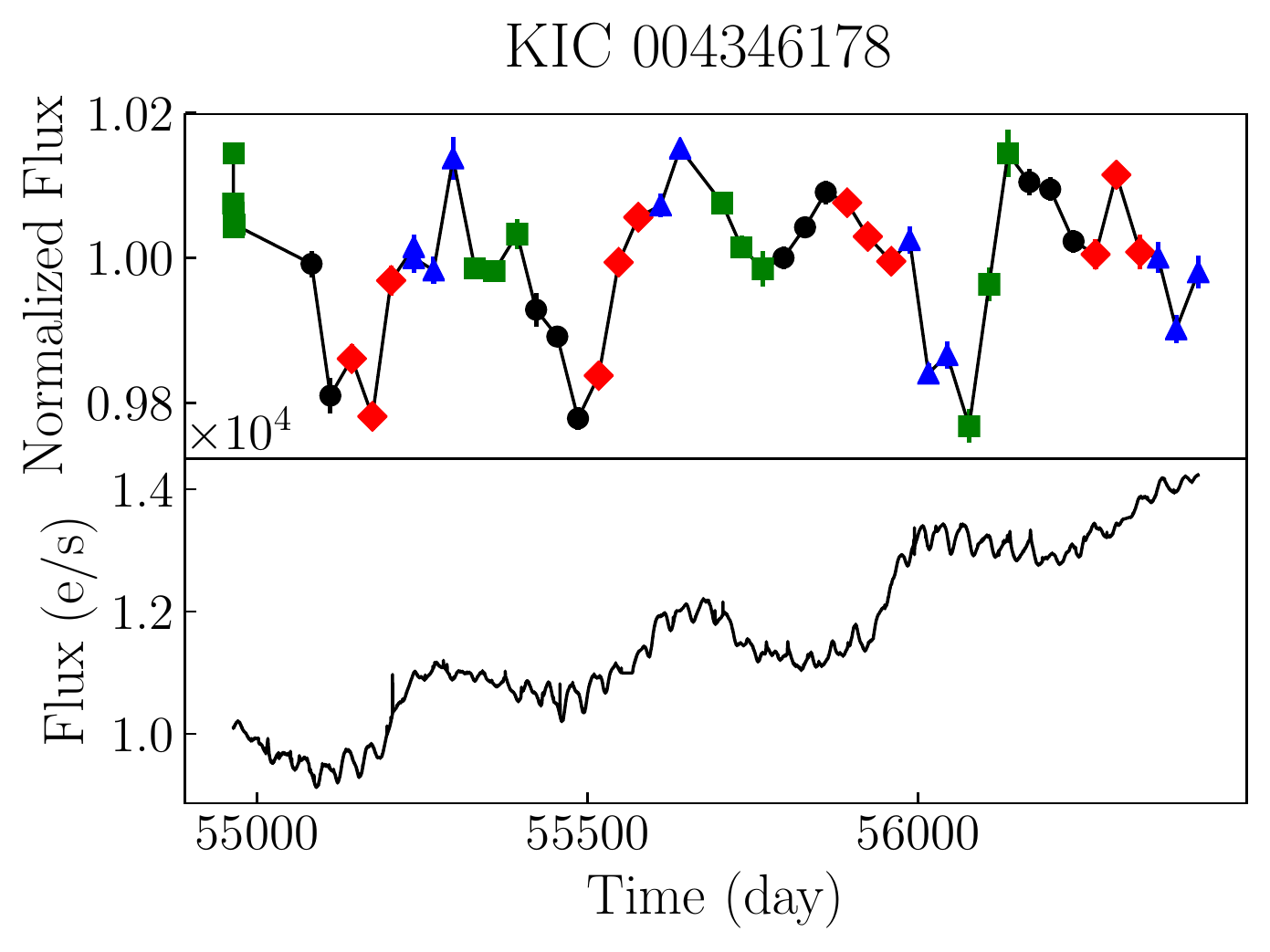}
	\caption{
		Four examples of the comparisons of the FFI and the SAP. In each sub-figure, the top panel is the FFI photometry, the bottom panel is the concatenated SAP LC. In the FFI photometry, the green squares, blue triangles, black circles, and red diamonds indicate the four quarters of a year, the same with those figures of \citet{Montet2017}.}
	\label{fig: ffi_results}
\end{figure*}

In addition, \citet{Montet2017} found some long-term magnetic cycles in \kepler through the FFI photometry. Although the FFI is not suitable for searches of rotation period because of its low duty cycle (one point per month), it might be helpful to distinguish some systematics with large variations. So, with the help of f3 \citep{Montet2017}\footnote{\url{https://github.com/benmontet/f3}}, we check the FFI photometry for our newly found objects with the default aperture. We find that the FFI could match some large variations in the SAP, but not the small or short period signals. Most of the one-year large variations of the SAP are different with those of the FFI, which confirms the long-term trends in the SAP are not astrophysical.
Four examples are shown in Figure \ref{fig: ffi_results}.

\section{Summary}\label{sec:summary}
We have carried out a light curve pre-processing work based on Butterworth filter and detect periods with Lomb-Scargle periodogram from the \kepler SAP data. 
The rotation period detection from the SAP LCs is so complicated that our exploratory method should be justified cautiously. To do so, we create a lot of simulated LCs with long-term trends to test the detection limits, and we found that our method could detect more long period objects but the lower amplitude the less detected. 
Then we used a selection process to distinguish the true or false positive periodical signals based on the channel sequence to identify the similarity of periods and LCs. 
Finally, we found more than 1000 rotators with the period longer than 30 days, including 165 newly detected objects. Most of our detection results are in agreement with previous studies. 
Considering the apertures are fixed in the SAP, we also inspect our newly found objects with the two photometry methods, and most of our results are confirmed.
Our method could detect long period signals effectively from the trend-dominated LCs, so with the data releasing of the TESS, we could also apply our method to the LCs in the TESS continuous viewing zones (CVZs) and find more long rotation period candidates to expand our sample for further studies.

\section*{Acknowledgements}
We thank both the editor and referee, for a number of helpful comments and corrections which greatly improved our article.
We thank all the people that have discussed with us about this work. This
includes but not limited to Hang Gong, Siming Liu, Han He, Shangbin Yang and all the people who are interested in my poster at IAU Symposium 340. 
We acknowledge support from the
Chinese Academy of Sciences (grant XDB09000000),  from the National Science Foundation of China
(NSFC, Nos. 11425313, 11603035, 11603038 and 11673035), and the National Program on Key Research and Development Project (Grant No. 2016YFA0400800).
The paper includes data collected by the \kepler mission. Funding for the \kepler mission is provided by theNASA Science Mission Directorate. 
All of the \kepler data presented in this paper were obtained from the MAST. STScI is operated by the Association of Universities for Research in Astronomy, Inc., under NASA contract NAS5-26555. 
Support for MAST for non-HST data is provided by the NASA Office of Space Science via grant NNX09AF08G and by other grants and contracts. 
Our parameters are revised by LAMOST (also named as Guoshoujing Telescope) AFGK Star Catalog. Guoshoujing Telescope (the Large Sky Area Multi-Object Fiber Spectroscopic Telescope LAMOST) is a National Major Scientific Project built by the Chinese Academy of Sciences. Funding for the project has been provided by the National Development and Reform Commission. LAMOST is operated and managed by the National Astronomical Observatories, Chinese Academy of Sciences.
This research made use of Lightkurve, a Python package for Kepler and TESS data analysis \citep{2018ascl.soft12013L}. LC visualisation is based on our Kepler Data Integration Platform (\url{http://kepler.bao.ac.cn}).
Data storage and computational resources are supported by Chinese Astronomical Data Center (CAsDC), Chinese Virtual Observatory (China-VO) and Astronomical 
Big Data Joint Research Center, co-founded by National Astronomical Observatories, Chinese Academy of Sciences and Alibaba Cloud. 




\bibliographystyle{mnras}
\bibliography{lrot_papers} 





\bsp	
\label{lastpage}
\end{document}